# Comparative Study of the Median Based Unit Rayleigh and its Generalized Form the Generalized Odd Median Based Unit Rayleigh

Iman M. Attia *

Imanattiathesis1972@gmail.com ,imanattia1972@gmail.com

*Department of Mathematical Statistics, Faculty of Graduate Studies for Statistical Research, Cairo University, Egypt

## Abstract

In the present paper, the author discusses the Generalized Odd Median Base Unit Rayleigh (GOMBUR) in relation to the Median Based Unit Rayleigh (MBUR) to evaluate the additive value of the new shape parameter on the estimation process as regards validity indices, goodness of fit statistics, estimated variances of the estimated parameters and their standard errors. This evaluation is conducted on real datasets. Each dataset is analyzed by fitting different competitor distributions in addition to MBUR and GOMBUR distributions. The parameter estimation is achieved by applying Maximum likelihood estimator (MLE) using Nelder Mead optimizer.

## Keywords

## Introduction

Generalizing a distribution by adding new parameters can undeniably enrich and amend the estimation process in numerous manners. First, adding parameters empowers the distribution to apprehend different data conducts such as skewness, kurtosis and different tail behaviors. For example, this is glorified when generalized gamma (J et al., 2019) distribution extends the gamma distribution to accommodate modeling wide range of data with different characteristics. The extra parameter in the generalized Pareto (Malik & Kumar, 2019) distribution controlling the tail behavior facilitates it to better analyze the extreme values. Second, the newly added parameters amplify the goodness of fit to empirical data and reduce the systematic bias. An example for such an effect, when the data exhibit extreme values; the four parameter Beta distribution (McGarvey et al., 2002) extending the standard beta distribution can better model these heavy tail features data.

Third, extending the distribution with new parameters augments it flexibility to more properly align with the basic data hence diminishing the bias in parameter estimation, obtaining more efficient estimators with minor variance and amend the implementation of MLE. The Generalized Weibull Distribution (Shama et al., 2023) enhances the estimates of failures rates in reliability studies. Fourth, the newly introduced parameters behave as regularizers that take control against over-fitting thus improving stability in estimation. The newly supplemented shape parameter in Exponentiated Weibull Distribution (Shama et al., 2023) assists to model both decreasing and increasing failure rates. The inserted skewness parameter in Skew-Normal Distribution (Arellano-Valle & Azzalini, 2022) promotes the distribution to better model the asymmetric data. Fifth, in real world analysis like Generalized Logistic Distribution (Aljarrah et al., 2020) implanting a shape parameter can regulate the rate of decay in growth models. The incorporated extra parameter in the Generalized Gamma (Gupta & Kundu, 1999) Distribution aids modeling diverse hazard rates.

The limitations of this generalization are increased computational complexity depending on the number of the embedded parameters which mandates more complex advanced optimization techniques. Increased number of parameters can lead to identifiability issues. Some parameters may have insufficient clear practical meaning. It is a matter of tradeoff between the benefits of adjoining new parameters and the anomalies that may arise with such approach.

The paper is structured as follows: Section 1 handles a quick reminder of the MBUR and GOMBUR version1 and 2. Section 2 points to the real datasets. Section 3 involves the exposing the results and tackling the discussion. Section 4 displays the conclusion. Section 5 contains the future work.

## Section1:

### 1.1 MBUR and GOMBUR

In previous work iman Attia (Iman M. Attia, 2024) discussed the MBUR. Y is a random variable distributed as MBUR with the following PDF, CDF respectively as shown in equation 1 and 2

$$f(y) = \frac{6}{\alpha^2}\left[1 - y^{\frac{1}{\alpha^2}}\right]y^{\left(\frac{2}{\alpha^2}-1\right)}, \quad 0 < y < 1, \quad \alpha > 0 \dots\dots\dots\dots\dots\dots\dots\dots\dots\dots\dots.(1)$$

$$F(y) = 3y^{\frac{2}{\alpha^2}} - 2y^{\frac{3}{\alpha^2}}, \quad 0 < y < 1, \quad \alpha > 0 \dots\dots\dots\dots\dots\dots\dots\dots\dots\dots\dots\dots\dots\dots.(2)$$

In previous work iman Attia (Iman M. Attia, 2025) discussed the GOMBUR. Y is a random variable distributed as GOMBUR version 1 and version 2 with the following PDFs respectively as shown in equation 3 and 4

$$f_y(y) = \frac{\Gamma(2n+2)}{\Gamma(n+1)\Gamma(n+1)} \frac{1}{\alpha^2} \left[1 - y^{\alpha^{-2}}\right]^n [y]^{\frac{n+1}{\alpha^2} - 1}, n \geq 0, \alpha > 0, 0 < y < 1 \ldots \ldots \ldots (3)$$

$$f_y(y) = \frac{\Gamma(n+1)}{\Gamma\left(\frac{n}{2} + \frac{1}{2}\right)\Gamma\left(\frac{n}{2} + \frac{1}{2}\right)} \frac{1}{\alpha^2} \left[1 - y^{\alpha^{-2}}\right]^{\frac{n-1}{2}} [y]^{\frac{n+1}{2\alpha^2} - 1}, n \geq 1, \alpha > 0, 0 < y < 1 \ldots \ldots (4)$$

And the following CDFs for version 1 and version 2 respectively as shown in equation 5

$$P(Y < y) = I_w(n+1, n+1) \text{ for version 1 \& } P(Y < y) = I_w\left(\frac{n+1}{2}, \frac{n+1}{2}\right) \text{ for version 2.} \ldots (5)$$

## 1.2 Maximum Likelihood Estimation

Let Y is a random variable having the PDF of GOMBUR-1. To derive the MLE for version 1, for one observation taking the log of equation 3 results into equation 6:

$$l(y; \alpha, n) =$$

$$\ln \Gamma(2n+2) - \ln \Gamma(n+1) - \ln \Gamma(n+1) + \ln \alpha^{-2} + n \ln\left[1 - y_i^{\alpha^{-2}}\right] + \left(\frac{n+1}{\alpha^2} - 1\right) \ln y_i \ldots (6)$$

Take the first derivative of equation 6 with respect to n and alpha parameter yields equation 7 and 8 resectively:

$$\frac{\partial l}{\partial n} = 2 * \psi(2n+2) - \psi(n+1) - \psi(n+1) + \ln\left[1 - y_i^{\alpha^{-2}}\right] + \alpha^{-2} \ln y_i \ldots \ldots (7)$$

$$\frac{\partial l}{\partial n} = \frac{-2}{\alpha} + \left(\frac{2n}{\alpha^3}\right) \frac{y_i^{\alpha^{-2}} \ln y_i}{1 - y_i^{\alpha^{-2}}} - \left(\frac{2[n+1]}{\alpha^3}\right) \ln y_i \ldots \ldots \ldots (8)$$

Let Y is a random variable having the PDF of GOMBUR-2. To derive the MLE for version 2, for one observation taking the log of equation 4 results into equation 9:

$$l(y; \alpha, n) =$$

$$\ln \Gamma(n+1) - \ln \Gamma\left(\frac{n+1}{2}\right) - \ln \Gamma\left(\frac{n+1}{2}\right) + \ln \alpha^{-2} + \left(\frac{n-1}{2}\right) \ln\left[1 - y_i^{\alpha^{-2}}\right] + \left(\frac{n+1}{2\alpha^2} - 1\right) \ln y_i \ldots (9)$$

Take the first derivative of equation 9 with respect to n and alpha parameter yields equation 10 and 11 respectively:

$$\frac{\partial l}{\partial n} = \psi(n+1) - \frac{1}{2}\psi\left(\frac{n+1}{2}\right) - \frac{1}{2}\psi\left(\frac{n+1}{2}\right) + \frac{1}{2}\ln\left[1 - y_i^{\alpha^{-2}}\right] + \frac{1}{2}\alpha^{-2} \ln y_i \ldots \ldots (10)$$

$$\frac{\partial l}{\partial n} = \frac{-2}{\alpha} + \left(\frac{n-1}{\alpha^3}\right) \frac{y_i^{\alpha^{-2}} \ln y_i}{1 - y_i^{\alpha^{-2}}} - \left(\frac{n+1}{\alpha^3}\right) \ln y_i \ldots \ldots \ldots (11)$$

For each version set the above equations to zero and since they are non-linear equations, numerical methods like quasi-Newton method can be used as a solution.

## Section 2

### Description of the Real Data:

The real data used in this paper can be found in the appendix A at the end of the paper; see Table (A). These are 14 datasets. The first five datasets were used by the author (Iman M. Attia, 2024) in previous work. The first dataset is the dwelling without basic facilities. The second one is the quality support network. The third one is the educational attainment dataset. The fourth one is the flood data. The fifth one is the time between failures dataset. The sixth dataset is the COVID-19 death rate in Canada previously discussed by (Nasiru et al., 2022) . The seventh dataset is the COVID-19 death rate in Spain previously discussed by (Ahmed Z. Afify et al., 2022) . The eighth dataset is the COVID-19 death rate in United Kingdom previously discussed by (Nasiru et al., 2022) . The ninth dataset is the 48 rock samples from a petroleum reservoir previously discussed by (Gauss Moutinho Cordeiro & Rejane dos Santos Brito, 2010) . The tenth dataset is the daily snowfall amounts of 30 observations measured in inches of water taken from non-seeded experimental units, which was conducted in the vicinity of Climax Colorado previously discussed by (Ishaq et al., 2023). The eleventh dataset is about the total milk production in the first birth 107 cows from SINDI race. These cows are property of the Carnauba farm located in Brazil previously mentioned by (Gauss Moutinho Cordeiro & Rejane dos Santos Brito, 2010). The twelfth dataset is the COVID-19 recovery rate in Spain previously mentioned by (Ahmed Z. Afify et al., 2022). The thirteen dataset is the voter turnout dataset previously mentioned by (Iman M.Attia, 2024) . The fourteen dataset is the unit capacity factors previously mentioned by (Maya et al., 2024) .

## Section 3:

### Real Data Analysis

The analysis of the data sets points to conclude how these sets support the following distributions: Beta, Topp Leone, Unit Lindely, Kumaraswamy. The fitting of these data sets will be compared to the fitting of the new MBUR distribution and its generalized forms GOMBUR-1 and GOMBUR-2. The metrics used for this evaluation include the following: LL (log-likelihood), Akaike Information Criterion (AIC), corrected AIC (CAIC), Bayesian Information Criterion (BIC), and Hannan-Quinn Information Criterion (HQIC). Additionally, the Kolmogorov-Smirnov (K-S) test will be conducted. The test's results will include its value, along with the outcome of the null hypothesis (H0), which proposes that the data set follows the investigated distribution; if this assumption is not encountered, the null hypothesis will be rejected. The P-value for the test will also be documented. Furthermore, the Cramér-von Mises test and the Anderson-Darling test will be implemented, with their respective values stated. Figures depicting the empirical cumulative distribution function (eCDF) and the theoretical cumulative distribution functions (CDF) of the distributions will be illustrated, each in its place. Finally, the values of the estimated parameters, along with their estimated variances and standard errors, will be reported. The competitors' distributions are:

1- Beta Distribution:

$$f(y;\alpha,\beta) = \frac{\Gamma(\alpha+\beta)}{\Gamma(\alpha)\Gamma(\beta)} y^{\alpha-1}(1-y)^{\beta-1}, 0 < y < 1, \alpha > 0, \beta > 0$$

2- Kumaraswamy Distribution:

$$f(y;\alpha,\beta) = \alpha\beta y^{\alpha-1}(1-y^\alpha)^{\beta-1}, 0 < y < 1, \quad \alpha > 0, \beta > 0$$

3- Median Based Unit Rayleigh:

$$f(y;\alpha) = \frac{6}{\alpha^2}\left[1 - y^{\frac{1}{\alpha^2}}\right] y^{\left(\frac{2}{\alpha^2}-1\right)}, \ 0 < y < 1, \quad \alpha > 0$$

4- Topp-Leone Distribution:

$$f(y;\theta) = \theta(2-2y)(2y-y^2)^{\theta-1}, 0 < y < 1, \quad \theta > 0$$

5- Unit-Lindley:

$$f(y;\theta) = \frac{\theta^2}{1+\theta}(1-y)^3 \exp\left(\frac{-\theta y}{1-y}\right), \quad 0 < y < 1, \quad \theta > 0$$

Comparison tools are: (k) is the number of parameter, (n) is the number of observations.

$$AIC = -2MLL + 2k \quad, \quad CAIC = -2MLL + \frac{2k}{n-k-1} \quad, \quad BIC = -2MLL + k\log(n)$$

$$HQIC = -2\log L + 2k * \ln[\ln(n)]$$

$$KS - test = Sup_n |F_n - F|, \quad F_n = \frac{1}{n}\sum_{i=1}^{n} I_{x_i < x}$$

$$Cramer - Von - Mises - test(CVM) = \frac{1}{12n} + \sum_{i=1}^{n}\left\{F(x_i) - \frac{2i-1}{2n}\right\}^2$$

$$Anderson - Darling - test(AD) = -n - \sum_{i=1}^{n}\left(\frac{2i-1}{n}\right)\{log[F(x_i)] + log[1 - F(x_{n-i+1})]\}$$

## 3.1 Descriptive statistics

Table (1) illustrates the descriptive statistics for the above datasets:

| Data number | min | mean | std | skewness | kurtosis | Q(25) | Q(50) | Q(75) | max |
|---|---|---|---|---|---|---|---|---|---|
| 1 | 0.001 | 0.0345 | 0.0560 | 2.5981 | 10.9552 | 0.0032 | 0.0070 | 0.0455 | 0.2590 |
| 2 | 0.7700 | 0.9005 | 0.0640 | -0.9147 | 2.6716 | 0.8650 | 0.9200 | 0.9500 | 0.9800 |
| 3 | 0.4200 | 0.7894 | 0.1504 | -1.3554 | 3.9461 | 0.7500 | 0.8400 | 0.8950 | 0.9400 |
| 4 | 0.2600 | 0.4225 | 0.1244 | 1.1625 | 4.2363 | 0.3300 | 0.4050 | 0.4650 | 0.7400 |
| 5 | 0.0062 | 0.1578 | 0.1931 | 1.4614 | 3.9988 | 0.0292 | 0.0614 | 0.2100 | 0.6560 |
| 6 | 0.1159 | 0.2305 | 0.0520 | -0.0897 | 2.7360 | 0.2011 | 0.2262 | 0.2678 | 0.3347 |
| 7 | 0.1372 | 0.2760 | 0.1086 | 0.7214 | 2.6670 | 0.2023 | 0.2467 | 0.3545 | 0.5714 |
| 8 | 0.0807 | 0.2888 | 0.1167 | 0.0502 | 1.9766 | 0.1761 | 0.2884 | 0.3879 | 0.5331 |
| 9 | 0.0903 | 0.2310 | 0.1299 | 3.3611 | 18.5254 | 0.1621 | 0.1989 | 0.2627 | 0.9103 |
| 10 | 0.0050 | 0.0962 | 0.1143 | 2.2227 | 8.1749 | 0.0200 | 0.0625 | 0.1100 | 0.4950 |
| 11 | 0.0168 | 0.4689 | 0.1920 | -0.3401 | 2.7292 | 0.3509 | 0.4741 | 0.5994 | 0.8781 |
| 12 | 0.4286 | 0.7240 | 0.1086 | -0.7214 | 2.6670 | 0.6455 | 0.7533 | 0.7977 | 0.8628 |
| 13 | 0.4500 | 0.6911 | 0.1251 | -0.0456 | 2.2538 | 0.6200 | 0.6800 | 0.7800 | 0.9200 |
| 14 | 0.0060 | 0.2881 | 0.3181 | 0.8223 | 2.0246 | 0.0308 | 0.1160 | 0.5310 | 0.8660 |

Figures 1-3 illustrates the boxplot of the data sets

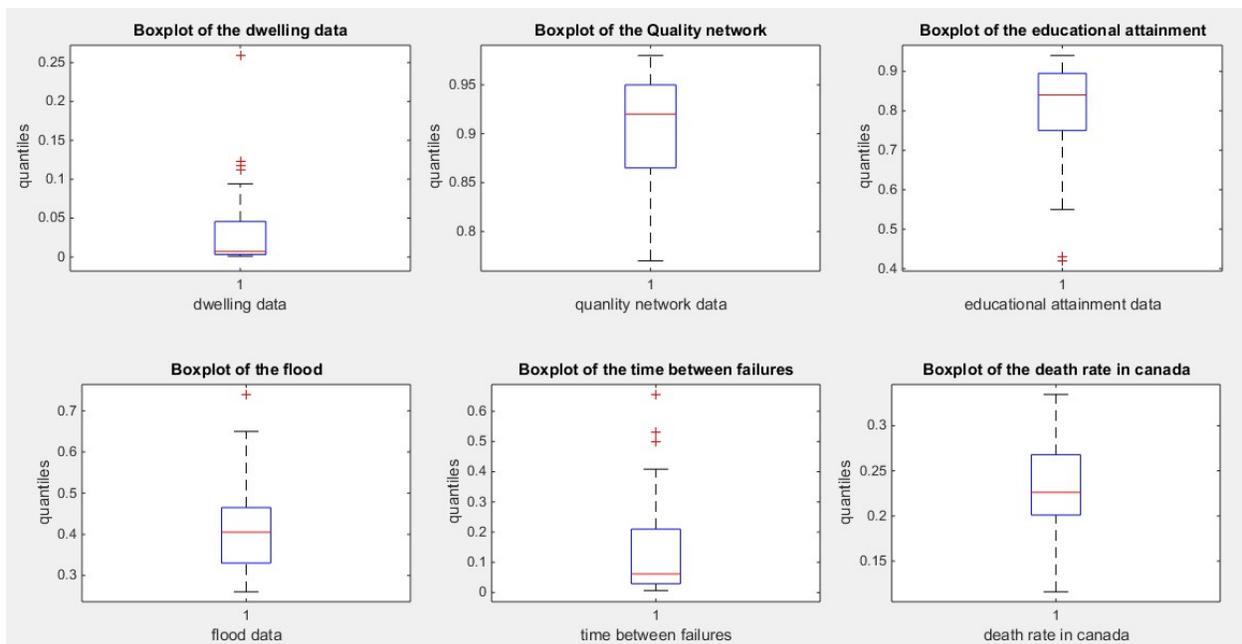

Fig. 1 illustrates the boxplot of the first 6 datasets.

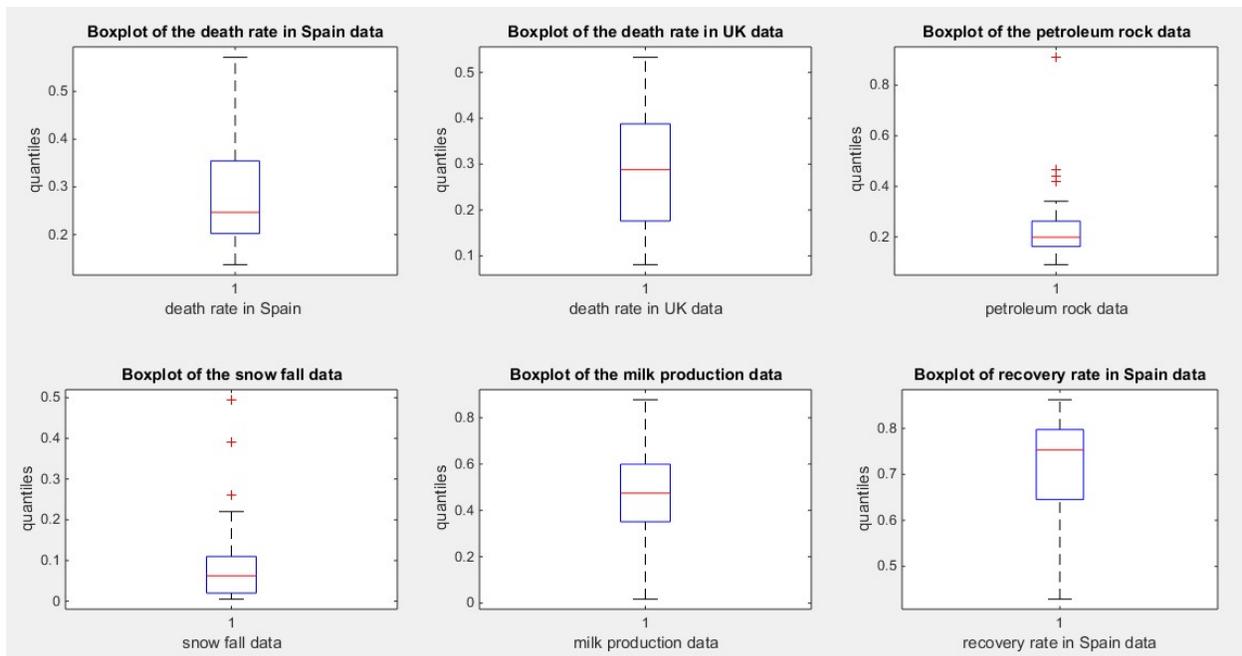
Fig. 2 illustrates the boxplot of the second 6 datasets

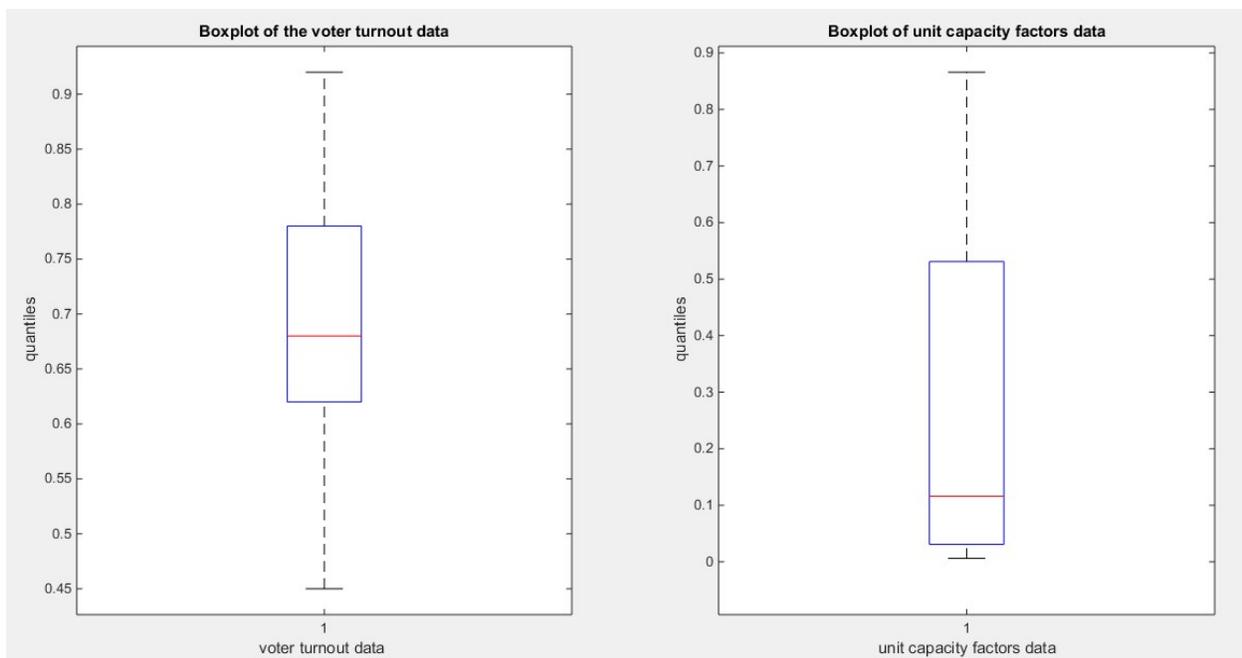
Fig. 3 illustrates the boxplot of the last 2 datasets

Table (1) and Figures 1-3 illustrate different shapes and characteristics of data.

## 3.2. Individual Dataset Analysis (Results & Discussion)

For each datasets a table showing the results of the fitted distribution and 6 figures highlight the fitted CDFs and PDFs for the tested distributions, the PP plots of these distributions and the MBUR, GOMBUR-1& 2 versions. There is an additional figure to compare the fitted CDFs and the PDFs of the two versions. Table (2) discusses the results of the dwelling without basic facilities datasets.

Table (2) : dwelling data set analysis

|  | Beta | | Kumaraswamy | | MBUR | Topp-Leone | Unit-Lindley |
|---|---|---|---|---|---|---|---|
| theta | $\alpha = 0.5086$ | | $\alpha = 0.6013$ | | 2.3519 | 0.2571 | 26.1445 |
|  | $\beta = 14.036$ | | $\beta = 8.5999$ | | | | |
| Var | .0323 | .6661 | .0086 | .2424 | 0.023 | 0.0021 | 20.5623 |
|  | .6661 | 22.1589 | .2424 | 9.228 | | | |
| SE(a) | 0.03227 | | 0.01666 | | 0.0272 | 0.0082 | 0.8144 |
| SE(b) | 0.8455 | | 0.5456 | | - | - | - |
| AIC | -153.5535 | | -155.8979 | | -143.3057 | -133.593 | -140.592 |
| CAIC | -153.1249 | | -155.4693 | | -143.1678 | -133.4551 | -140.454 |
| BIC | -150.6855 | | -153.0299 | | -141.8717 | -132.159 | -139.158 |
| HQIC | -152.6186 | | -154.963 | | -142.8382 | -133.1255 | -140.1243 |
| LL | 78.7767 | | 79.9489 | | 72.6528 | 67.7965 | 71.2959 |
| K-S Value | 0.2052 | | 0.1742 | | 0.2034 | 0.2818 | 0.3762 |
| $H_0$ | Fail to reject | | Fail to reject | | Fail to reject | reject | reject |
| P-value | 0.1271 | | 0.271 | | 0.1336 | 0.0114 | 0.000189 |
| AD | 1.288 | | 0.9566 | | 2.8789 | 4.6021 | 7.439 |
| CVM | 0.2419 | | 0.1659 | | 0.4976 | 0.8749 | 1.1683 |
| Determinant | 0.2728 | | 0.0205 | | - | - | - |

Fitting GOMBUR-1 & GOMBUR-2 introduces a reduction in the covariance between the parameters alpha and n which indicates less association between the 2 parameters. This is in comparison with the high association seen between the 2 parameters of the Beta and Kumaraswamy distributions relative to that seen in the GOMBUR1 & GOMBUR-2

Table (2) to be continued

|  | GOMBUR-1 | | GOMBUR-2 | |
|---|---|---|---|---|
| theta | $n = 5.7248$ | | $n = 12.4496$ | |
|  | $\alpha = 2.4988$ | | $\alpha = 2.4988$ | |
| Var | 2.7974 | 0.0255 | 11.1896 | 0.0509 |
|  | 0.0255 | 0.008 | 0.0509 | 0.008 |
| SE(n) | 0.3004 | | 0.6008 | |
| SE(a) | 0.0161 | | 0.0161 | |
| AIC | -158.1462 | | -158.1462 | |
| CAIC | -157.7176 | | -157.7176 | |
| BIC | -155.2782 | | -155.2782 | |
| HQIC | -157.2113 | | -157.2113 | |
| LL | 81.0731 | | 81.0731 | |
| K-S Value | 0.1654 | | 0.1654 | |
| $H_0$ | Fail to reject | | Fail to reject | |
| P-value | 0.3279 | | 0.3279 | |
| AD | 0.8727 | | 0.8727 | |
| CVM | 0.1515 | | 0.1515 | |
| Determinant | 0.0218 | | 0.0874 | |
| Significant(n) | P<0.001 | | P<0.001 | |
| Significant(a) | P<0.001 | | P<0.001 | |

Table 2 shows marked improvement in the indices for the fitted GOMBUR-1 & GOMBUR-2 with increased negativity of AIC, CAIC, BIC and HQIC to levels more than that of beta, Kumaraswamy and MBUR. Log-likelihood increased to levels above that of the beta, kumaraswamy and MBUR. Tests statistics of KS, AD and CVM show lowered values than those of beta, kumaraswamy and MBUR. The determinants of both GOMBUR-1 & GOMBUR-2 are less than that of the Beta distribution but higher than that of the Kumaraswamy distribution. The overall assessment of generalization shows that both GOMBUR-1 & GOMBUR-2 outperform the beta, Kumaraswamy and MBUR distribution. Figure 1 shows how the fitted GOMBUR-1 CDF aligns better with the fitted CDFs of both Kumaraswamy and Beta distributions than the fitted CDF of BMUR. Figure 2 also shows how the fitted GOMBUR-1 PDF aligns better than with the fitted PDFs of both Beta and Kuamraswamy distributions than the fitted PDF of the MBUR. The determinant of GOMBUR-1 is less than that of the GOMBUR-2 which points to the fact that the GOMBUR-1 is more efficient than GOMBUR-2 in fitting the data for the same level of other indices like AIC & BIC and tests like KS, AD & CVM. GOMBUR-1 has higher AIC & BIC and log-likelihood than that of the Kumaraswamy, although Kumaraswamy has less determinant than that of GOMBUR-1. It is a tradeoff between the biasness and efficiency between the 2 distributions.

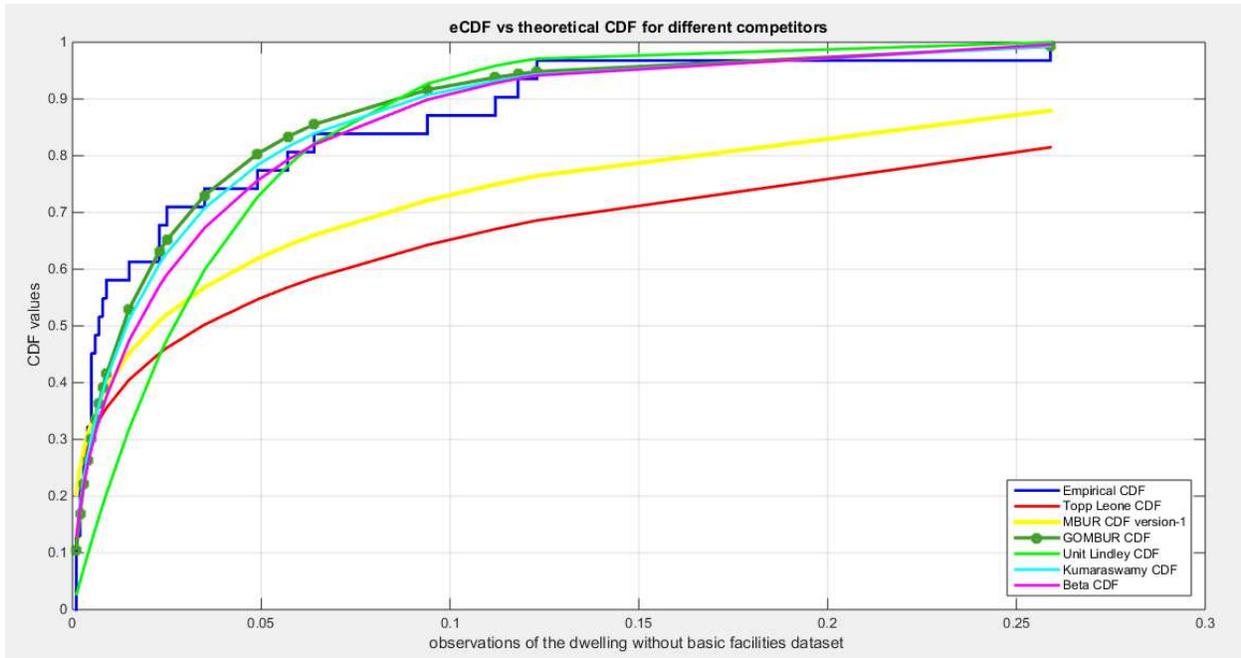

Fig. 1 shows the e-CDFs and the theoretical CDFs for the fitted distributions of dwelling data.

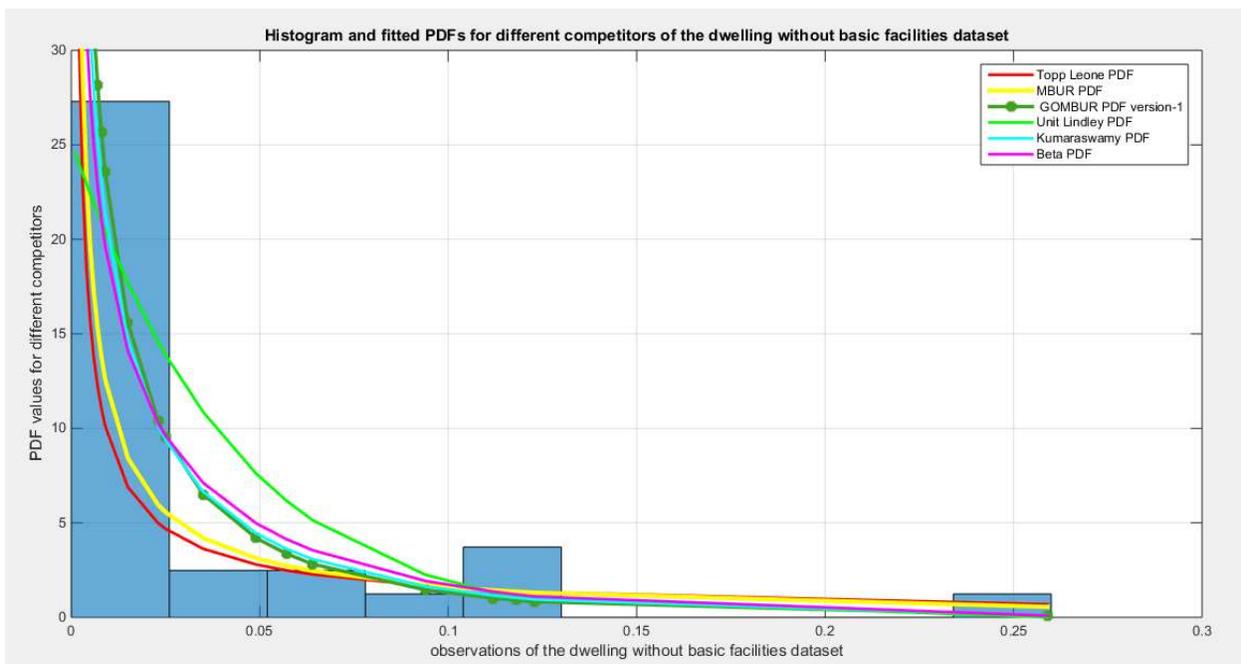

Fig.2 shows the histogram of the dwelling data and the theoretical PDFs for the fitted distributions

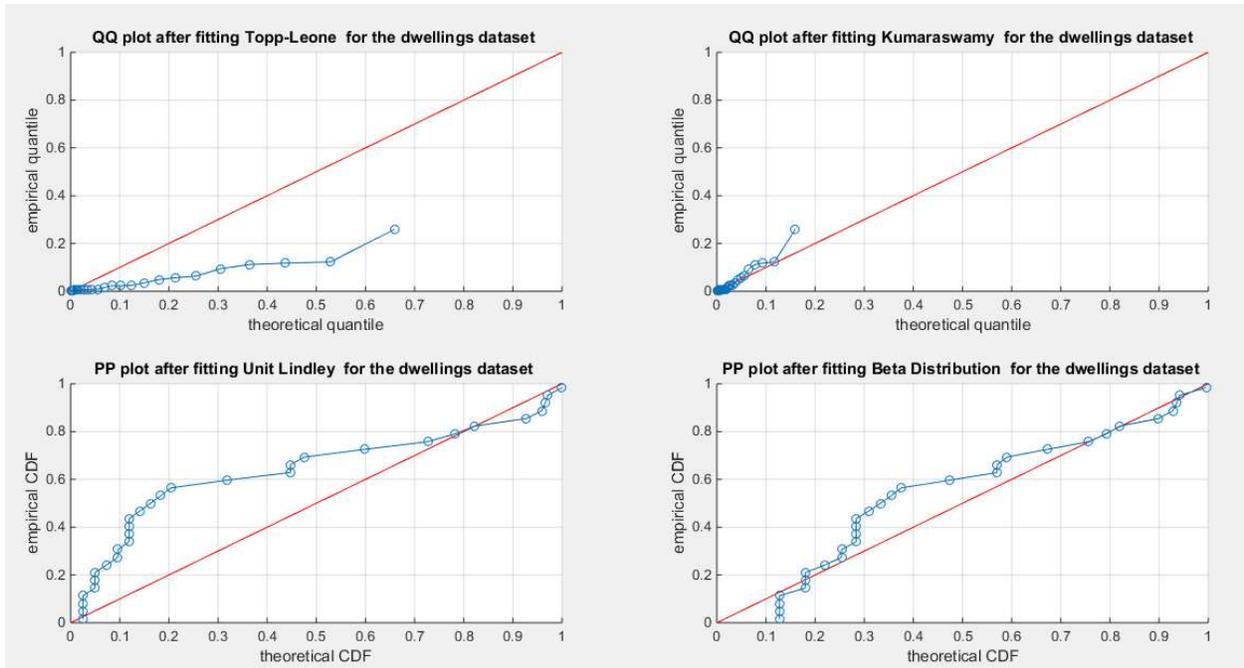

Fig. 3 shows the QQ plot for the fitted Topp Leone & Kumaraswamy distributions and the PP plot for the fitted Unit Lindley and Beta distribution as regards dwelling dataset.

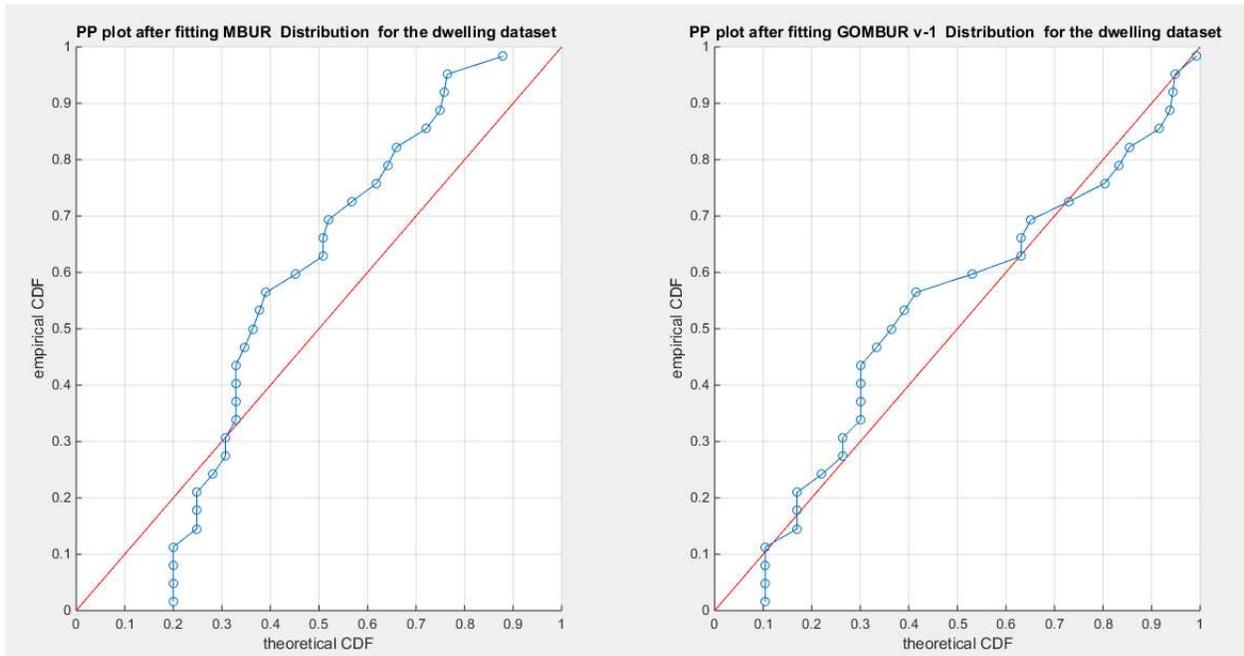

Fig. 4 shows the PP plot for the fitted MBUR & GOMBUR-1 as regards dwelling dataset. GOMBUR-1 on the right side are better aligned with the diagonal especially at the upper and lower tails than the center, in comparison with the MBUR distribution on the left subplot. This is attributed to the better fit of the generalized version than the fit of the MBUR distribution.

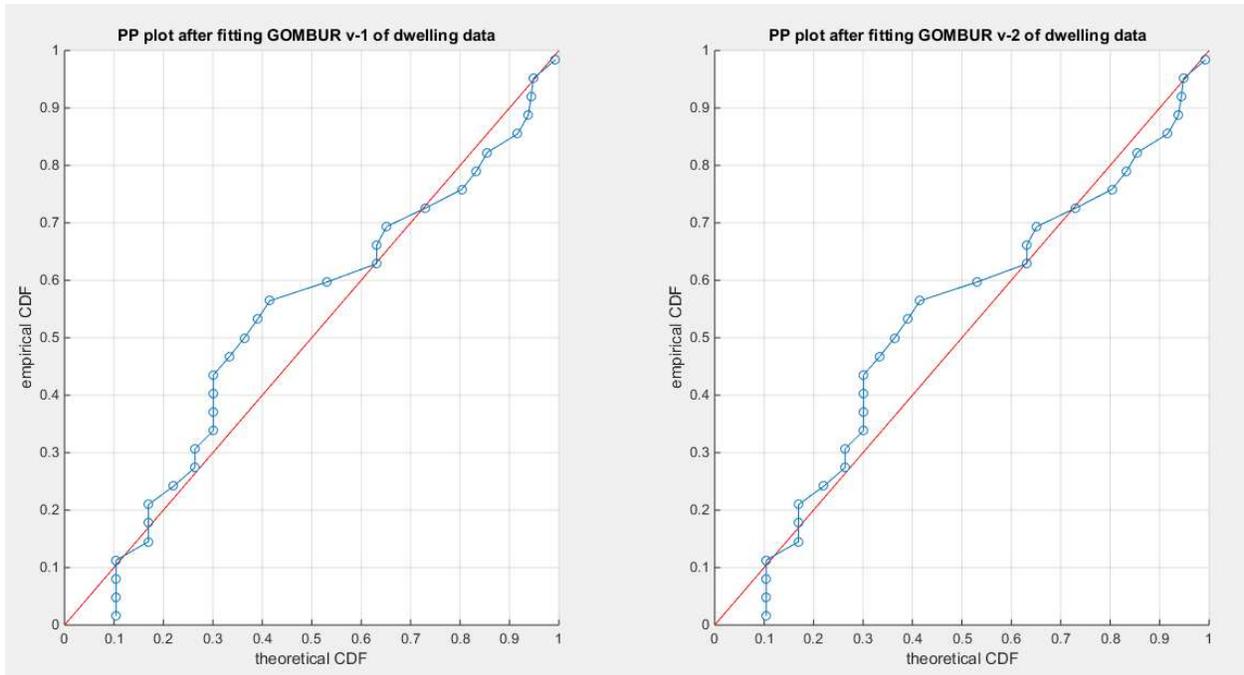

Fig.5 shows the PP plot for the fitted GOMBUR-1 & GOMBUR-2 as regards dwelling dataset. They are identical.

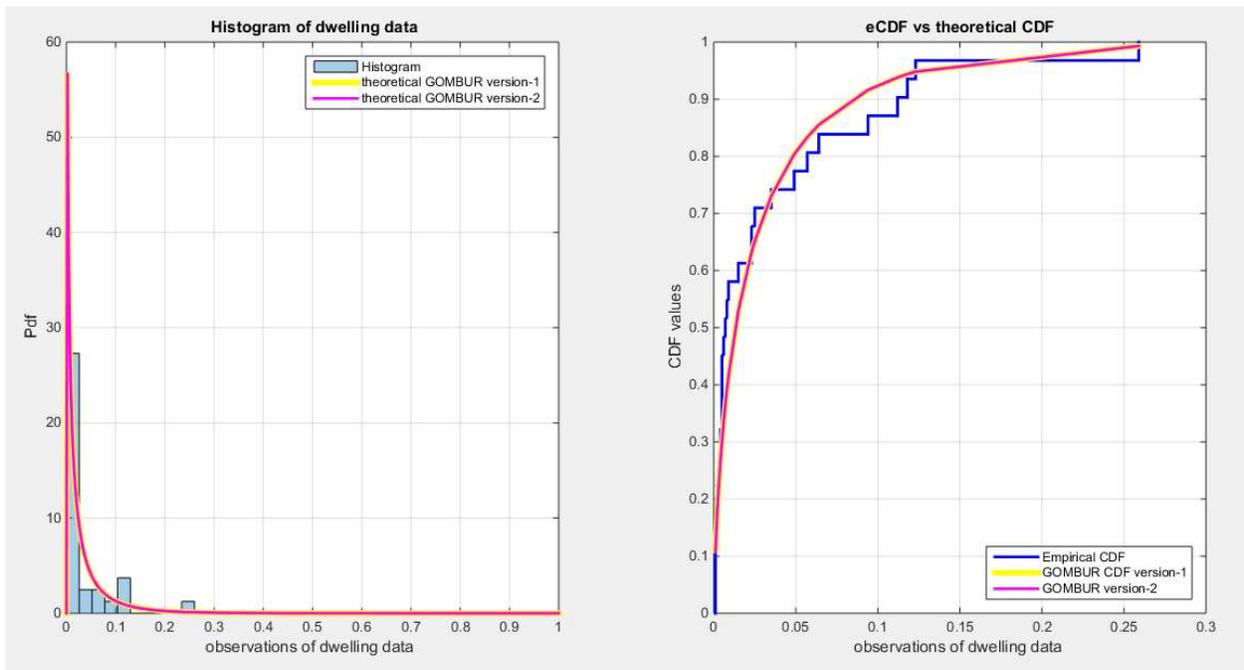

Fig.6 shows on the left subplot the histogram of the dwelling data and the fitted PDFs of both GOMBUR-1 & GOMBUR-2 and on the right subplot the e-CDFs and the theoretical CDFs for both distributions. Both the fitted CDFs and the fitted PDFs of both versions are identical.

Figure 3 shows the PP plot of the fitted Beta distribution and the QQ plot for the fitted Kumaraswamy. Figure 4-5 show PP plot comparison between the fitted MBUR, GOMBUR-1

& GOMBUR-2 distributions. Figure 6 shows a comparison between the theoretical CDF and PDF of the fitted GOMBUR 1 & 2 for the dwelling dataset. The fitted CDFs and PDFs of both versions are identical and this is reflected on the indices obtained by estimation. Table 3 illustrates the Quality support network analysis results. Table 3 shows the results of quality network analysis.

Table (3): analysis of the quality support network

|  | Beta | | Kumaraswamy | | MBUR | Topp-Leone | Unit-Lindley |
|---|---|---|---|---|---|---|---|
| theta | $\alpha = 21.7353$ | | $\alpha = 16.5447$ | | 0.3591 | 71.2975 | 0.1334 |
|  | $\beta = 2.4061$ | | $\beta = 2.772$ | | | | |
| Var | 86.461 | 9.0379 | 15.7459 | 3.2005 | 0.000837 | 254.1667 | 0.00045 |
|  | 9.0379 | 1.0646 | 3.2005 | 1.0347 | | | |
| SE(a) | 2.079 | | 0.8873 | | 0.0063 | 3.565 | 0.0047 |
| SE(B) | 0.231 | | 0.2275 | | | | |
| AIC | -56.5056 | | -56.7274 | | -58.079 | -56.6796 | -57.3746 |
| CAIC | -55.7997 | | -56.0215 | | -57.8567 | -56.4574 | -57.1523 |
| BIC | -54.5141 | | -54.7359 | | -57.0832 | -55.6839 | -56.3788 |
| HQIC | -56.1168 | | -56.3386 | | -57.8846 | -56.4852 | -57.1802 |
| LL | 30.2528 | | 30.3637 | | 30.0395 | 29.3398 | 29.6873 |
| K-S | 0.0974 | | 0.0995 | | 0.1309 | 0.1327 | 0.1057 |
| H$_0$ | Fail to reject | | Fail to reject | | Fail to reject | Fail to reject | Reject to reject |
| P-value | 0.9416 | | 0.9513 | | 0.8399 | 0.4627 | 0.954 |
| AD | 0.3828 | | 0.3527 | | 0.3184 | 0.9751 | 0.2749 |
| CVM | 0.0566 | | 0.0498 | | 0.0407 | 0.1719 | 0.0261 |
| determinant | 10.3649 | | 6.0494 | | - | - | - |

Table 4 shows that the 5 distributions fit the data and the MBUR outperform all of them followed by the Unit Lindley. Figure 7 shows more or less perfect alignment of the theoretical CDFs of Beta, Kumaraswamy and GOMBUR-1 distributions, although all the fitted distributions align well with each other. Figure 8 shows the perfect alignment of the graphs of the PDFs of the three distributions: the Beta, Kumaraswamy and GOMBUR-1 distributions, this can be a reflection of having nearly equaled AIC, CAIC, BIC & HQIC. Figure 9-11 shows varieties of graphs of PP plots and QQ plots for the competitor distributions. Figure 12 illustrates the comparison between the fitted CDFs and fitted PDFs of both GOMBUR-1 and 2.

Table (3) to be continued:

|  | GOMBUR-1 | | GOMBUR-2 | |
|---|---|---|---|---|
| theta | $n = 1.4765$ | | $n = 3.9529$ | |
|  | $\alpha = 0.3651$ | | $\alpha = 0.3651$ | |
| Var | 0.5554 | 0.0056 | 2.2216 | 0.0113 |
|  | 0.0056 | 0.00075961 | 0.0113 | 0.00075961 |
| SE(n) | 0.1666 | | 0.3333 | |
| SE(a) | 0.0062 | | 0.0062 | |
| AIC | -56.5514 | | -56.5514 | |
| CAIC | -55.8455 | | -55.8455 | |
| BIC | -54.5599 | | -54.5599 | |
| HQIC | -56.1626 | | -56.1626 | |
| LL | 30.2757 | | 30.2757 | |
| K-S Value | 0.0977 | | 0.0977 | |
| H₀ | Fail to reject | | Fail to reject | |
| P-value | 0.9458 | | 0.9458 | |
| AD | 0.3728 | | 0.3728 | |
| CVM | 0.0545 | | 0.0545 | |
| Determinant | 0.00039 | | 0.0016 | |
| Significant(n) | 0 | | 0 | |
| Significant(a) | 0 | | 0 | |

Analysis shows that although the metrics (AIC, CAIC, BIC & HQIC) for MBUR and Unit Lindley are slightly higher than those of the GOMBUR-1 but the determinant for the latter is less than the variance of the MBUR and the variance of the Unit Lindley indicating more efficiency. Both MBUR and the generalized form fit the data better than the Beta distribution. The determinants of both Beta and Kumaraswamy are higher than the determinants of the generalized forms of MBUR. The determinant for GOMBUR-1 is lesser than the determinant for the GOMBUR-2 which suggests better efficiency of GOMBUR-1 over the GOMBUR-2. Fitting both GOMBUR-1 & GOMBUR-2 yields the same estimate of alpha, variance and standard error. Both parameters of the generalized forms are highly significant.

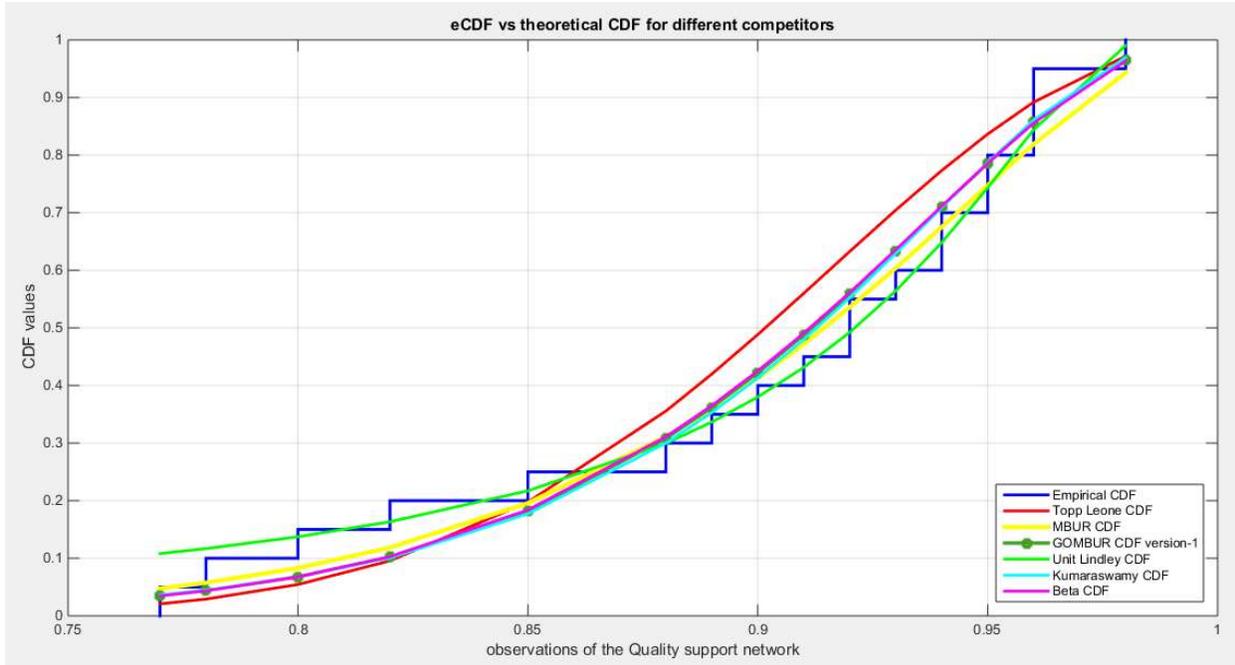

Fig. 7 shows the e-CDFs and the theoretical CDFs for the fitted distributions of quality data.

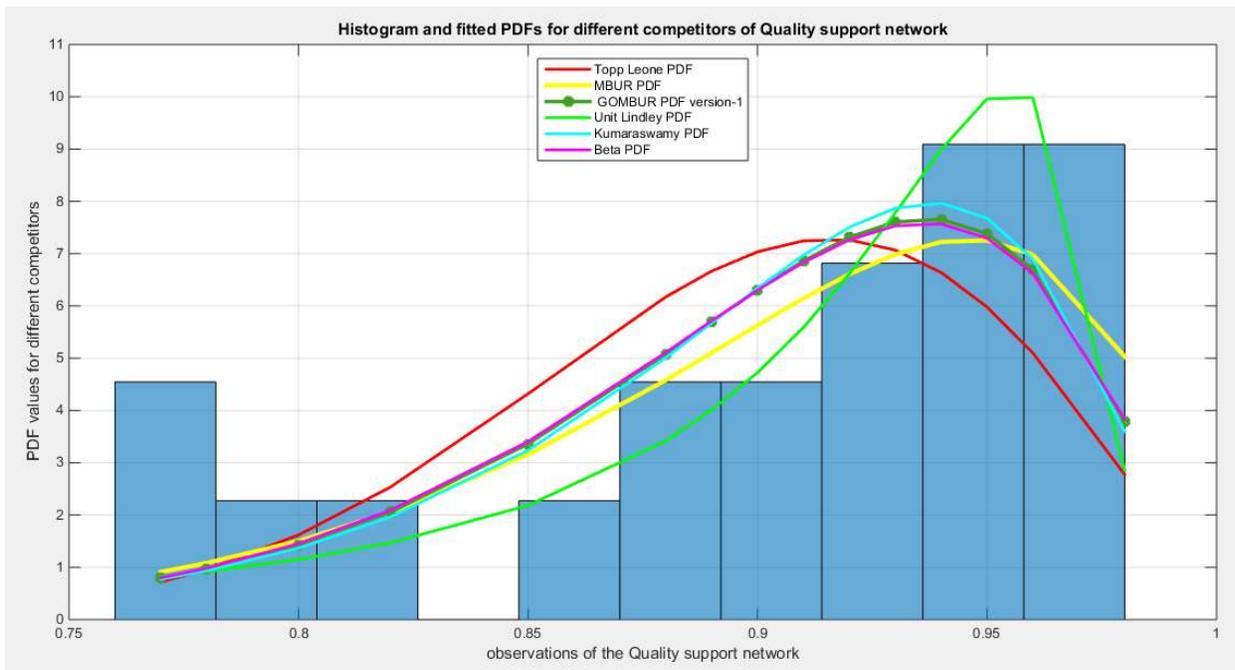

Fig. 8 shows the histogram of the quality data and the theoretical PDFs for the fitted distributions

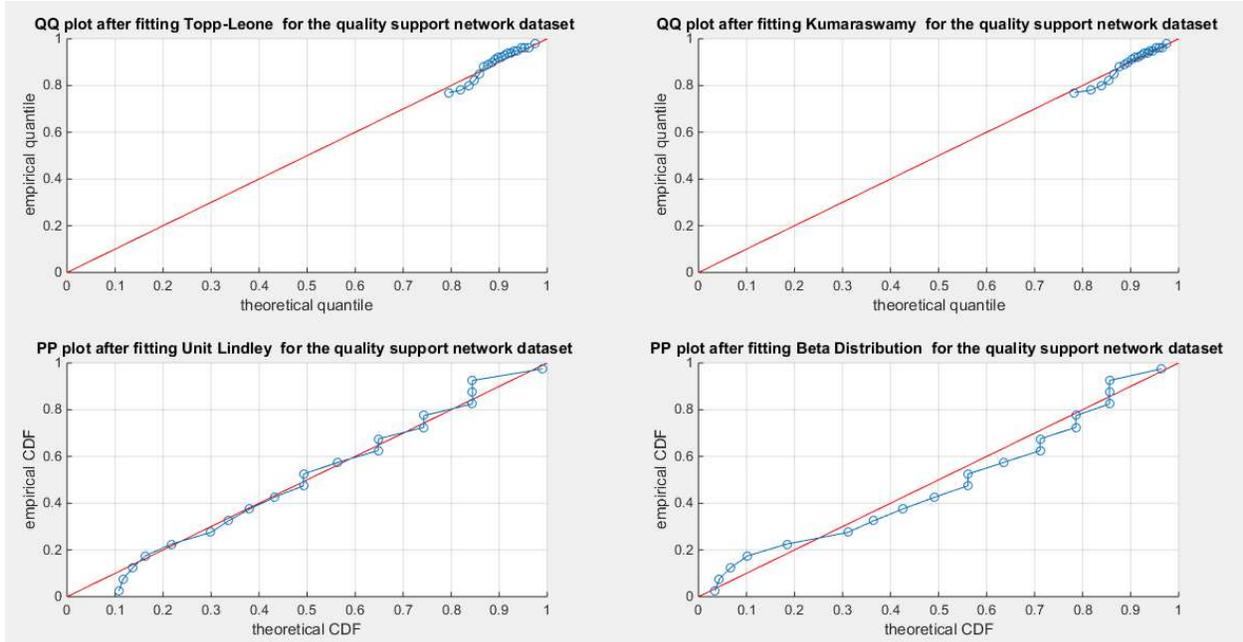

Fig. 9 shows the QQ plot for the fitted Topp Leone & Kumaraswamy distributions and the PP plot for the fitted Unit Lindley and Beta distribution as regards quality dataset.

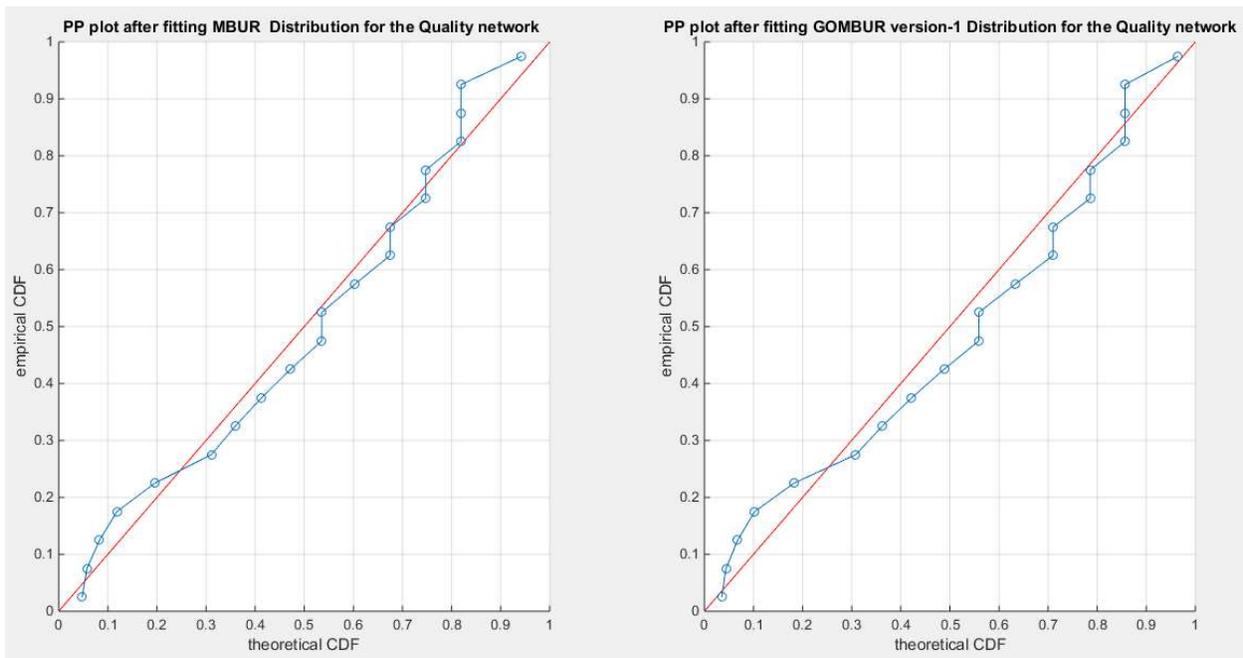

Fig. 10 shows the PP plot for the fitted MBUR & GOMBUR-1 as regards quality dataset

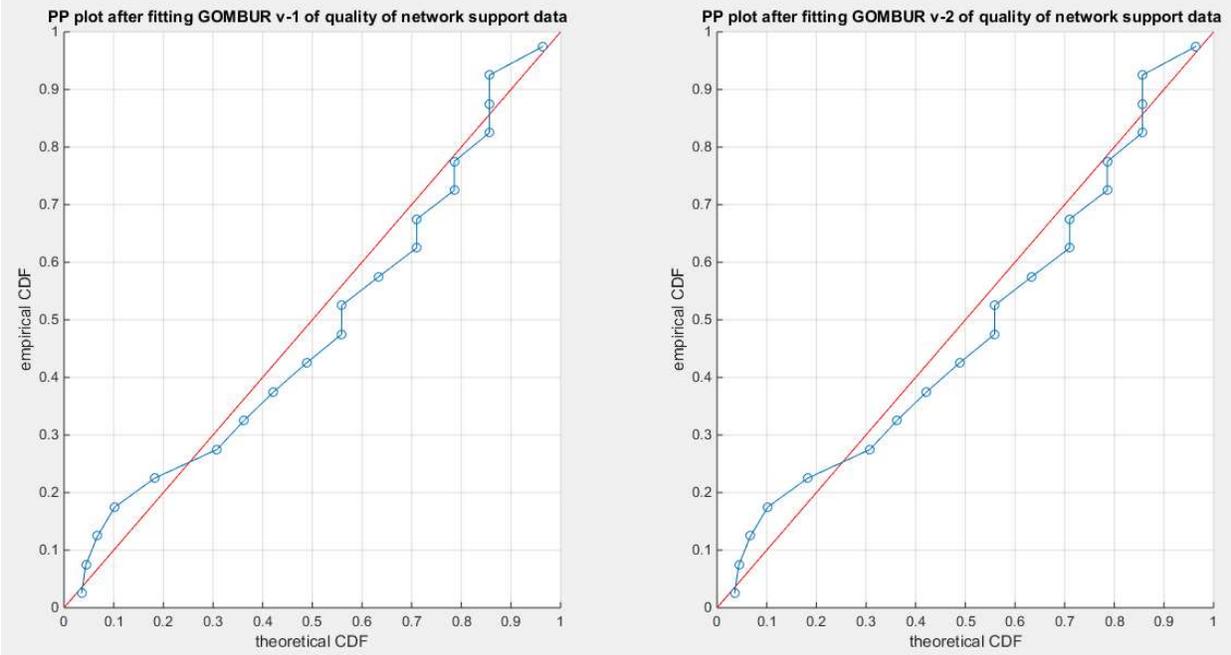

Fig.11 shows the PP plot for the fitted GOMBUR-1 & GOMBUR-2 as regards quality dataset. They are identical.

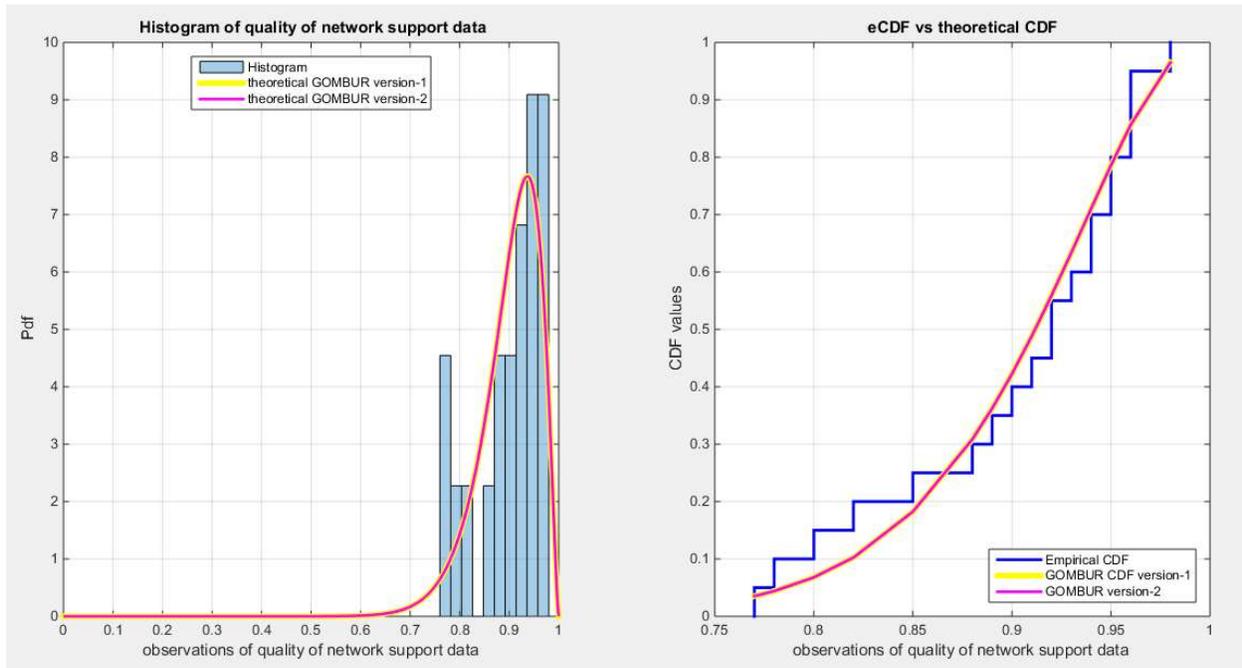

Fig.12 shows on the left subplot the histogram of the quality data and the fitted PDFs of both GOMBUR-1 & GOMBUR-2 and on the right subplot the e-CDFs and the theoretical CDFs for both distributions. Both the fitted CDFs and the fitted PDFs of both versions are identical.

Table 4 shows the result of analysis of the educational attainment dataset.

Table (4): Educational attainment data analysis results

|  | Beta | | Kumaraswamy | | MBUR | Topp-Leone | Unit-Lindley |
|---|---|---|---|---|---|---|---|
| theta | $\alpha = 6.7222$ | | $\alpha = 6.0746$ | | 0.5556 | 13.4254 | 0.2905 |
| | $\beta = 1.8405$ | | $\beta = 2.1284$ | | | | |
| Variance | 3.4283 | 1.0938 | 1.3854 | 0.5232 | 0.0011 | 5.0067 | 0.0012 |
| | 1.0938 | 0.416 | 0.5232 | 0.3234 | | | |
| SE | 0.3086 | | 0.1962 | | 0.0055 | 0.373 | 0.0058 |
| | 0.1075 | | 0.0948 | | - | - | - |
| AIC | -46.6152 | | -47.5937 | | -48.8713 | -40.5725 | -56.9322 |
| CAIC | -46.2516 | | -47.23 | | -48.7537 | -40.4548 | -56.8145 |
| BIC | -43.4482 | | -44.4266 | | -47.2878 | -38.9889 | -55.3487 |
| HQIC | -45.5098 | | -46.4883 | | -48.3186 | -40.0198 | -56.3795 |
| LL | 25.3076 | | 25.7968 | | 25.4357 | 21.2862 | 29.4661 |
| K-S | 0.1453 | | 0.1390 | | 0.1468 | 0.2493 | 0.0722 |
| $H_0$ | Fail to reject | | Fail to reject | | Fail to reject | Reject | Fail to reject |
| P-value | 0.2055 | | 0.2411 | | 0.1979 | 0.0062 | 0.8300 |
| AD | 1.2191 | | 1.1724 | | 1.2694 | 4.0059 | 0.2977 |
| CVM | 0.2041 | | 0.192 | | 0.2133 | 0.7501 | 0.0430 |
| determinant | 0.2299 | | 0.1743 | | - | - | - |

The analysis shows that the Unit Lindley fits the data well, followed by the MBUR then the Kumaraswamy and lastly comes the Beta distribution. Fitting GOMBUR-1 and 2 comes before the Beta distribution for fitting the data. The striking news is that the determinants for both generalized forms are less than the variance of any distribution fitting the data which makes the generalized distributions more efficient for fitting the data to these distributions. Also, as in the previous data analysis, the determinant for GOMBUR-1 is less than the determinant for the GOMBUR-2. Alpha estimate after fitting MBUR distribution is nearly equal to its value after estimating the GOMBUR-1 & 2 . Figure 13 show the theoretical CDFs of the competitors while Figure 14 shows the fitted PDFs. Figure 15-17 shows the PP plot and QQ plot for the tested distributions. Figure 18 shows the CDFs and PDFs of GOMBUR-1 and GOMBUR-2.

Table (4) to be continued

|  | GOMBUR-1 | | GOMBUR-2 | |
|---|---|---|---|---|
| theta | $n = 0.9014$ | | $n = 2.8027$ | |
|  | $\alpha = 0.5531$ | | $\alpha = 0.5531$ | |
| Variance | 0.1775 | 0.0046 | 0.7102 | 0.0092 |
|  | 0.0046 | 0.0013 | 0.0092 | 0.0013 |
| SE(n) | 0.0702 | | 0.1405 | |
| SE(a) | 0.006 | | 0.006 | |
| AIC | -46.9242 | | -46.9242 | |
| CAIC | -46.5606 | | -46.5606 | |
| BIC | -43.7572 | | -43.7572 | |
| HQIC | -45.8189 | | -45.8189 | |
| LL | 25.4621 | | 25.4621 | |
| K-S Value | 0.1421 | | 0.1421 | |
| H$_0$ | Fail to reject | | Fail to reject | |
| P-value | 0.2228 | | 0.2228 | |
| AD | 1.1827 | | 1.1827 | |
| CVM | 0.1957 | | 0.1957 | |
| Determinant | 0.00020661 | | 0.00082644 | |
| Significant(n) | P<0.001 | | P<0.001 | |
| Significant(a) | P<0.001 | | P<0.001 | |

The CDF of the GOMBUR-1 perfectly aligns with the CDFs of the Beta, the Kumaraswamy and the MBUR distributions; this can be deduced from the approximate values of AIC, CAIC, BIC and HQIC. This is also true for the fitted PDFs as shown in Figure 14. Fitted CDFs and fitted PDF s of both GOMBURT-1 and GOMBURT-2 are identical as illustrated in Figure 18 and this has a reflection on the equal indices ( AIC, CAIC, BIC, HQIC, AD, CVM, and KS statistics) obtained from the analysis. Also, the alignment seen in PP plot depicted in Figure 17 for both versions is identical.

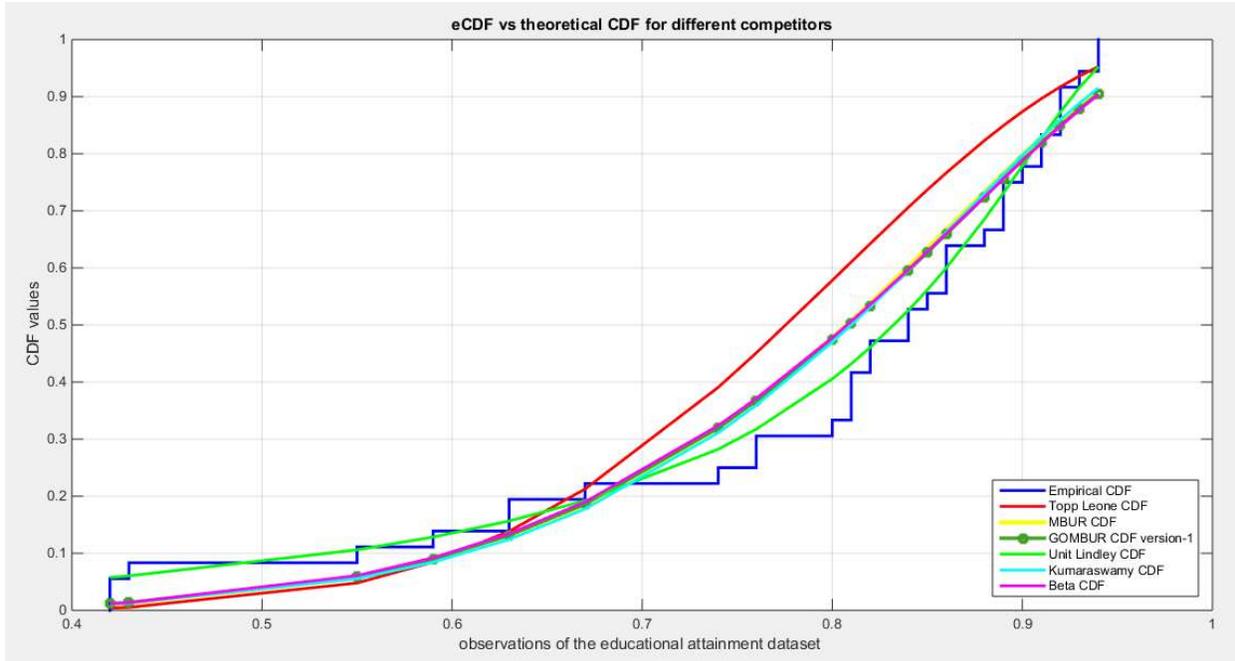

Fig. 13 shows the e-CDFs and the theoretical CDFs for the fitted distributions of educational attainment data.

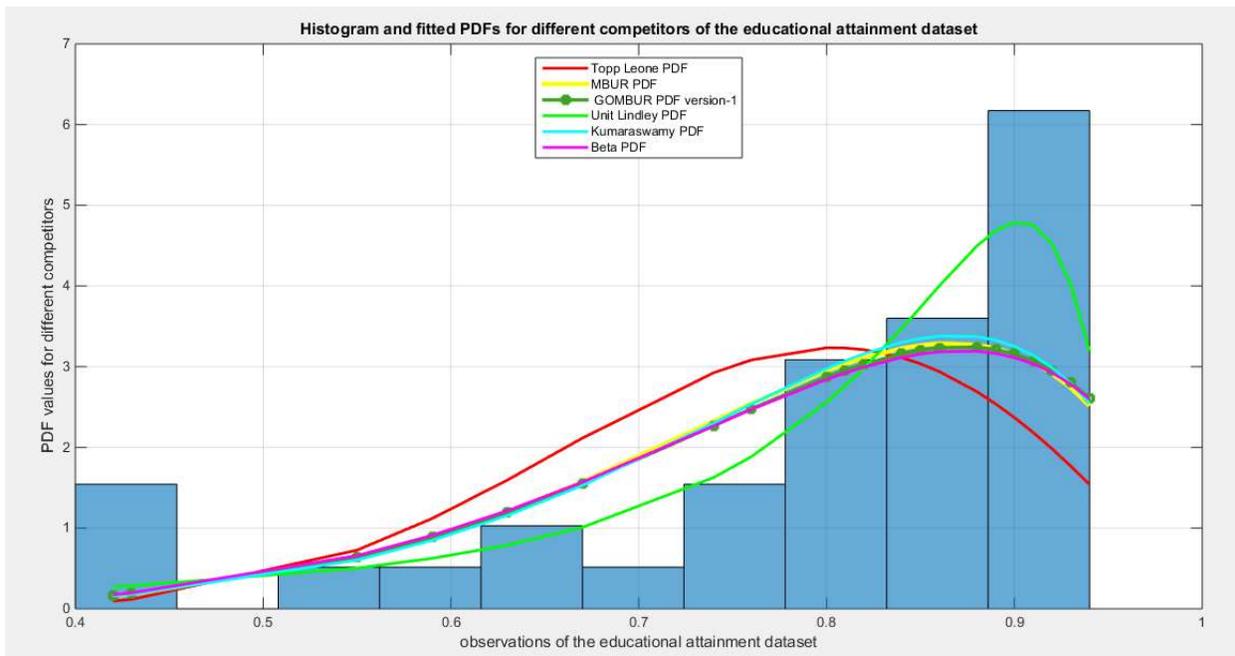

Fig. 14 shows the histogram of the educational attainment data and the theoretical PDFs for the fitted distributions

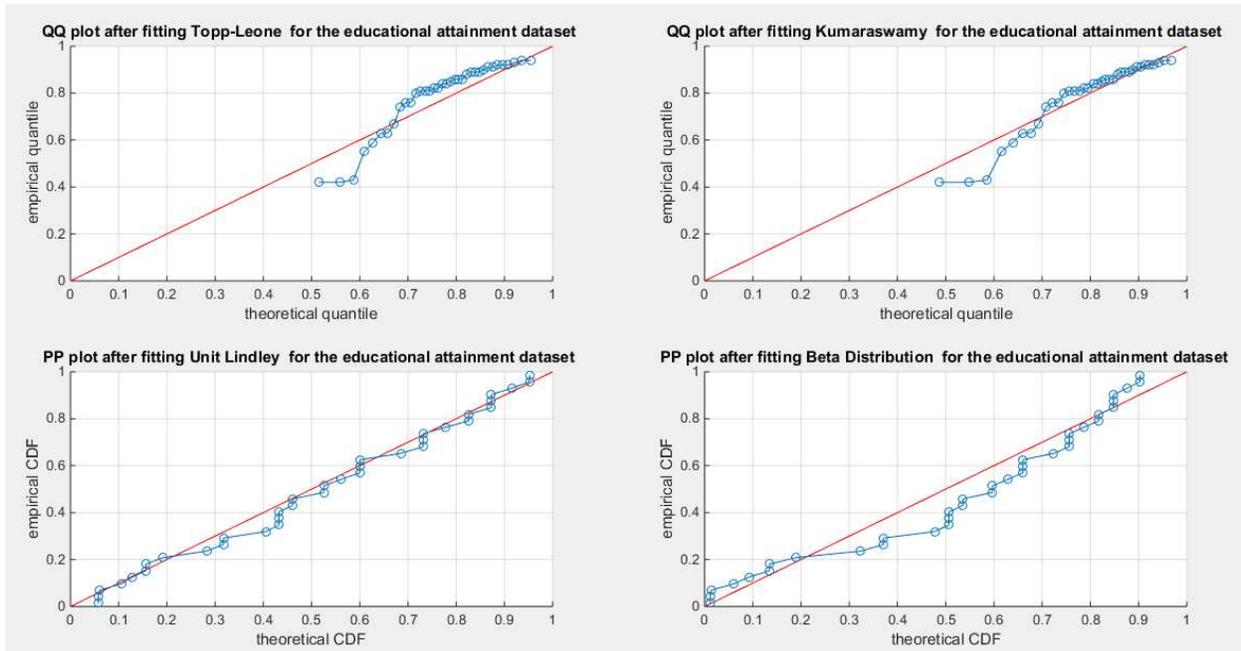

Fig. 15 shows the QQ plot for the fitted Topp Leone & Kumaraswamy distributions and the PP plot for the fitted Unit Lindley and Beta distribution as regards educational attainment dataset.

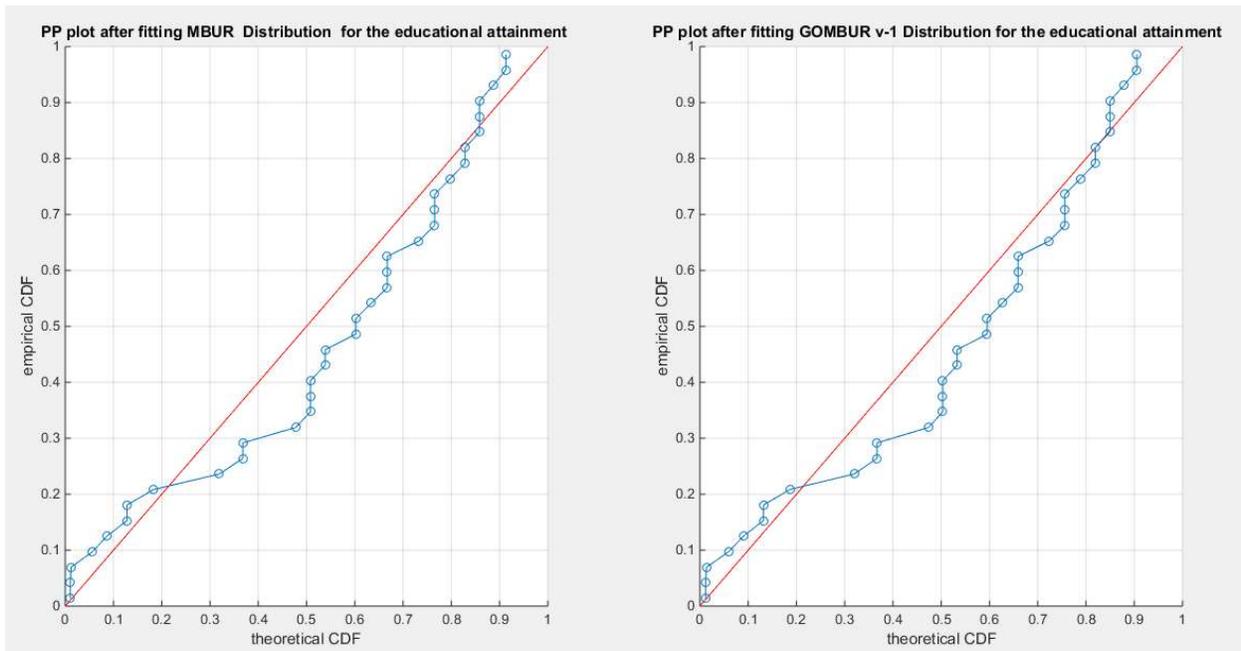

Fig. 16 shows the PP plot for the fitted MBUR & GOMBUR-1 as regards educational attainment dataset. The alignments with the diagonal are nearly identical all through the course.

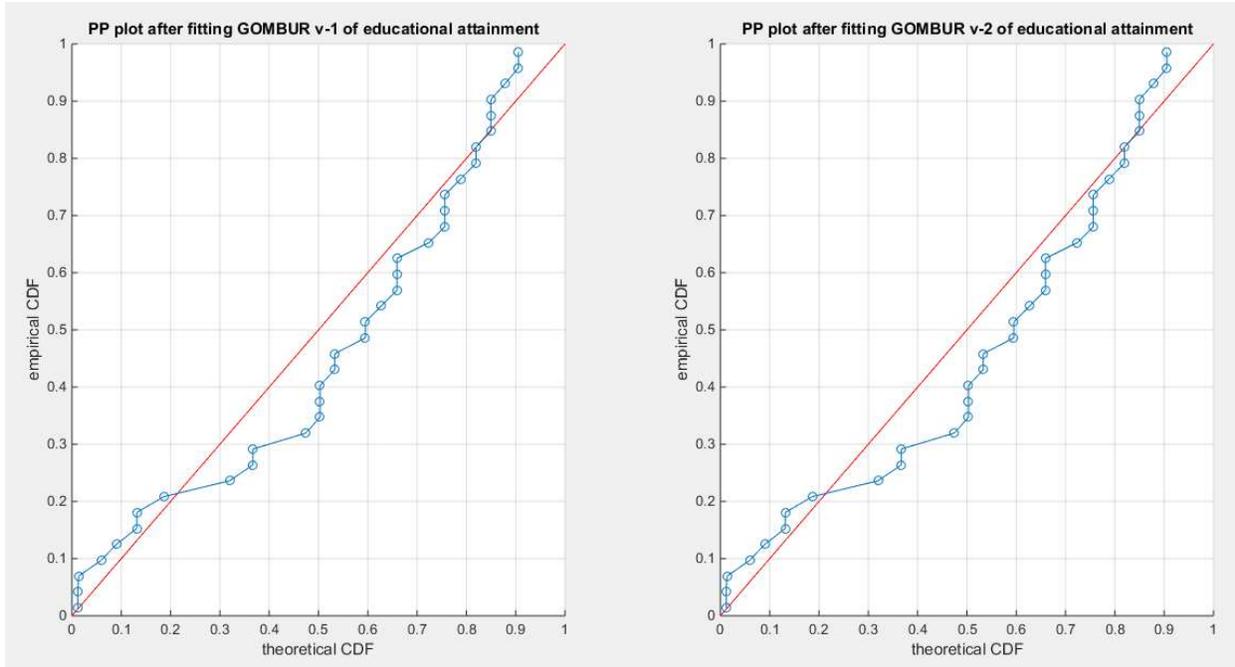

Fig.17 shows the PP plot for the fitted GOMBUR-1 & GOMBUR-2 for educational attainment dataset. They are identical.

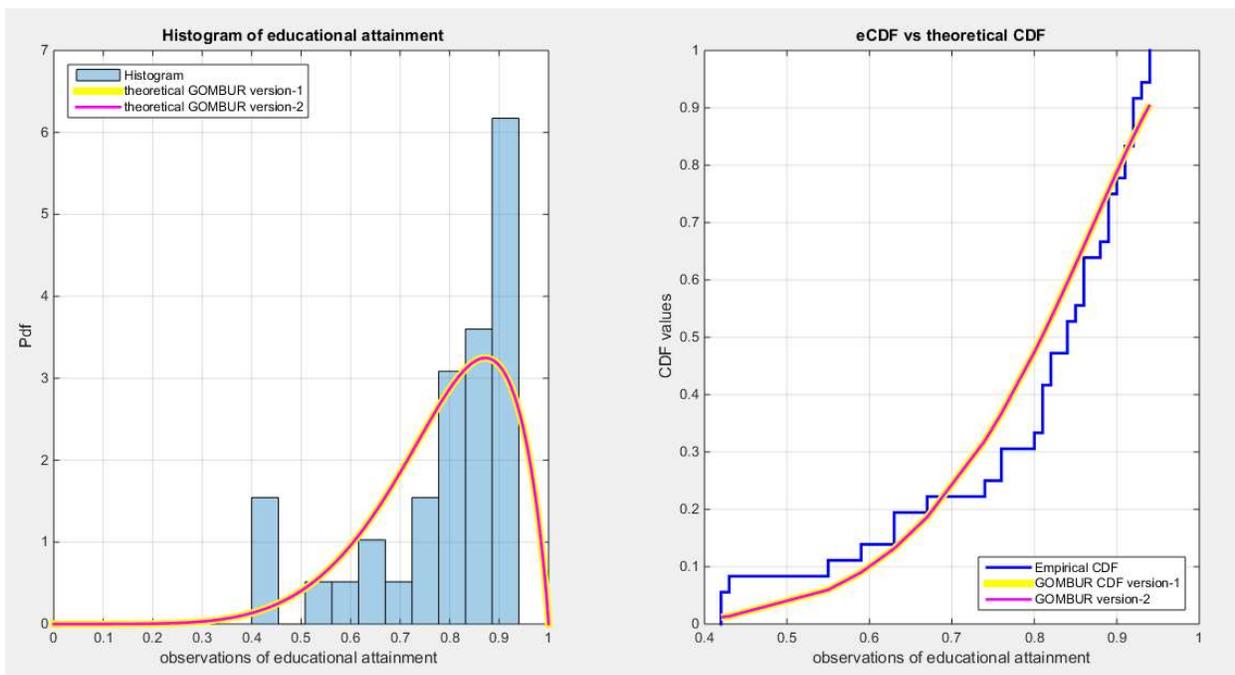

Fig. 18 shows on the left subplot the histogram of the educational attainment data and the fitted PDFs of both GOMBUR-1 & GOMBUR-2 and on the right subplot the e-CDFs and the theoretical CDFs for both distributions. Both the fitted CDFs and the fitted PDFs of both versions are identical.

Table (5) shows the result of flood data analysis

Table (5): results of the flood data analysis

|  | Beta | | Kumaraswamy | | MBUR | Topp-Leone | Unit-Lindley |
|---|---|---|---|---|---|---|---|
| theta | $\alpha = 6.8318$ | | $\alpha = 3.3777$ | | 1.0443 | 2.2413 | 1.6268 |
|  | $\beta = 9.2376$ | | $\beta = 12.0057$ | | | | |
| Var | 7.22 | 7.2316 | 0.3651 | 2.8825 | 0.007 | 0.2512 | 0.0819 |
|  | 7.2316 | 8.0159 | 2.8825 | 29.963 | | | |
| SE(a) | 0.6008 | | 0.1351 | | 0.0187 | 0.1121 | 0.0639 |
| SE(b) | 0.6331 | | 1.2239 | | - | - | - |
| AIC | -24.3671 | | -21.9465 | | -10.9233 | -12.7627 | -12.3454 |
| CAIC | -23.6613 | | -21.2407 | | -10.7011 | -12.5405 | -12.1231 |
| BIC | -22.3757 | | -19.9551 | | -9.9276 | -11.767 | -11.3496 |
| HQIC | -23.9784 | | -21.5578 | | -10.7289 | -12.584 | -12.151 |
| LL | 14.1836 | | 12.9733 | | 6.4617 | 7.3814 | 7.1727 |
| K-S | 0.2063 | | 0.2175 | | 0.3202 | 0.3409 | 0.2625 |
| $H_0$ | Fail to reject | | Fail to reject | | Fail to reject | Reject | Fail to reject |
| P-value | 0.3174 | | 0.2602 | | 0.0253 | 0.0141 | 0.0311 |
| AD | 0.7302 | | 0.9365 | | 2.7563 | 2.9131 | 2.3153 |
| CVM | 0.1242 | | 0.1653 | | 0.531 | 0.5857 | 0.4428 |
| determinant | 5.5784 | | 2.6314 | | - | - | - |

The analysis shows that Beta distribution fits the data better than any other distribution. The Topp Leone did not fit the distribution. Generalization of MBUR using the GOMBUR-1 improves the fitting up to the level of the Beta distribution and slightly exceeding it. Marked increases in the negativity levels of AIC, CAIC, BIC & HQIC are obtained. The level of Log-likelihood shows marked improvement. Marked reduction in the levels of AD and CVM statistics are obvious. The variance of alpha shows marked reduction after fitting the GOMBUR-1. The determinant of GOMBUR-1&2 is far less than the determinant of the Beta distribution and Kumaraswamy distribution demonstrating more efficiency. GOMBUR-1 has lesser determinant than the GOMBUR-2. The estimates of alpha value, their variances and standard errors are identical for both versions. While the estimates of the n parameter, its variance and standard error obtained after fitting GOMBUR-2 is higher than the levels obtained after fitting GOMBUR-1 distribution. Figure 19 shows near perfect alignment of the theoretical CDF of GOMBUR-1 distribution with the Beta distribution. This is also evident in Figure 20 which portrays near perfect alignment of fitted PDFs for both distributions.

Table (5) to be continued

|  | GOMBUR-1 | | GOMBUR-2 | |
|---|---|---|---|---|
| theta | $n = 8.1044$ | | $n = 17.2087$ | |
| | $\alpha = 1.1168$ | | $\alpha = 1.1168$ | |
| Variance | 8.0302 | 0.0177 | 32.1208 | 0.0354 |
| | 0.0177 | 0.0018 | 0.0354 | 0.0018 |
| SE(n) | 0.6336 | | 1.2673 | |
| SE(a) | 0.0095 | | 0.0095 | |
| AIC | -24.4562 | | -24.4562 | |
| CAIC | -23.7503 | | -23.7503 | |
| BIC | -22.4647 | | -22.4647 | |
| HQIC | -24.0674 | | -24.0674 | |
| LL | 14.2281 | | 14.2281 | |
| K-S Value | 0.204 | | 0.204 | |
| H₀ | Fail to reject | | Fail to reject | |
| P-value | 0.3297 | | 0.3297 | |
| AD | 0.7153 | | 0.7153 | |
| CVM | 0.1205 | | 0.1205 | |
| Determinant | 0.0143 | | 0.0573 | |
| Significant(n) | P<0.001 | | P<0.001 | |
| Significant(a) | P<0.001 | | P<0.001 | |

Both Figures 19 & 20 reveal how MBUR exhibits near perfect alignment with the beta distribution after using the generalized version. Figure 21 elucidates the PP plot and QQ plot for the competitor distributions. The plots for Beta and Kumaraswamy suggest better fit of the data. Figure 22 demonstrates the amendment in alignment of the diagonal after fitting the GOMBUR-1 in dissimilarity to the alignment with the fitted MBUR. Figure 23 exemplifies the PP plot of the GOMBUR-1 & 2. They are identical. Moreover, Figure 24 illuminates the fitted CDFs and the fitted PDFs for both versions of the generalization expounding identity of the curves.

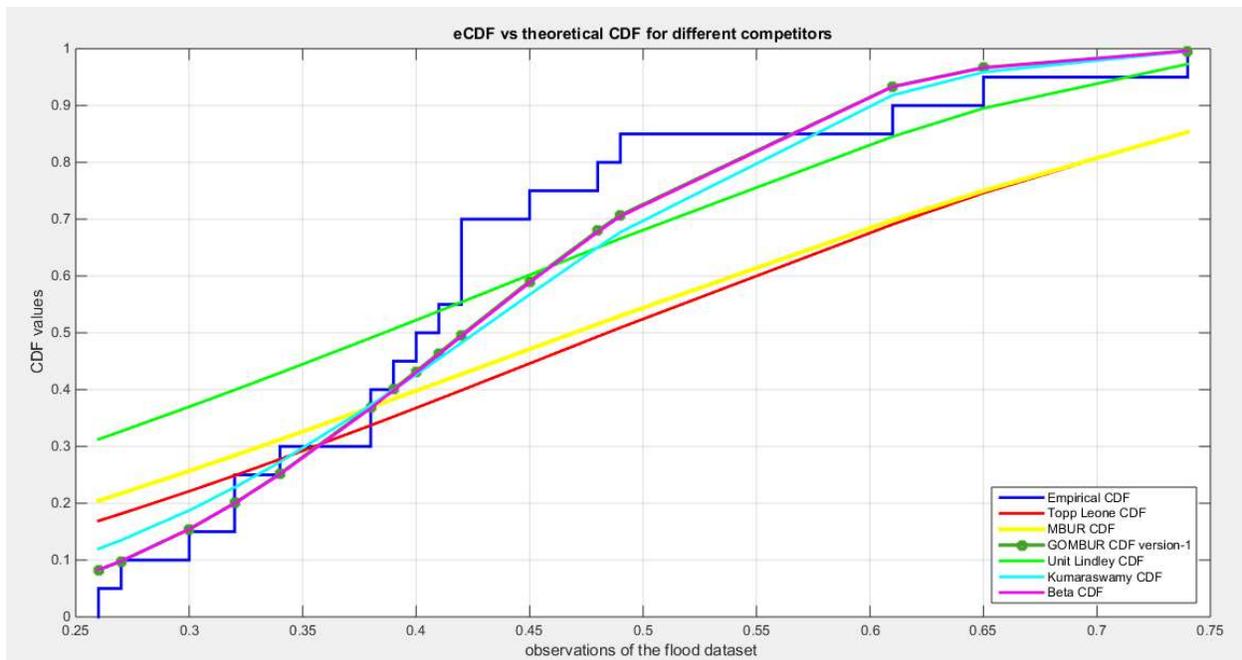

Fig. 19 shows the e-CDFs and the theoretical CDFs for the fitted distributions of flood data.

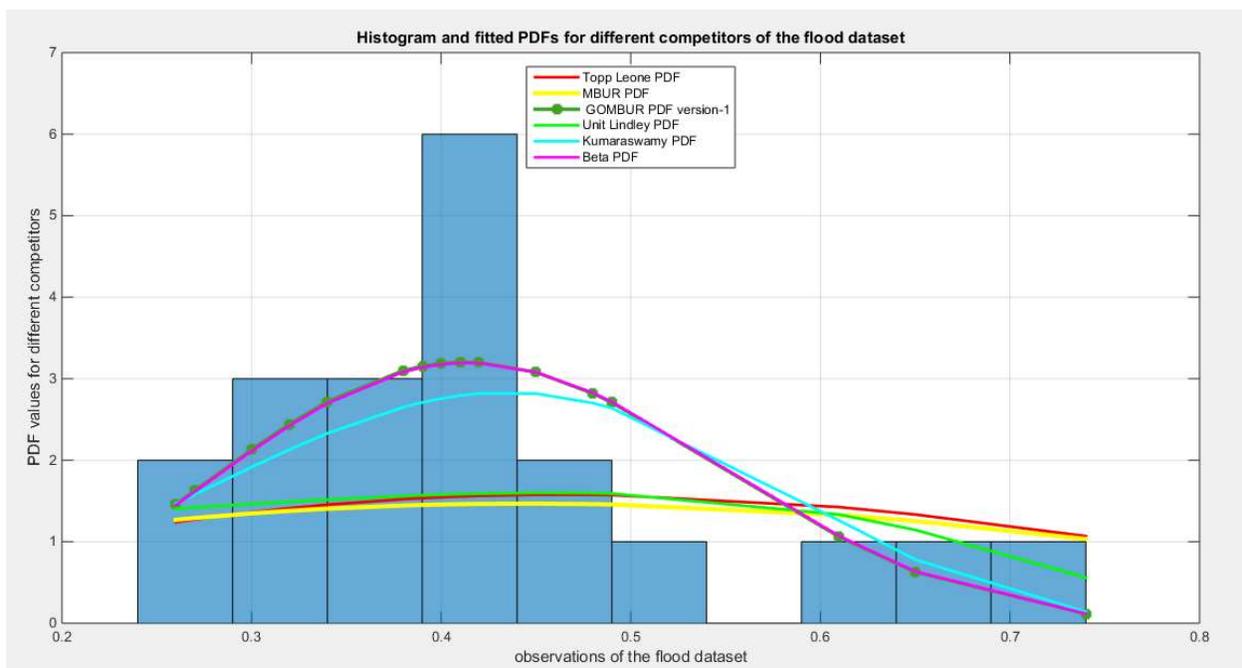

Fig. 20 shows the histogram of the flood data and the theoretical PDFs for the fitted distributions. The GOMBUR-1 perfectly aligns with the Beta distribution.

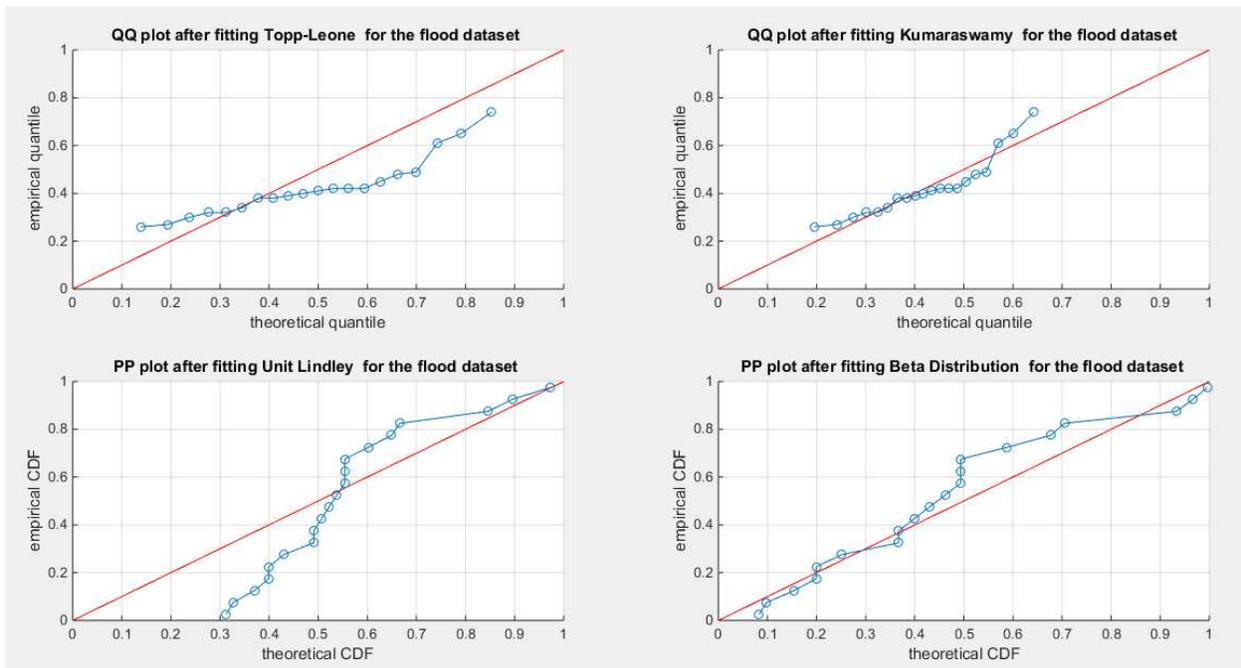

Fig. 21 shows the QQ plot for the fitted Topp Leone & Kumaraswamy distributions and the PP plot for the fitted Unit Lindley and Beta distribution for the flood dataset.

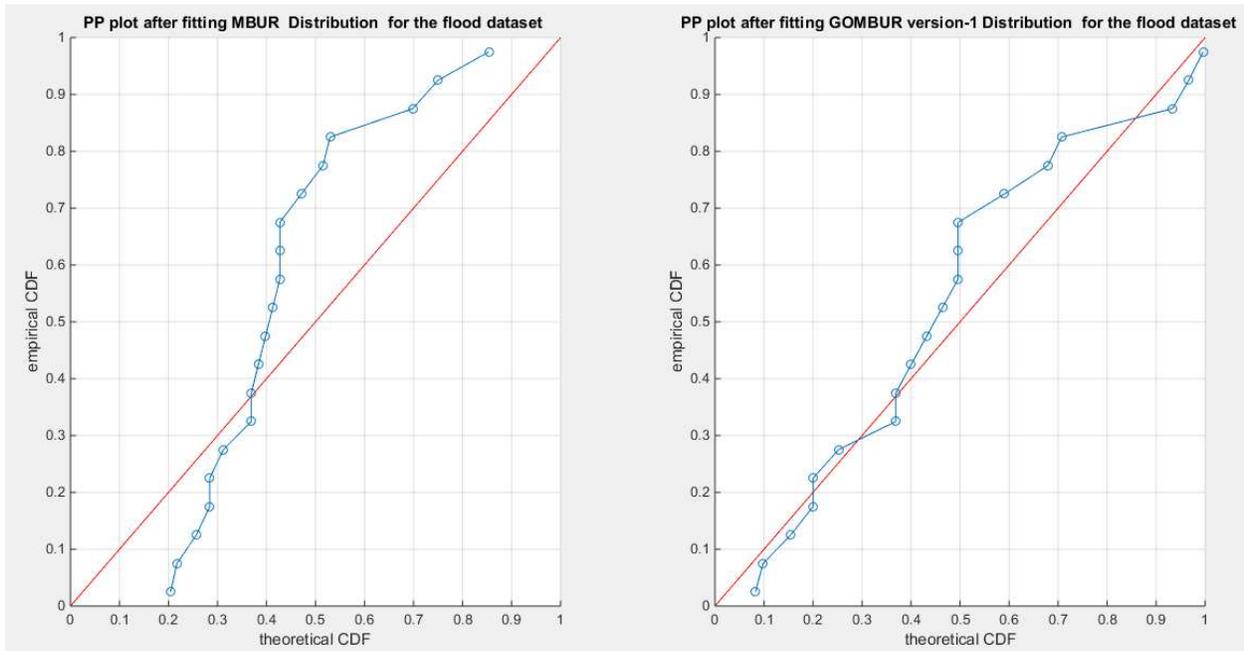

Fig. 22 shows the PP plot for the fitted MBUR & GOMBUR-1 for the flood dataset. The alignments with the diagonal shows marked improvement after fitting the GOMBUR-1 than the alignment with MBUR especially at the lower and upper tails.

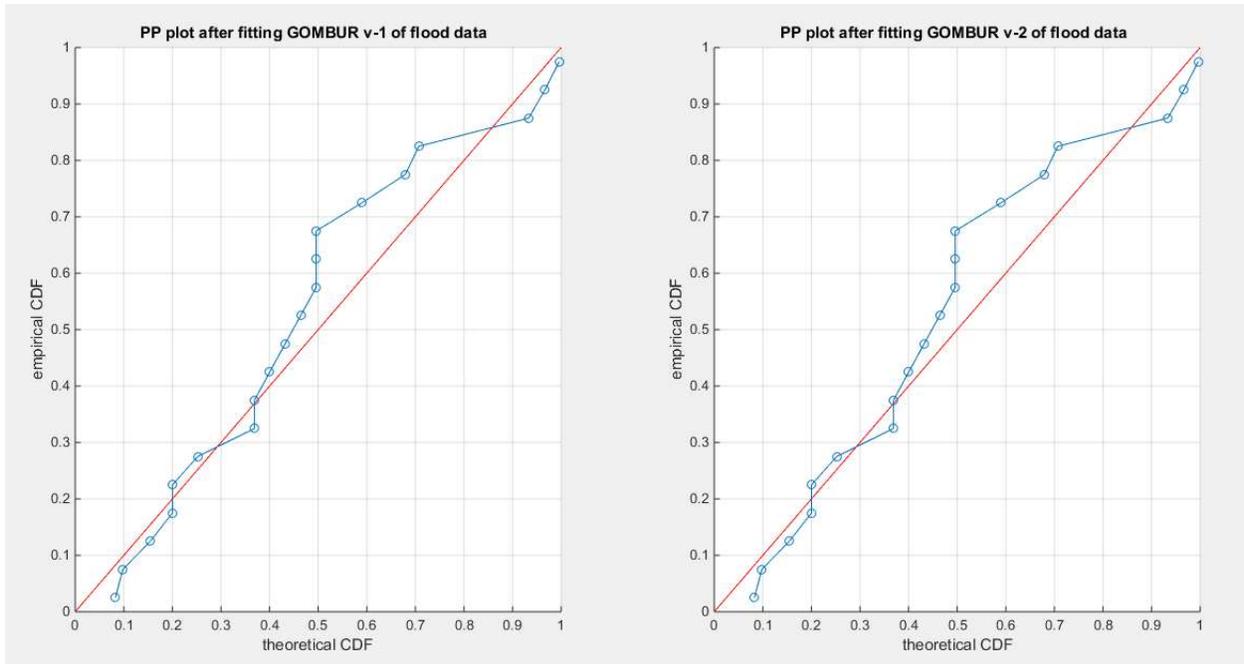

Fig. 23 shows the PP plot for the fitted GOMBUR-1 & GOMBUR-2 for flood dataset. They are identical.

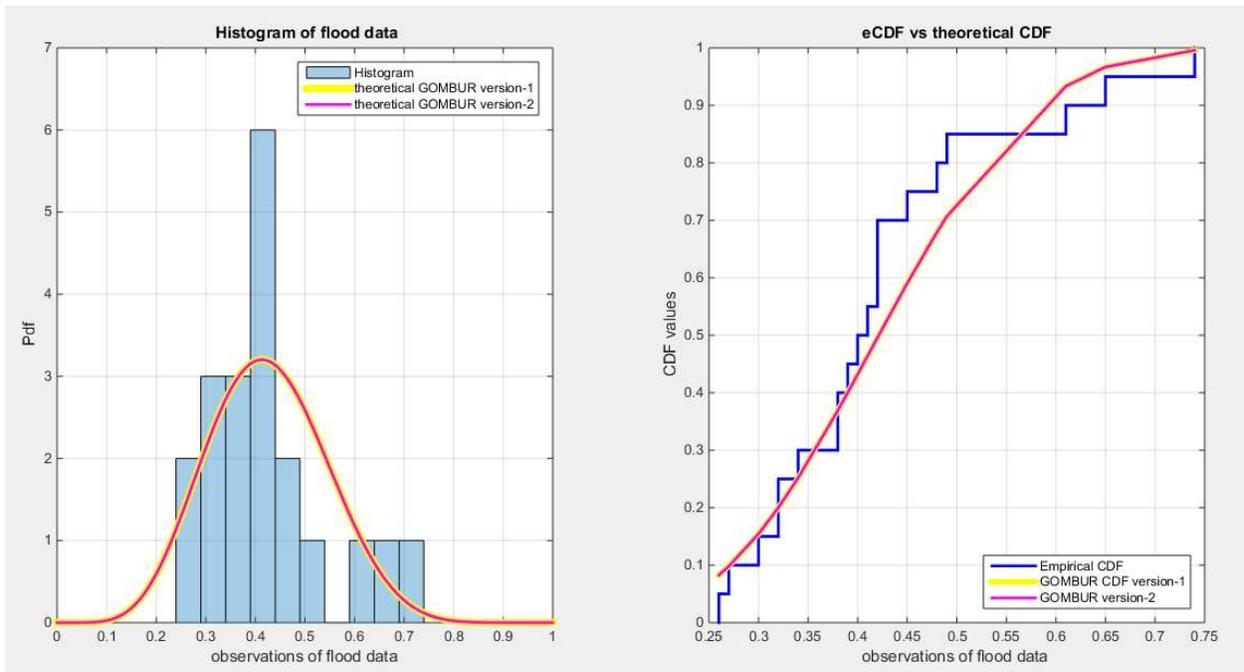

Fig. 24 shows on the left subplot the histogram of the flood data and the fitted PDFs of both GOMBUR-1 & GOMBUR-2 and on the right subplot the e-CDFs and the theoretical CDFs for both distributions. Both the fitted CDFs and the fitted PDFs of both versions are identical.

Table (6) shows the results of time between failures data analysis

Table (6) the results of the analysis of time between failures:

| | Beta | | Kumaraswamy | | MBUR | Topp-Leone | Unit-Lindley |
|---|---|---|---|---|---|---|---|
| theta | $\alpha = 0.6307$ | | $\alpha = 0.6766$ | | 1.7886 | 0.4891 | 4.1495 |
| | $\beta = 3.2318$ | | $\beta = 2.936$ | | | | |
| Var | 0.071 | 0.2801 | 0.0198 | 0.1033 | 0.018 | 0.0104 | 0.5543 |
| | 0.2801 | 1.647 | 0.1033 | 0.9135 | | | |
| SE | 0.0555 | | 0.0293 | | 0.0279 | 0.0213 | 0.1552 |
| | 0.2676 | | 0.1993 | | - | - | - |
| AIC | -36.0571 | | -36.6592 | | -37.862 | -35.5653 | -27.007 |
| CAIC | -35.4571 | | -36.0592 | | -37.6712 | -35.3749 | -26.8165 |
| BIC | -33.7861 | | -34.3882 | | -36.7262 | -34.4298 | -25.8715 |
| HQIC | -35.4859 | | -36.0881 | | -37.5764 | -35.2798 | -26.7214 |
| LL | 20.0285 | | 20.3296 | | 19.9310 | 18.7827 | 14.5035 |
| K-S | 0.1541 | | 0.1393 | | 0.1584 | 0.1962 | 0.3274 |
| $H_0$ | Fail to reject | | Fail to reject | | Fail to reject | Fail to Reject | Reject |
| P-value | 0.5918 | | 0.7123 | | 0.5575 | 0.2982 | 0.0107 |
| AD | 0.6886 | | 0.5755 | | 0.6703 | 1.1022 | 4.7907 |
| CVM | 0.1264 | | 0.0989 | | 0.1253 | 0.2149 | 0.8115 |
| determinant | 0.0385 | | 0.0074 | | - | - | - |

The analysis shows that the all the proposed distributions fit the data except the Unit Lindley distributions. The GOMBURT-1 fits the data with slight negativity reduction in some indices like AIC, BIC & HQIC and enhanced reduction in AD,CVM and KS tests which points a good remark for adding a mentionable improvement in fitting. The Log-likelihood value obtained with generalization is higher than that obtained after fitting MBUR and exceeding slightly the levels obtained after fitting the Beta and Kumaraswamy distributions. Also, the variance of the alpha after fitting the GOMBUR-1 is less than the variance obtained after fitting the MBUR and all other distributions. The Kumaraswamy distribution has the least determinant among kumaraswamy, GOMBUR-1 & 2. But other indices like AIC, CAIC, BIC & HQIC make the generalization a better alternative. It is a matter of tradeoff between efficiency and biasness. The determinant of the variance-covariance matrix obtained after fitting the Beta distribution is slightly larger than that obtained after GOMBUR-2 fitting. Moreover, the latter has better indices than the indices of the Beta distribution. All this explains the enhanced parameter estimation when generalizing the MBUR distribution by adding a new shape parameter.

Table (6) to be continued

|  | GOMBUR-1 | | GOMBUR-2 | |
| --- | --- | --- | --- | --- |
| theta | $n = 1.8592$ | | $n = 4.7184$ | |
|  | $\alpha = 1.8362$ | | $\alpha = 1.8362$ | |
| Variance | 0.6516 | 0.0251 | 2.6064 | 0.0502 |
|  | 0.0251 | 0.143 | 0.0502 | 0.0143 |
| SE(n) | 0.1683 | | 0.3366 | |
| SE(a) | 0.0249 | | 0.0249 | |
| AIC | -37.3005 | | -37.3005 | |
| CAIC | -36.7005 | | -36.7005 | |
| BIC | -35.0295 | | -35.0295 | |
| HQIC | -36.7293 | | -36.7293 | |
| LL | 20.6502 | | 20.6502 | |
| K-S Value | 0.1355 | | 0.1355 | |
| H₀ | Fail to reject | | Fail to reject | |
| P-value | 0.7424 | | 0.7424 | |
| AD | 0.5399 | | 0.5399 | |
| CVM | 0.0897 | | 0.0897 | |
| Determinant | 0.0087 | | 0.0348 | |
| Significant(n) | P<0.001 | | P<0.001 | |
| Significant(a) | P<0.001 | | P<0.001 | |

Figure 25 shows a proximate alignment of the GOMBUR-1 with the Beta, Kumaraswamy, MBUR, and Topp Loene distributions especially at the ends of the distributions and most of the center of these distributions. No such alignment is well-defined with the Unit Lindley distribution as the latter does not essentially fit the data. Figure 26 displays the same outcome. Figure 27 spectacles the QQ plot and PP plot for different competitors. Figure 28 exposes how the GOMBUR-1 enhances the lower tail alignment with the diagonal over the MBUR. Figure 29-30 clear up that GOMBUR-1 &2 have the same PP plots, the same fitted CDFs and the same fitted PDFs with right skewness.

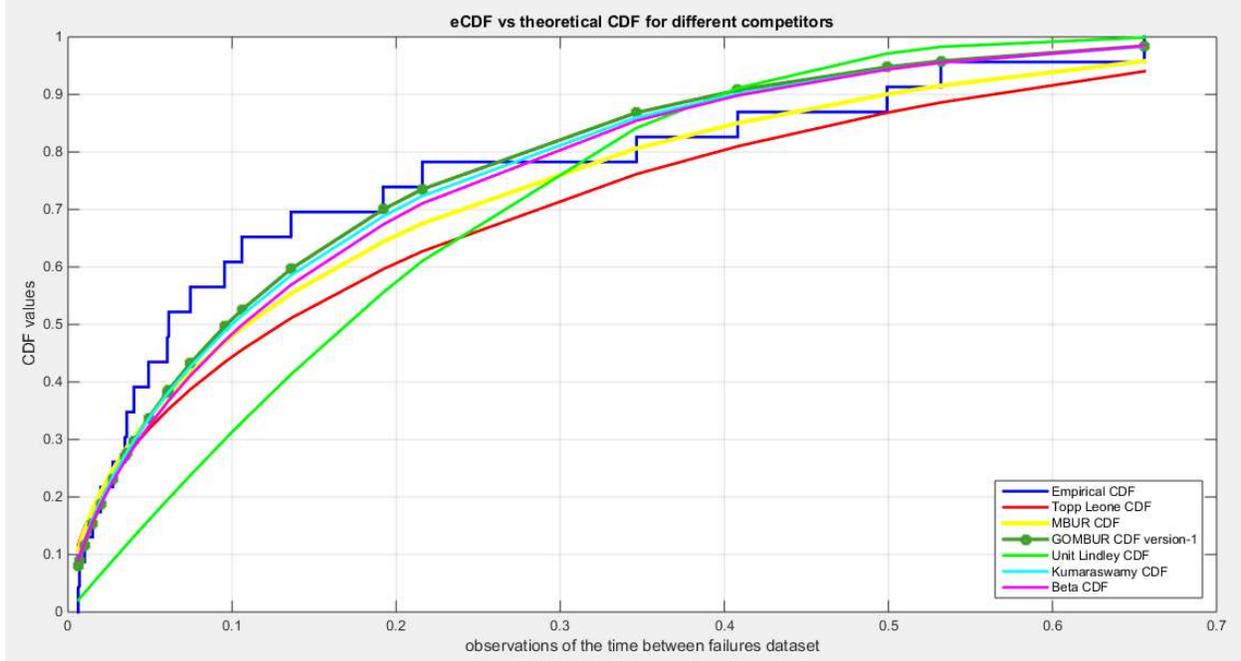

Fig. 25 shows the e-CDFs and the theoretical CDFs for the fitted distributions of time between failures data.

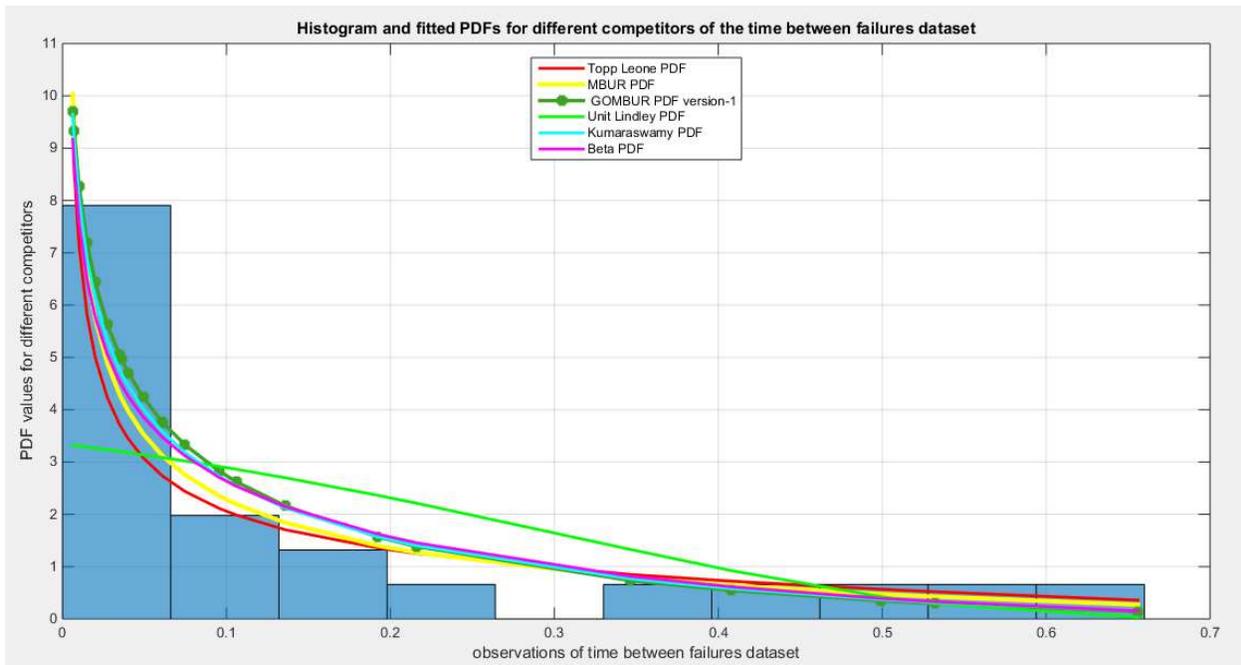

Fig. 26 shows the histogram of the time between failures data and the theoretical PDFs for the fitted distributions. The GOMBUR-1 shows near perfect alignments with all distributions at the boundaries except with Unit Lindley which does not fit the data.

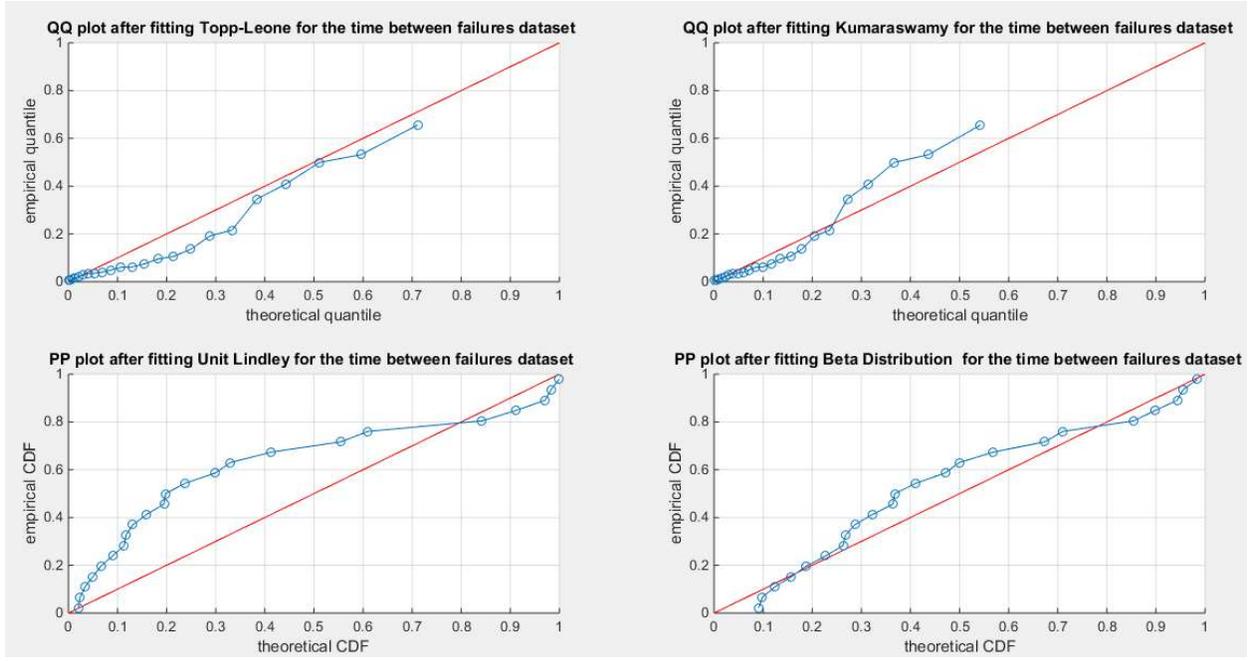

Fig. 27 shows the QQ plot for the fitted Topp Leone & Kumaraswamy distributions and the PP plot for the fitted Unit Lindley and Beta distribution for the time between failures dataset.

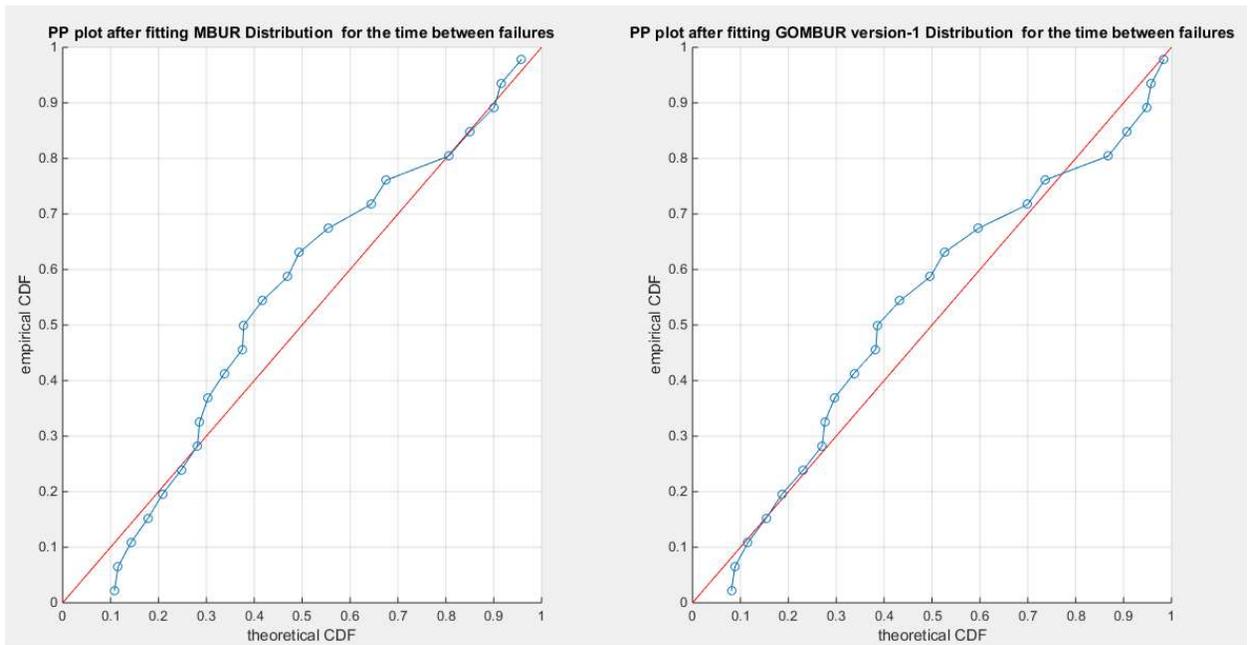

Fig. 28 shows the PP plot for the fitted MBUR & GOMBUR-1 for the time between failures dataset. The alignment with the diagonal shows improvement after fitting the GOMBUR-1 than the alignment with MBUR especially at the lower tail.

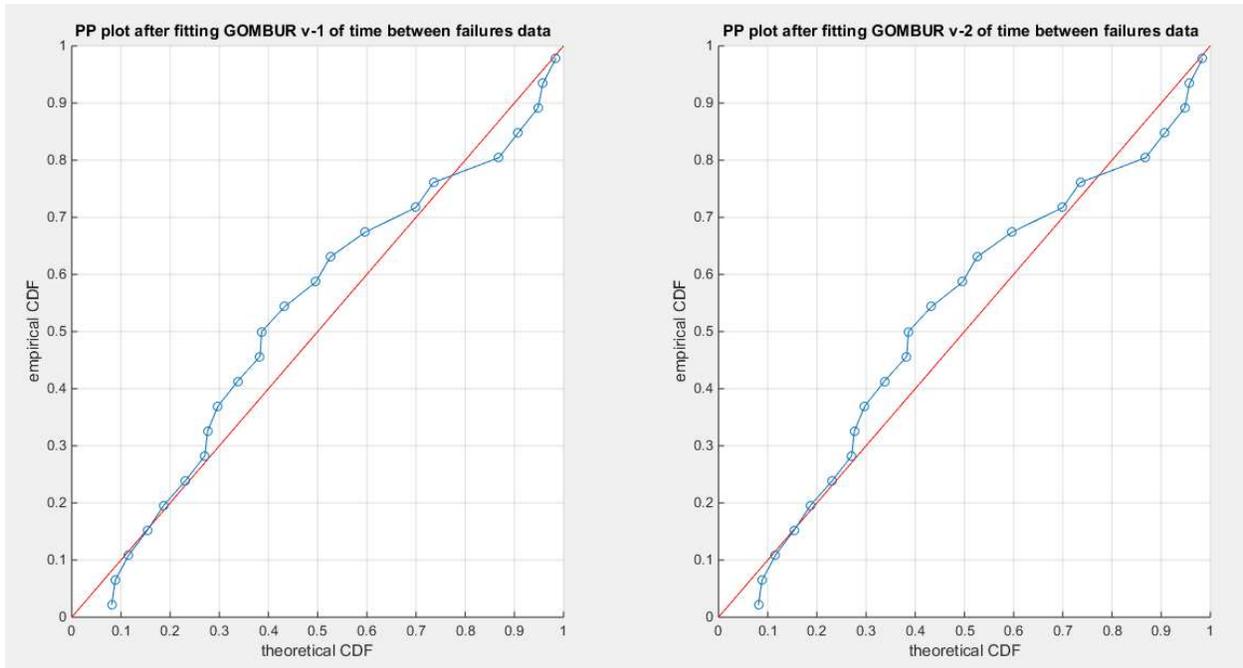

Fig. 29 shows the PP plot for the fitted GOMBUR-1 & GOMBUR-2 for time between failure dataset. They are identical.

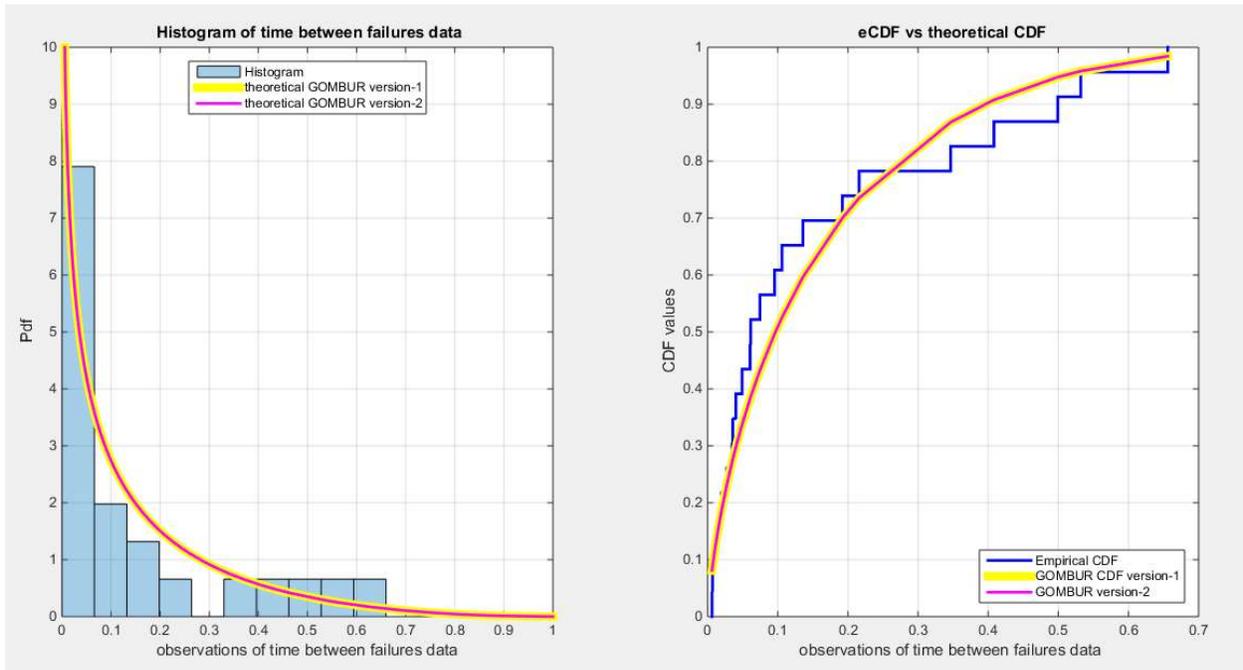

Fig. 30 shows on the left subplot the histogram of the time between failures data and the fitted PDFs of both GOMBUR-1 & GOMBUR-2 and on the right subplot the e-CDFs and the theoretical CDFs for both distributions. Both the fitted CDFs and the fitted PDFs of both versions are identical.

Table (7) shows the results of analysis of COVID-19 death rate analysis in Canada.

Table (7) shows the results of analysis of COVID-19 death rate analysis in Canada.

| | Beta | | Kumaraswamy | | MBUR | Topp-Leone | Unit-Lindley |
|---|---|---|---|---|---|---|---|
| theta | $\alpha = 14.5128$ | | $\alpha = 5.0309$ | | 1.3479 | 1.0814 | 3.9381 |
| | $\beta = 48.4899$ | | $\beta = 1049.6$ | | | | |
| Variance | 7.839 | 27.531 | 0.2719 | 370.2768 | 0.0042 | 0.0209 | 0.203 |
| | 27.531 | 100.6504 | 370.2768 | 523950 | | | |
| SE(a) | 0.3741 | | 0.0697 | | 0.0086 | 0.0193 | 0.0602 |
| SE(b) | 1.3406 | | 96.7273 | | - | - | - |
| AIC | -167.88 | | -169.2 | | -48.8337 | -46.3748 | -80.2707 |
| CAIC | -167.6536 | | -168.9736 | | -48.7596 | -46.3008 | -80.1966 |
| BIC | -163.8293 | | -165.1493 | | -46.8083 | -44.3495 | -78.2453 |
| HQIC | -166.3096 | | -167.6296 | | -48.0485 | -45.5896 | -79.4855 |
| LL | 85.94 | | 86.6 | | 25.4168 | 24.1874 | 41.1353 |
| K-S | 0.0754 | | 0.1029 | | 0.429 | 0.4685 | 0.359 |
| $H_0$ | Fail to reject | | Fail to reject | | reject | reject | reject |
| P-value | 0.6802 | | 0.5583 | | 0 | 0 | 0 |
| AD | 0.4398 | | 0.369 | | 14.0394 | 15.8748 | 12.7087 |
| CVM | 0.0692 | | 0.0686 | | 2.8621 | 3.3539 | 2.5936 |
| determinant | 31.0432 | | 5348.9 | | - | - | - |

The analysis shows that both the Beta distribution and the Kumaraswamy distribution fit the data well while MBUR, Topp Leone and Unit Lindley distributions did not fit the data. The GOMBUR-1&2 fit the data with levels of AIC, CAIC, BIC & HQIC far exceeding the MBUR, although slightly less than the Kumaraswamy and Beta distribution. The advantages of fitting the generalized form of MBUR is that generalization reduces the variance as illustrated by the markedly reduced values of the determinants of the variance-covariance matrices obtained after fitting the GOMBUR-1&2 distributions. The estimated variance for the alpha is markedly less than the variance of the estimated alpha obtained from Beta and Kumaraswamy fitting . The estimated alpha levels, their variances and standard errors are the same after fitting both versions of GOMBUR distributions. Figure 31-32 show perfect alignment of Beta distribution and GOMBUR-1. Kumaraswamy fitting has more negative values of AIC, BIC & HQIC indicating outperformance over both Beta and GOMBUR-1 data fitting, but in the same time it has a high value of the determinant of the variance-covariance matrix which yields a less efficient fitting of data.

Table (7) to be continued

|  | GOMBUR-1 | | GOMBUR-2 | |
|---|---|---|---|---|
| theta | $n = 41.02961$ | | $n = 83.0593$ | |
|  | $\alpha = 1.4623$ | | $\alpha = 1.4623$ | |
| Variance | 62.6379 | 0.0083 | 250.5451 | 0.0166 |
|  | 0.0083 | 0.0002376 | 0.0166 | 0.00023762 |
| SE(n) | 1.0576 | | 2.1152 | |
| SE(a) | 0.0021 | | 0.0021 | |
| AIC | -167.5661 | | -167.5661 | |
| CAIC | -167.3396 | | -167.3396 | |
| BIC | -163.5154 | | -163.5154 | |
| HQIC | -165.9956 | | -165.9956 | |
| LL | 85.783 | | 85.783 | |
| K-S Value | 0.0781 | | 0.0781 | |
| H₀ | Fail to reject | | Fail to reject | |
| P-value | 0.6461 | | 0.6461 | |
| AD | 0.4653 | | 0.4653 | |
| CVM | 0.0728 | | 0.0728 | |
| Determinant | 0.0148 | | 0.0593 | |
| Significant(n) | P<0.001 | | P<0.001 | |
| Significant(a) | P<0.001 | | P<0.001 | |

Figure 33 shows graphs of PP plot and QQ plot for the competitor distributions. Figure 34 illuminates how generalization of MBUR markedly enhances the diagonal alignment of GOMBUR-1 along both ends and along the center of the distribution. Figure 35 shows the PP plots of both versions of the distribution being identical. Figure 36 depicts the fitted CDFs and the fitted PDFs of both versions. The shapes of the PDFs are symmetric and identical for both versions. And this is anticipated from the large values of the estimated n, the larger the estimated n is, the more symmetrical the distribution is.

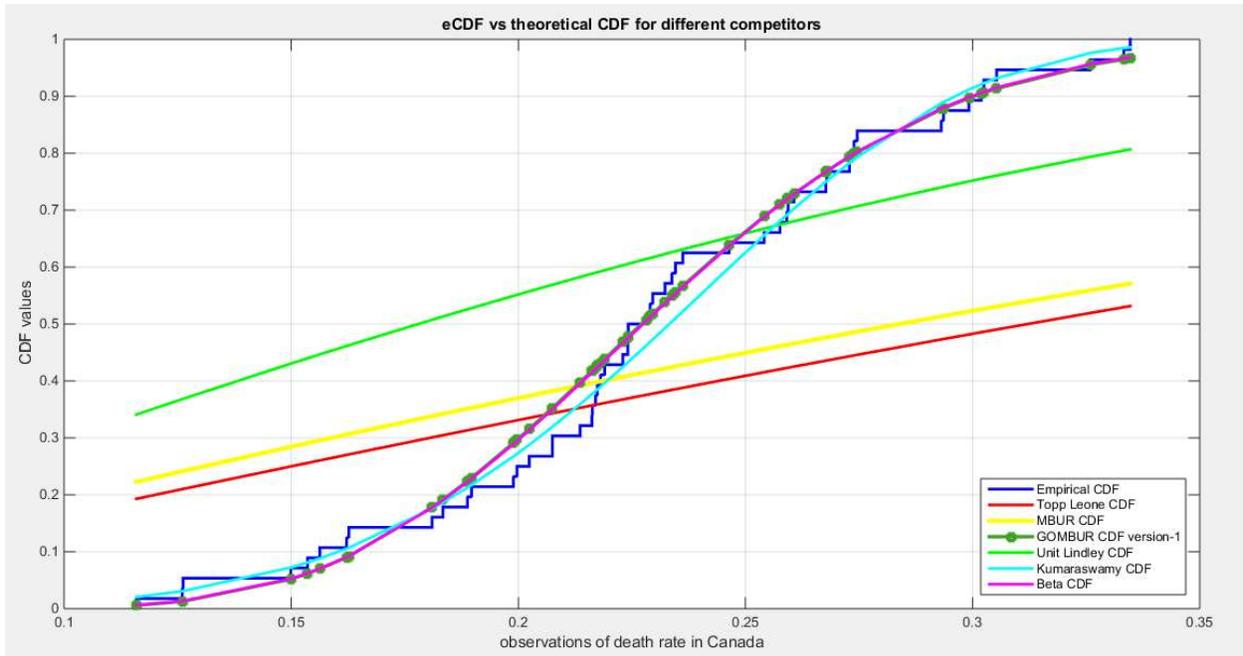

Fig. 31 shows the e-CDFs and the theoretical CDFs for the fitted distributions of death rate in Canada data.

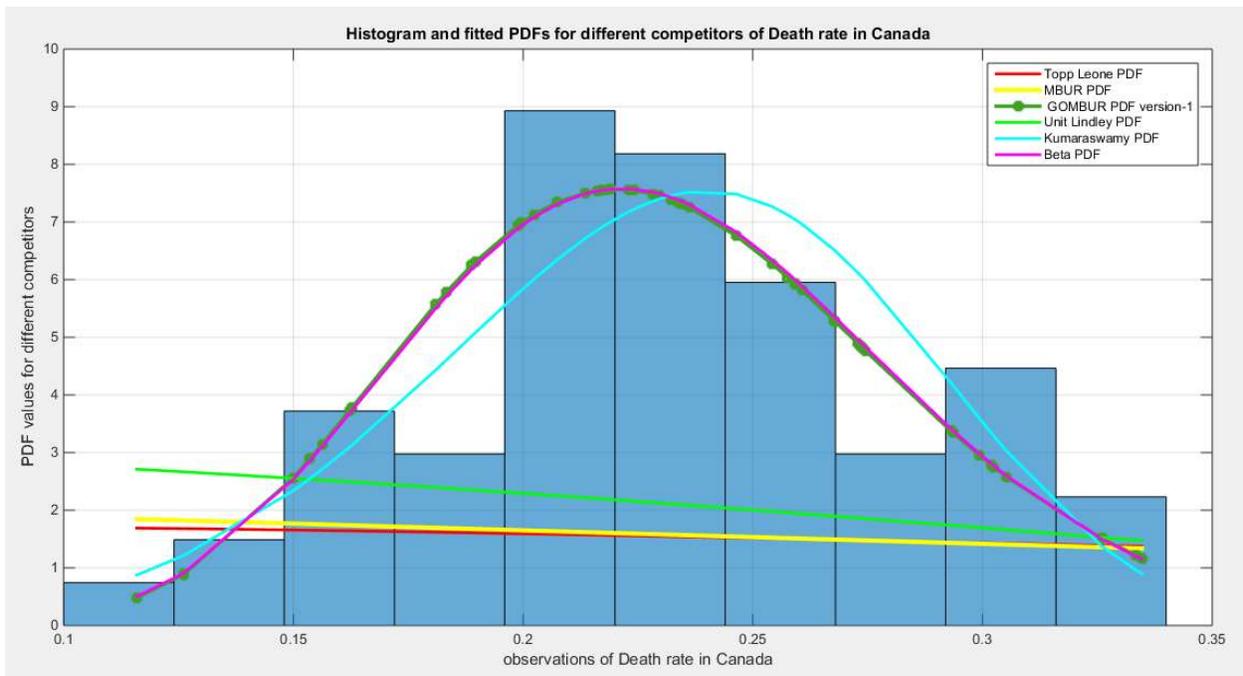

Fig. 32 shows the histogram of the COVID-19 death rate in Canada data and the theoretical PDFs for the fitted distributions. The GOMBUR-1 shows near perfect alignments with Beta distribution. Kumaraswamy distribution fits the data. BMUR , Topp Leone, Unit Lindley distributions do not fit the data.

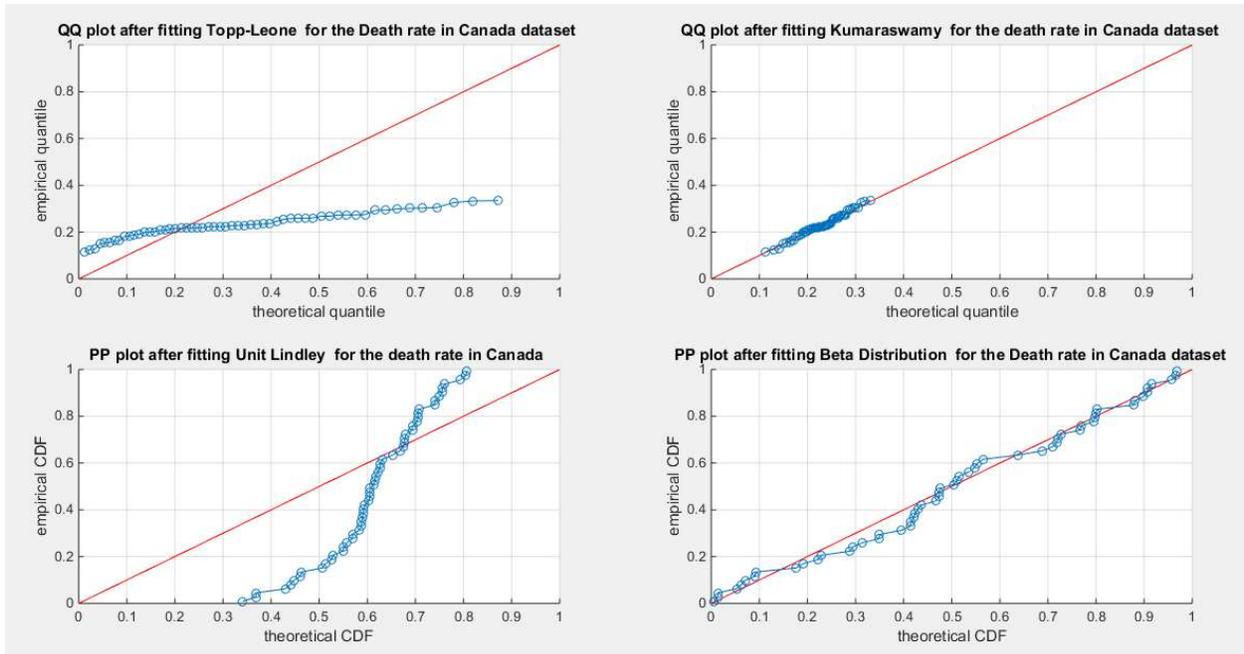

Fig. 33 shows the QQ plot for the fitted Topp Leone & Kumaraswamy distributions and the PP plot for the fitted Unit Lindley and Beta distribution for the death rate in Canada dataset.

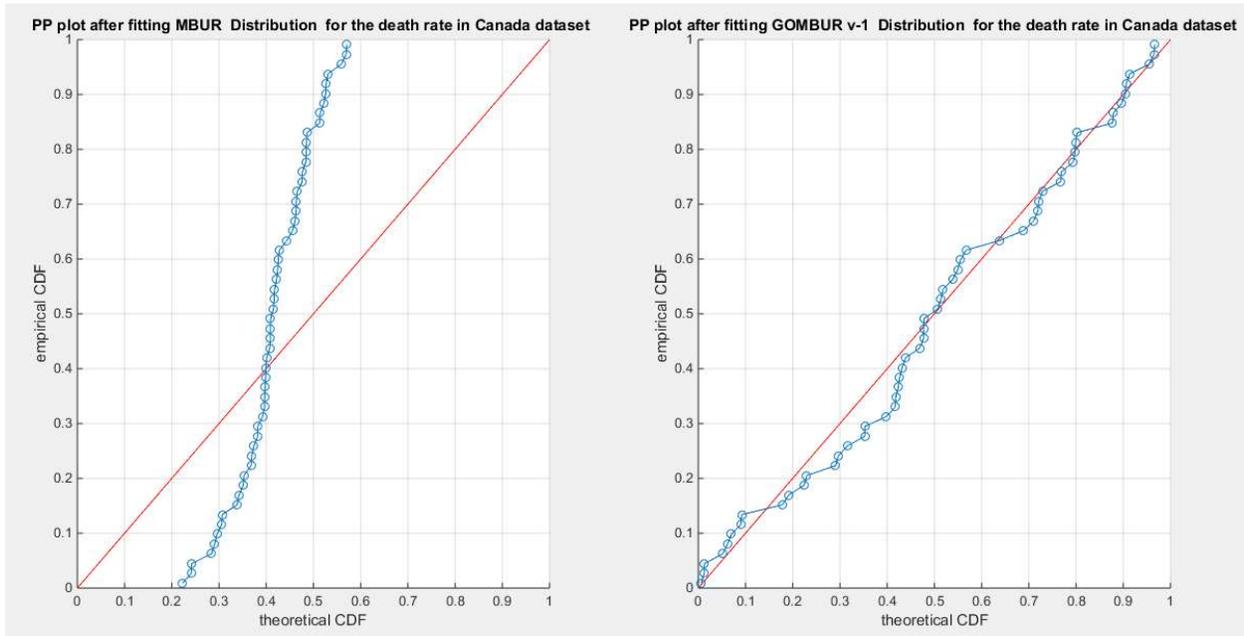

Fig. 34 shows the PP plot for the fitted MBUR & GOMBUR-1 for the death rate in Canada dataset. The Generalization of MBUR enhances the diagonal alignment of GOMBUR-1 along both ends and along the center of the distribution.

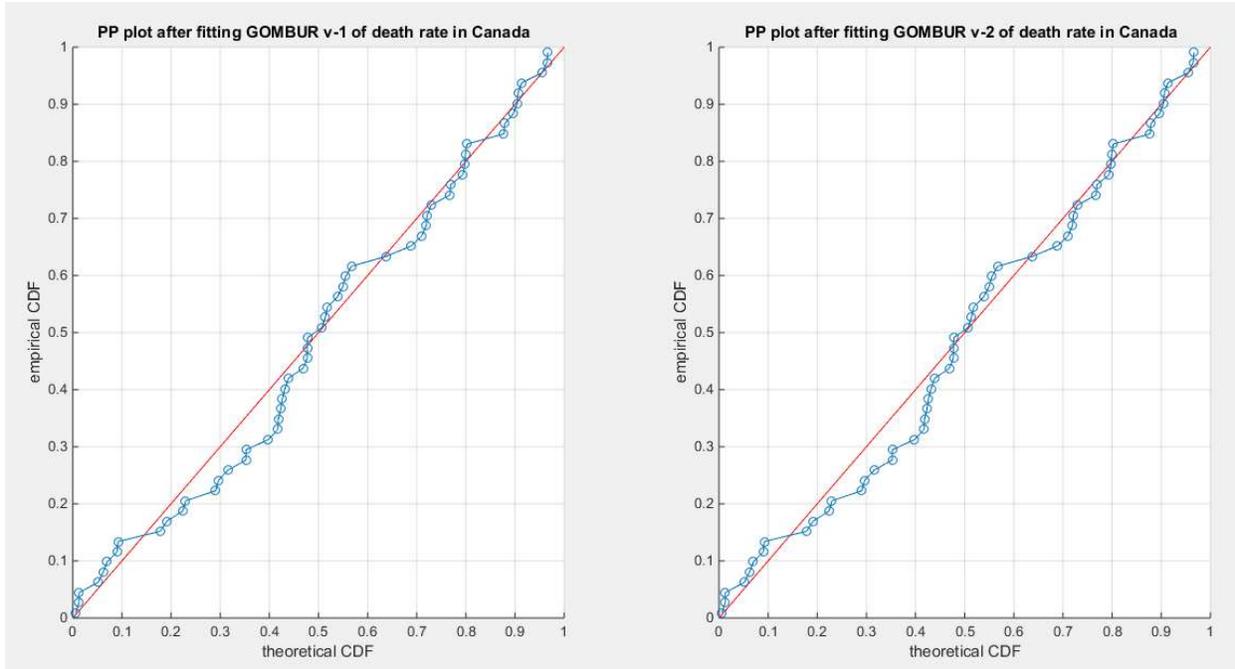

Fig. 35 shows the PP plot for the fitted GOMBUR-1 & GOMBUR-2 for death rate in Canada dataset. They are identical.

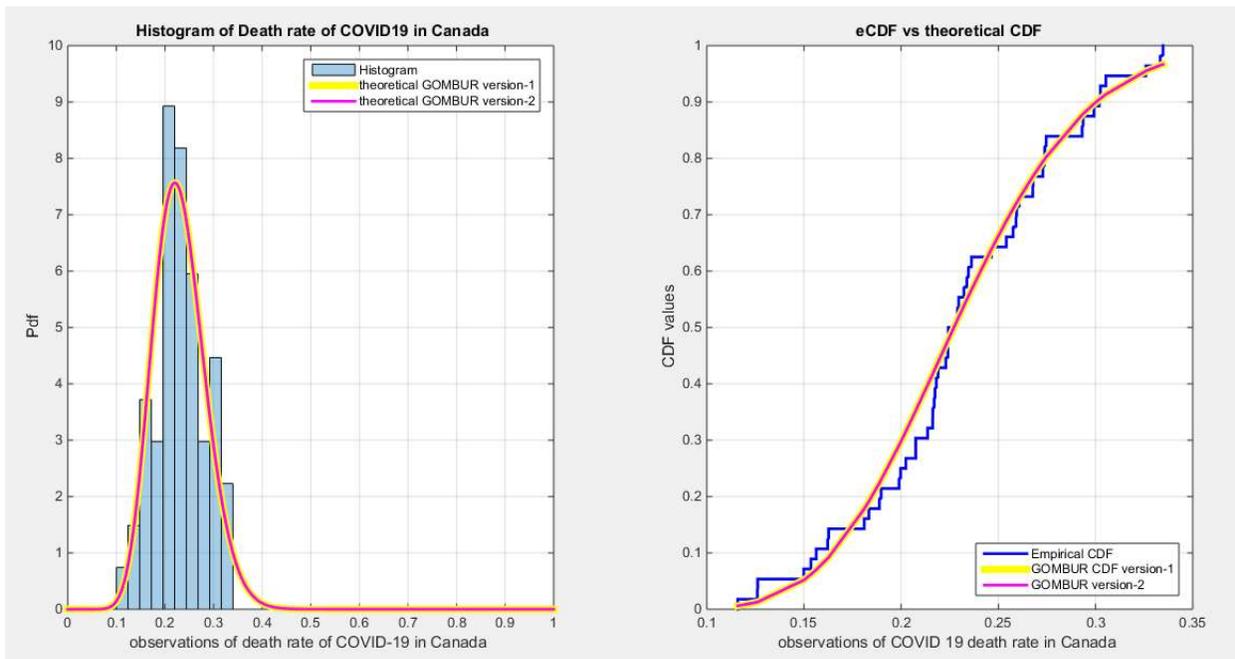

Fig. 36 shows on the left subplot the histogram of the death rate in Canada data and the fitted PDFs of both GOMBUR-1 & GOMBUR-2 and on the right subplot the e-CDFs and the theoretical CDFs for both distributions. Both the fitted CDFs and the fitted PDFs of both versions are identical.

Table (8) shows the results of COVID-19 death rates in Spain

Table (8): the results of analysis of COVID-19 death rates in Spain

|  | Beta | | Kumaraswamy | | MBUR | Topp-Leone | Unit-Lindley |
|---|---|---|---|---|---|---|---|
| theta | $\alpha = 4.8994$ | | $\alpha = 2.6847$ | | 1.2852 | 1.2218 | 2.9943 |
|  | $\beta = 12.7943$ | | $\beta = 22.1262$ | | | | |
| Variance | 1.4278 | 3.2731 | 0.071 | 1.6267 | 0.0032 | 0.0226 | 0.0945 |
|  | 3.2731 | 8.088 | 1.6267 | 44.6833 | | | |
| SE(a) | 0.1471 | | 0.0328 | | 0.007 | 0.0185 | 0.0378 |
| SE(b) | 0.3501 | | 0.8228 | | - | - | - |
| AIC | -1111486 | | -106.4226 | | -48.5631 | -47.7193 | -72.4535 |
| CAIC | -110.9581 | | -106.2321 | | -48.5006 | -47.6568 | -72.391 |
| BIC | -106.7692 | | -102.0433 | | -46.3734 | -45.5297 | -70.2639 |
| HQIC | -109.4181 | | -104.6921 | | -47.6978 | -46.8541 | -71.5883 |
| LL | 57.5743 | | 55.2113 | | 25.2815 | 24.8597 | 37.2268 |
| K-S | 0.1148 | | 0.1271 | | 0.2734 | 0.3038 | 0.2896 |
| $H_0$ | Fail to reject | | Fail to reject | | reject | reject | reject |
| P-value | 0.3243 | | 0.2172 | | 0 | 0 | 0 |
| AD | 1.052 | | 1.2437 | | 9.1314 | 10.5385 | 6.2515 |
| CVM | 0.0692 | | 0.0686 | | 2.8621 | 3.3539 | 2.5936 |
| determinant | 0.8345 | | 0.5267 | | - | - | - |

The analysis demonstrates that the Beta and Kumaraswamy distributions fit the data well as shown from the AIC, CAIC, BIS &HQIC. The MBUR, Topp Loene and Unit Lindley distributions did not fit the data. But the generalized form of MBUR in the form of GOMBUR-1 & 2 FIT THE DATA BETTER with more negative values for all the mentioned indices, lowered values for AD, CVM & KS statistics, and a larger Log-Likelihood value. This in addition with lower values of the determinants of the variance-covariance matrices in comparison with the values obtained of fitting Beta and Kumaraswamy distributions. Generalization renders the GOMBUR-1 &2 fitting the data more efficient than the Beta or Kumaraswamy fitting. Figure 37 illuminates the alignments of the theoretical CDFs of Beta and GOMBUR-1, both run in parallel with Kumaraswamy CDF. Figure 38 shows the same orientation. Figure 39 displays the PP plot and the QQ plot of the competitors. Figure 40 shows the improvement of the diagonal alignment of GOMBUR as an effect of generalizing the BMUR distribution which exhibit improper alignment. Figure 41 presents the identical PP plots for both versions of the GOMBUR. Figure 42 illustrates the similarity of the CDFs and PDFs of both versions. The PDF curve shows mild right skewness.

Table(8) : to be continued

| | GOMBUR-1 | | GOMBUR-2 | |
|---|---|---|---|---|
| theta | $n = 10.7533$ | | $n = 22.5065$ | |
| | $\alpha = 1.3801$ | | $\alpha = 1.3801$ | |
| Variance | 4.0831 | 0.0066 | 16.3323 | 0.0133 |
| | 0.0066 | 0.00065056 | 0.0133 | 0.00065056 |
| SE(n) | 0.2487 | | 0.4975 | |
| SE(a) | 0.0031 | | 0.0031 | |
| AIC | -111.7925 | | -111.7925 | |
| CAIC | -111.6020 | | -111.6020 | |
| BIC | -107.4132 | | -107.4132 | |
| HQIC | -110.062 | | -110.062 | |
| LL | 57.8962 | | 57.8962 | |
| K-S Value | 0.1105 | | 0.1105 | |
| H₀ | Fail to reject | | Fail to reject | |
| P-value | 0.3689 | | 0.3689 | |
| AD | 1.0013 | | 1.0013 | |
| CVM | 0.168 | | 0.168 | |
| Determinant | 0.0026 | | 0.0104 | |
| Significant(n) | P<0.001 | | P<0.001 | |
| Significant(a) | P<0.001 | | P<0.001 | |

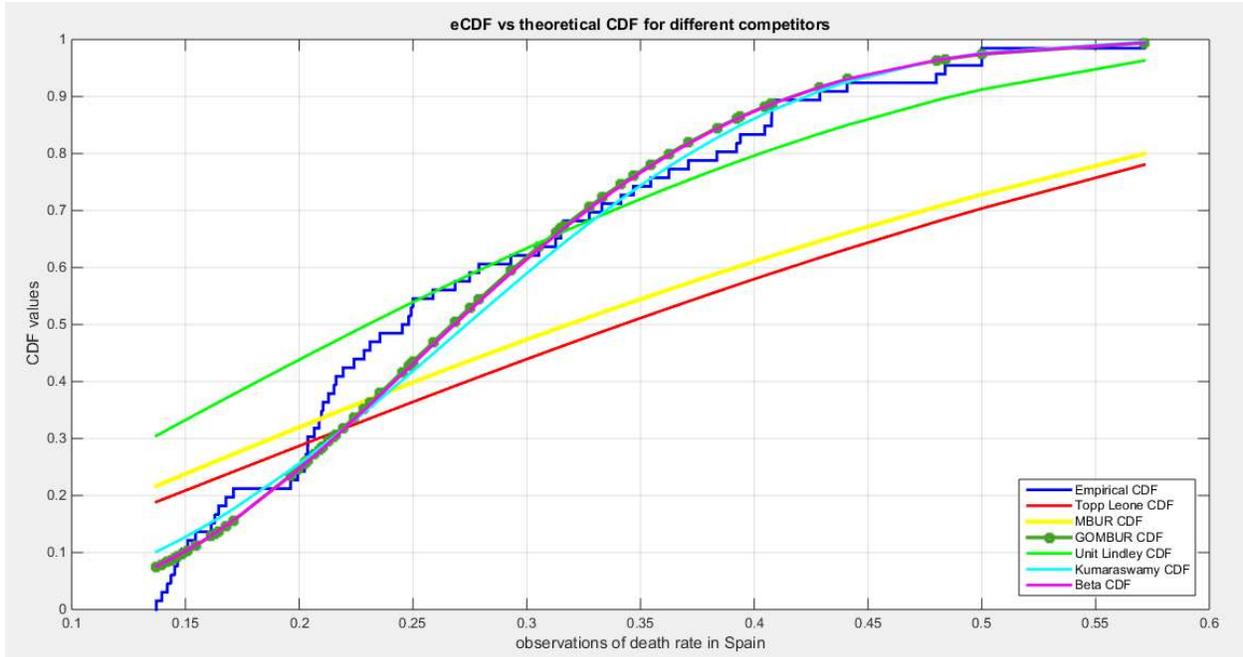

Fig. 37 shows the e-CDFs and the theoretical CDFs for the fitted distributions of death rate in Spain data.

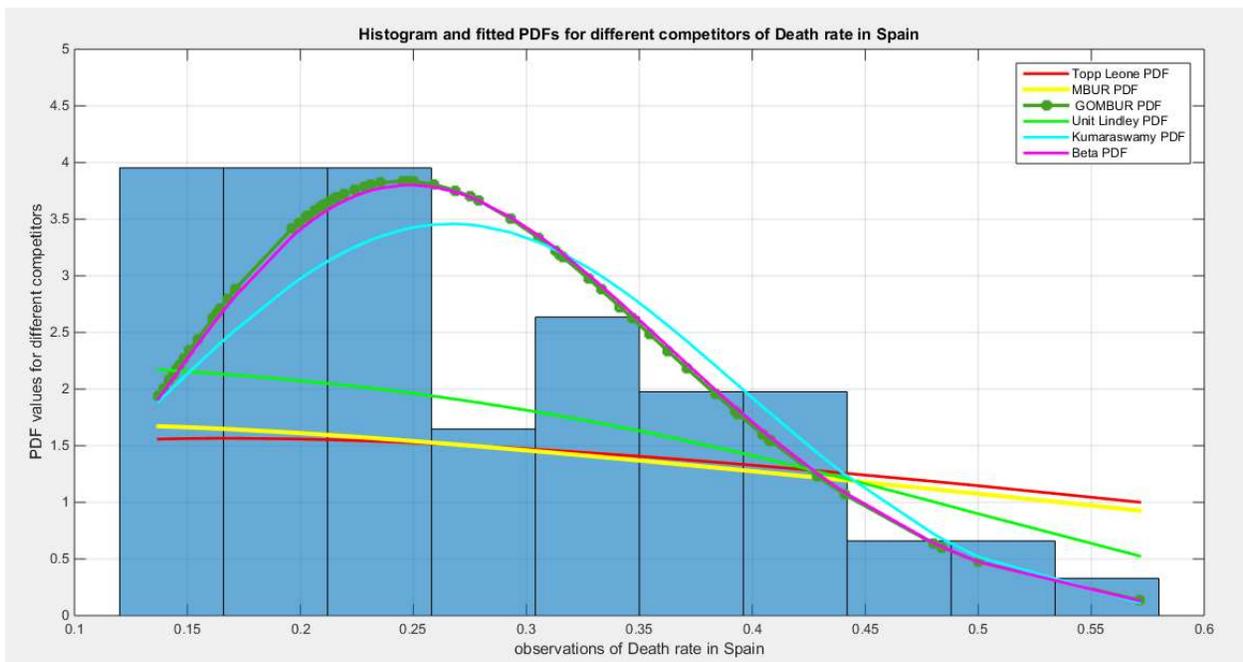

Fig. 38 shows the histogram of the COVID-19 death rate in Spain data and the theoretical PDFs for the fitted distributions. The GOMBUR-1 shows near perfect alignments with Beta distribution. Kumaraswamy distribution fits the data. BMUR , Topp Leone, Unit Lindley distributions do not fit the data.

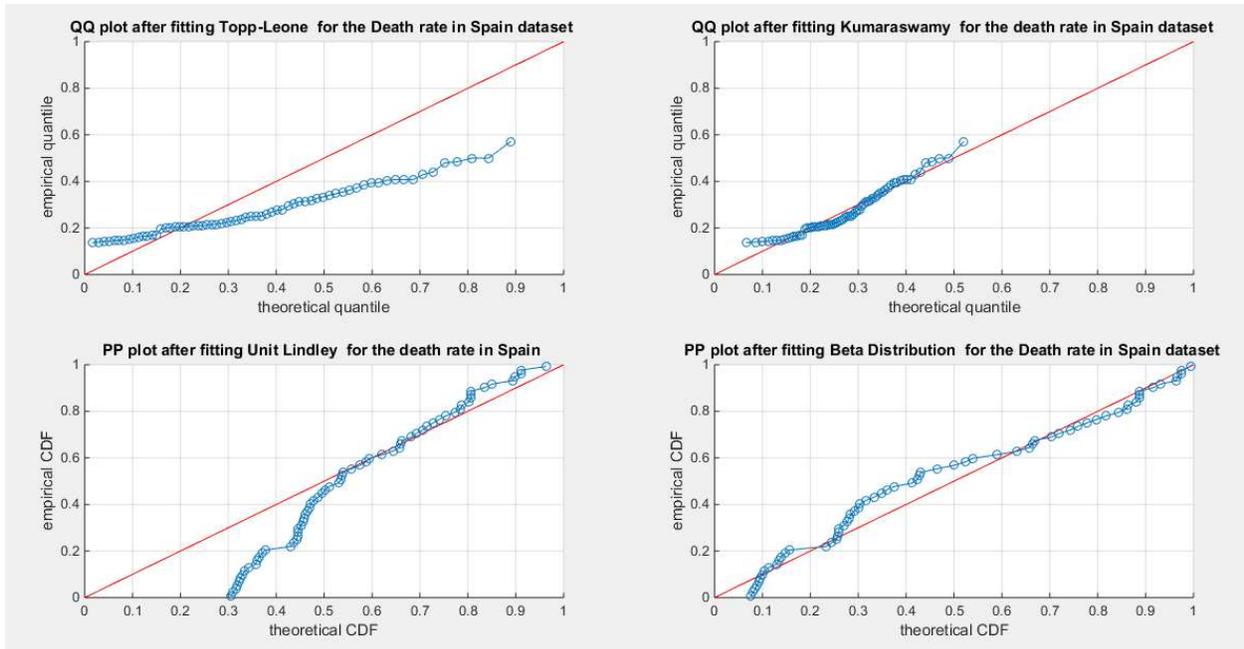

Fig. 39 shows the QQ plot for the fitted Topp Leone & Kumaraswamy distributions and the PP plot for the fitted Unit Lindley and Beta distribution for the death rate in Spain dataset.

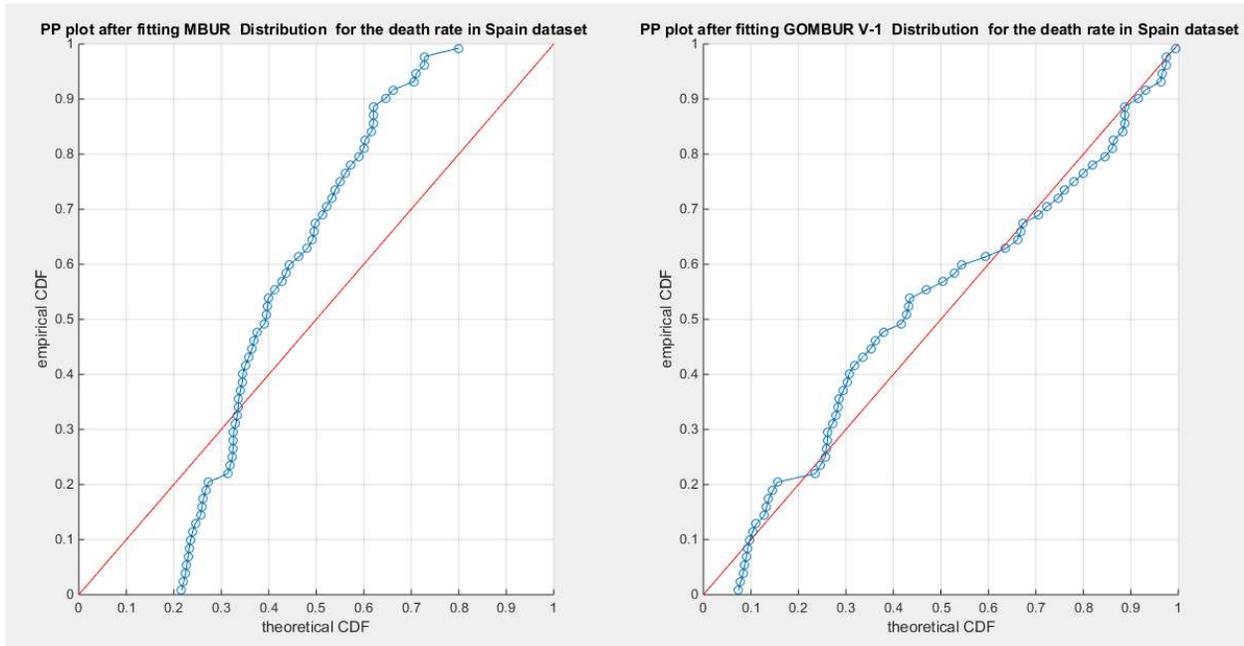

Fig. 40 shows the PP plot for the fitted MBUR & GOMBUR-1 for the death rate in Spain dataset. The Generalization of MBUR enhances the diagonal alignment of GOMBUR-1 along both ends and along the center of the distribution.

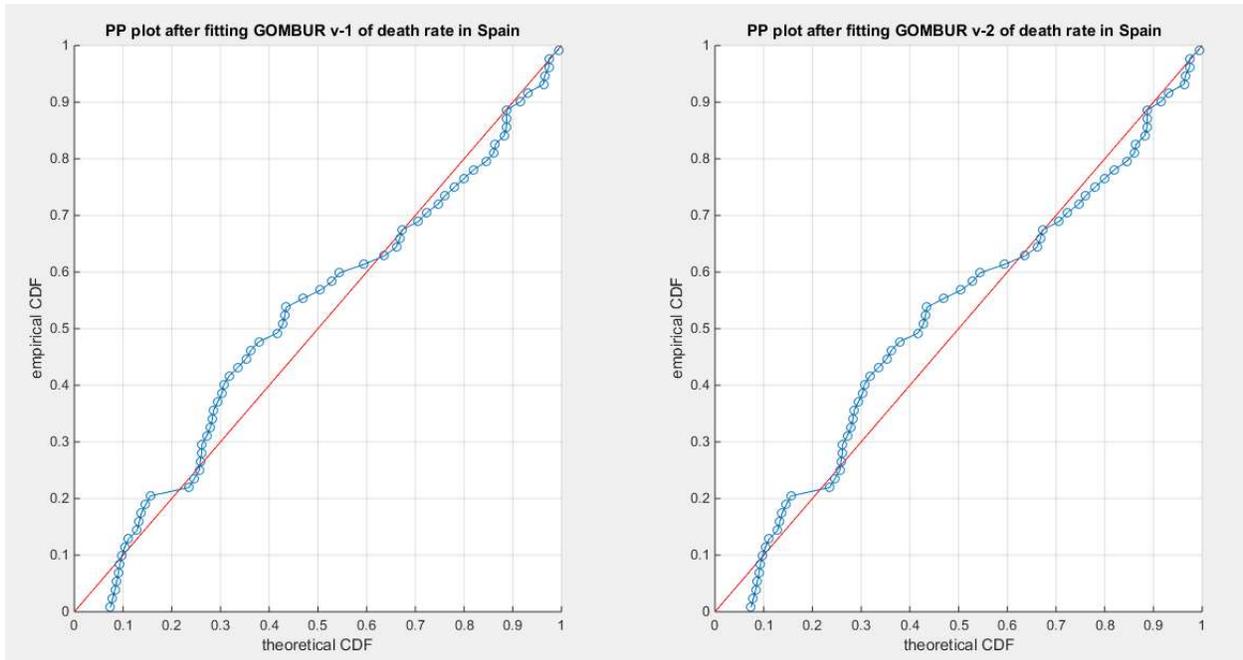

Fig. 41 shows the PP plot for the fitted GOMBUR-1 & GOMBUR-2 for death rate in Spain dataset. They are identical.

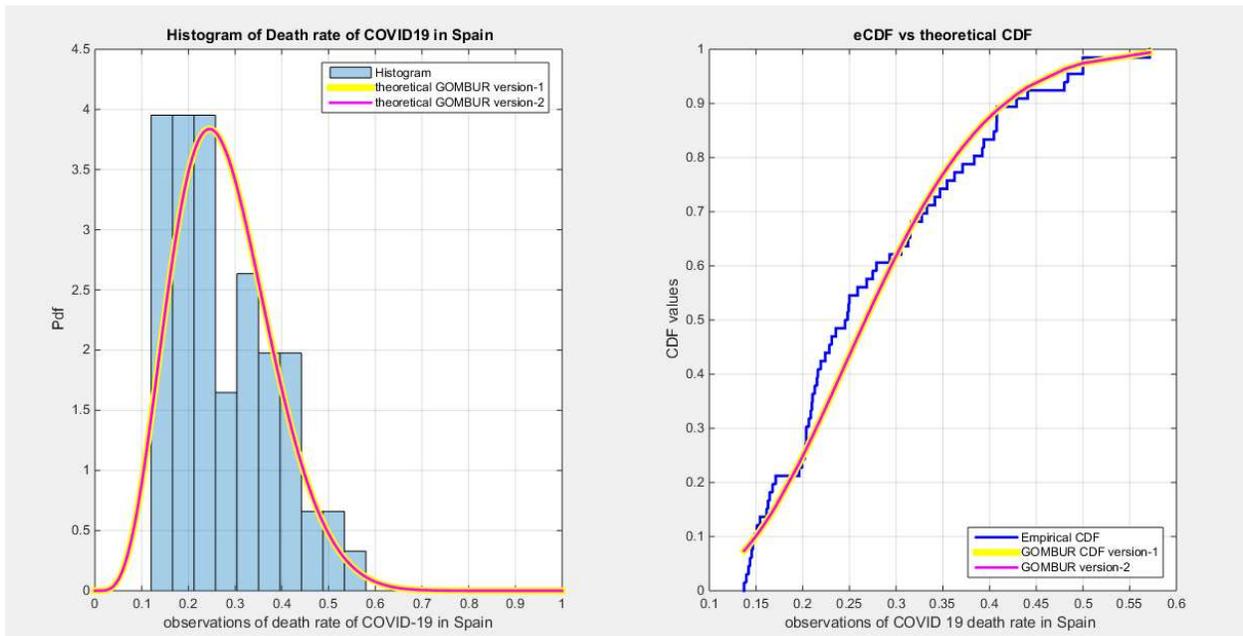

Fig. 42 shows on the left subplot the histogram of the death rate in Spain data and the fitted PDFs of both GOMBUR-1 & GOMBUR-2 and on the right subplot the e-CDFs and the theoretical CDFs for both distributions. Both the fitted CDFs and the fitted PDFs of both versions are identical. The PDF curve is slightly right skewed.

Table (9) shows the result of analysis of the COVID-19 death rates in United Kingdom

Table (9) shows the result of analysis of the COVID-19 death rates in United Kingdom

|  | Beta | | Kumaraswamy | | MBUR | Topp-Leone | Unit-Lindley |
|---|---|---|---|---|---|---|---|
| theta | $\alpha = 4.0503$ | | $\alpha = 2.6806$ | | 1.2733 | 1.2462 | 2.8294 |
|  | $\beta = 10.0137$ | | $\beta = 19.5873$ | | | | |
| Variance | 0.9001 | 2.3667 | 0.0905 | 1.8019 | 0.0035 | 0.0259 | 0.0918 |
|  | 2.3667 | 6.8204 | 1.8019 | 42.2606 | | | |
| SE(a) | 0.1225 | | 0.0388 | | 0.0035 | 0.0259 | 0.0918 |
| SE(b) | 0.3372 | | 0.8393 | | | | |
| AIC | -86.7989 | | -87.7288 | | -41.7679 | -41.3478 | -62.7546 |
| CAIC | -86.5884 | | -87.5183 | | -41.6989 | -41.2788 | -62.6856 |
| BIC | -82.6102 | | -83.5401 | | -39.6735 | -39.2534 | -60.6602 |
| HQIC | -85.1605 | | -86.0904 | | -40.9487 | -40.5285 | -61.9354 |
| LL | 45.3995 | | 45.8644 | | 21.8839 | 21.6739 | 32.3773 |
| K-S | 0.0982 | | 0.1054 | | 0.2786 | 0.3062 | 0.1894 |
| H₀ | Fail to reject | | Fail to reject | | reject | reject | reject |
| P-value | 0.5752 | | 0.4847 | | 0.00013 | 0.0000017 | 0.0103 |
| AD | 0.7351 | | 0.6298 | | 6.4895 | 7.4679 | 4.4878 |
| CVM | 0.1279 | | 0.1047 | | 1.1285 | 1.3921 | 0.7574 |
| determinant | 0.5377 | | 0.5789 | | - | - | - |

The analysis illustrates that the Kumaraswamy and Beta fit the data well while the MBUR, Topp Leaone and Unit Lindley do not fit the data. The GOMBUR-1& 2 fit the data better than the MBUR. It is the third distribution after the Beta and Kumaraswamy as if has the third negative values of AIC, CAIC, BIC & HQIC after Kumaraswamy and Beta distribution. But it has the lesser values of the determinant of the variance covariance matrix obtained after estimating the distribution in comparison with fitting the Beta and the Kumaraswamy distribution. Figure 43 shows the theoretical CDFs of the fitted distribution. The CDF of the GOMBUR-1 is perfectly aligned with the CDF of the Beta distribution. Figure 44 shows similar orientation but for the fitted PDFs. Figure 45 shows the PP plot and QQ plot for the competitor distributions. Figure 45 demonstrates the enhanced effects of generalization and how it improves the diagonal alignment of the GOMBUR-1 &2 than the alignment of MBUR. Figure 46 illuminates the similar PP plot for the two versions and lastly Figure 48 shows the similar fitted CDF and PDF for the two versions.

Table ( 9) to be continued

|  | GOMBUR-1 | | GOMBUR-2 | |
|---|---|---|---|---|
| theta | $n = 8.1471$ | | $n = 17.2952$ | |
|  | $\alpha = 1.3616$ | | $\alpha = 1.3616$ | |
| Variance | 2.7020 | 0.0072 | 10.8082 | 0.0144 |
|  | 0.0072 | 0.00090006 | 0.0144 | 0.00090006 |
| SE(n) | 0.2122 | | 0.4244 | |
| SE(a) | 0.0039 | | 0.0039 | |
| AIC | -86.6033 | | -86.6033 | |
| CAIC | -86.3928 | | -86.3928 | |
| BIC | -82.4146 | | -82.4146 | |
| HQIC | -84.9649 | | -84.9649 | |
| LL | 45.3017 | | 45.3017 | |
| K-S Value | 0.0971 | | 0.0971 | |
| $H_0$ | Fail to reject | | Fail to reject | |
| P-value | 0.5896 | | 0.5896 | |
| AD | 0.7562 | | 0.7562 | |
| CVM | 0.1319 | | 0.1319 | |
| Determinant | 0.0024 | | 0.0095 | |
| Significant(n) | P<0.001 | | P<0.001 | |
| Significant(a) | P<0.001 | | P<0.001 | |

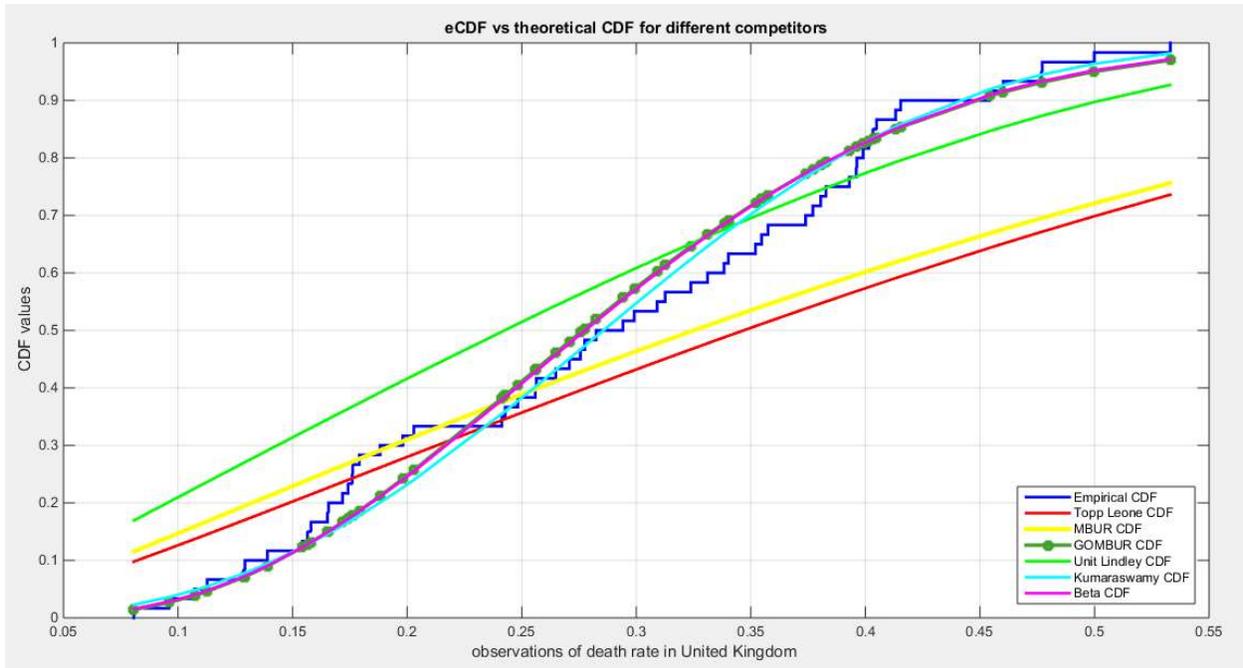

Fig. 43 shows the e-CDFs and the theoretical CDFs for the fitted distributions of death rate in UK data.

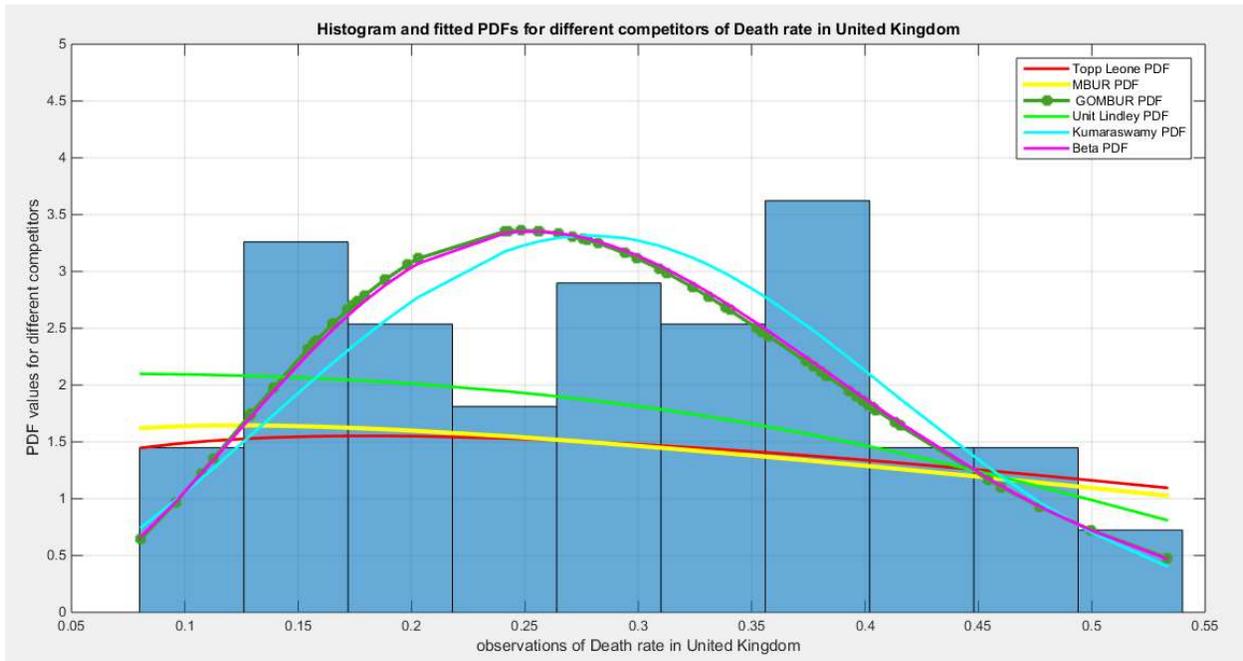

Fig. 44 shows the histogram of the COVID-19 death rate in UK data and the theoretical PDFs for the fitted distributions. The GOMBUR-1 shows near perfect alignments with Beta distribution. Kumaraswamy distribution fits the data. BMUR, Topp Leone, Unit Lindley distributions did not fit the data.

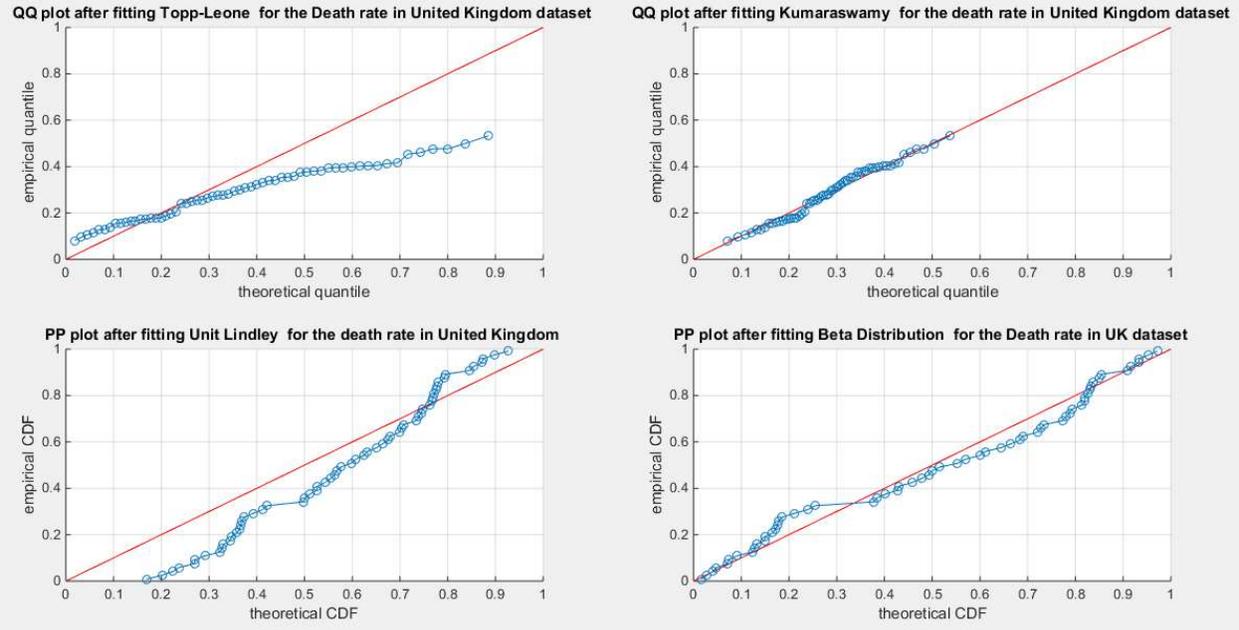

Fig. 45 shows the QQ plot for the fitted Topp Leone & Kumaraswamy distributions and the PP plot for the fitted Unit Lindley and Beta distribution for the death rate in UK dataset.

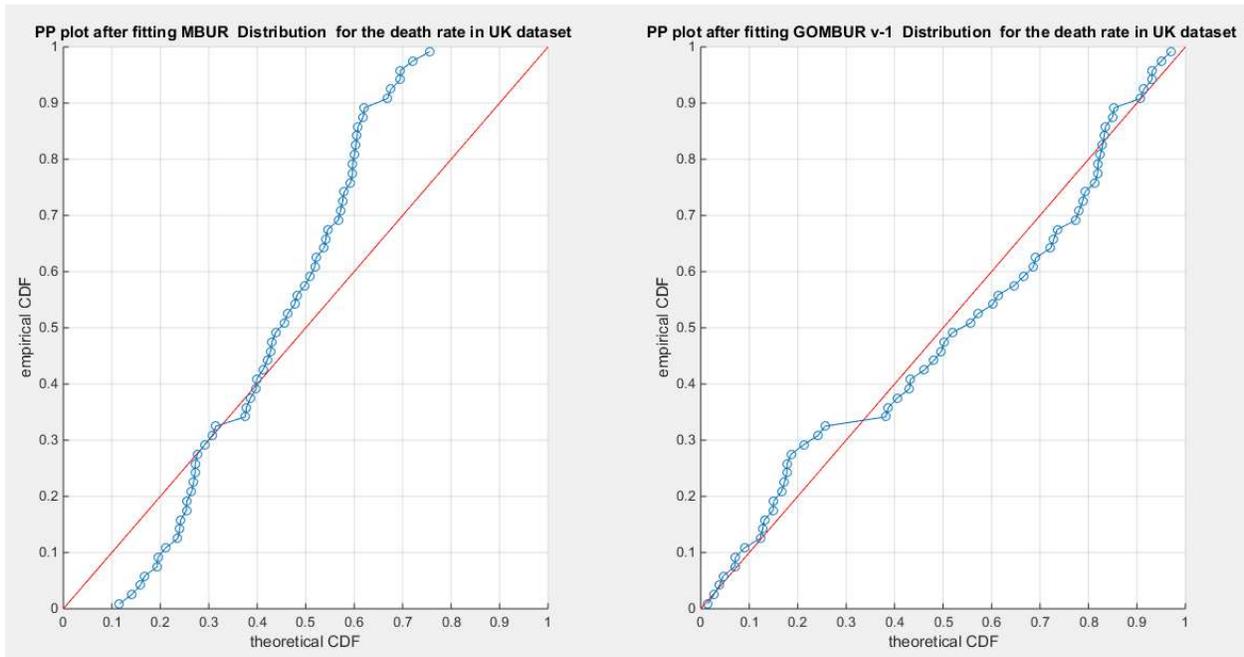

Fig. 46 shows the PP plot for the fitted MBUR & GOMBUR-1 for the death rate in UK dataset. The Generalization of MBUR enhances the diagonal alignment of GOMBUR-1 along both ends and along the center of the distribution.

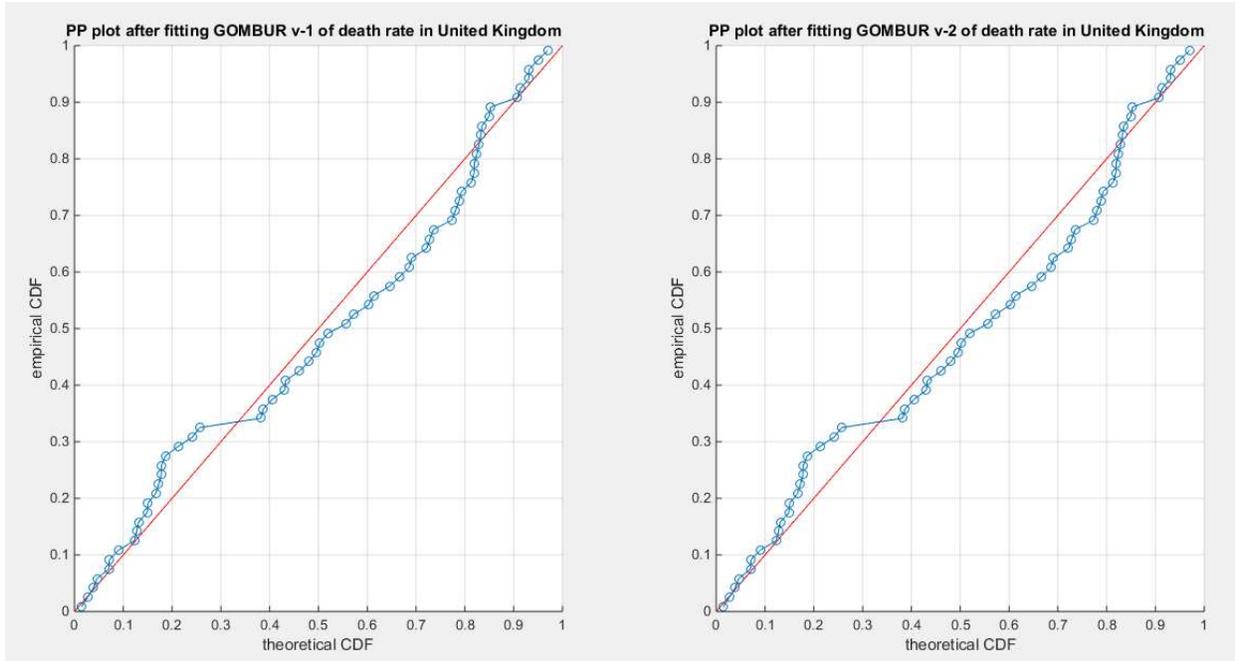

Fig. 47 shows the PP plot for the fitted GOMBUR-1 & GOMBUR-2 for death rate in UK dataset. They are identical

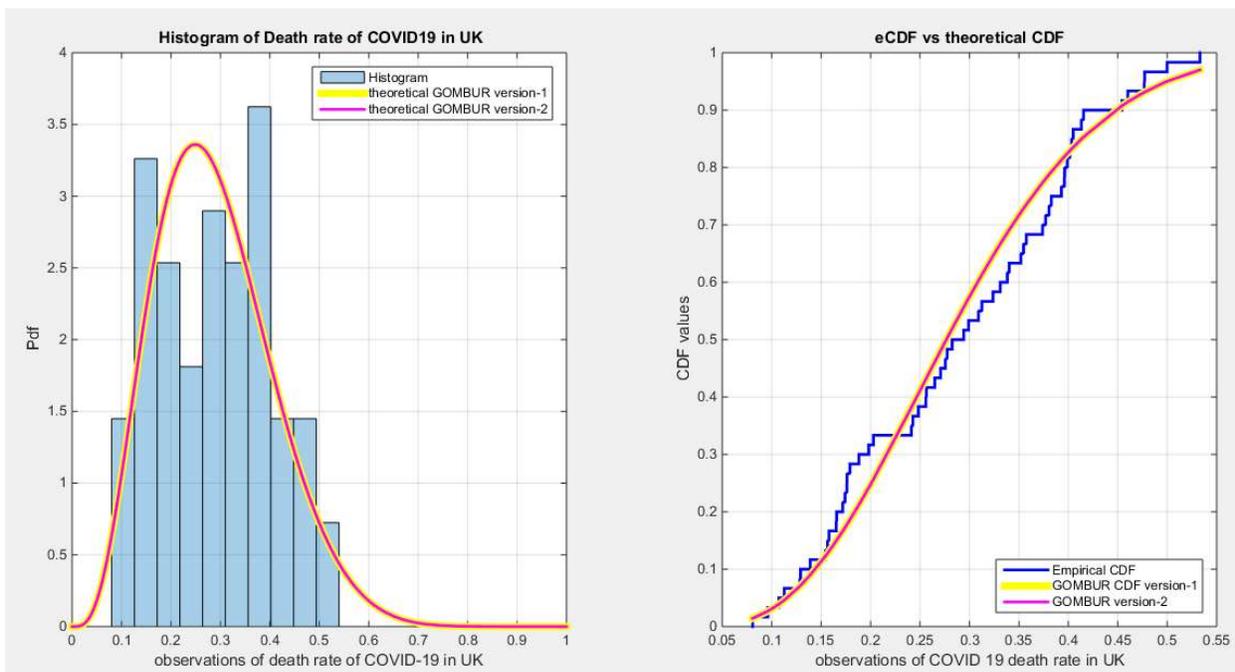

Fig. 48 shows on the left subplot the histogram of the death rate in UK data and the fitted PDFs of both GOMBUR-1 & GOMBUR-2 and on the right subplot the e-CDFs and the theoretical CDFs for both distributions. Both the fitted CDFs and the fitted PDFs of both versions are identical. The PDF curve is slightly right skewed.

Table (10) shows the result of analysis of the petroleum rock dataset

Table (10): result of analysis of the petroleum rock dataset

|  | Beta | | Kumaraswamy | | MBUR | Topp-Leone | Unit-Lindley |
|---|---|---|---|---|---|---|---|
| theta | $\alpha = 2.6515$ | | $\alpha = 1.6272$ | | 1.3776 | 1.0037 | 2.5672 |
|  | $\beta = 8.2085$ | | $\beta = 7.4583$ | | | | |
| Variance | 0.3055 | 0.2942 | 0.0329 | 0.2688 | 0.0051 | 0.021 | 0.0926 |
|  | 0.2942 | 0.4759 | 0.2688 | 3.3557 | | | |
| SE(a) | 0.0798 | | 0.0262 | | 0.0103 | 0.0209 | 0.0439 |
| SE(b) | 0.0996 | | 0.2644 | | - | - | - |
| AIC | -65.142 | | -58.3691 | | -38.7526 | -36.1927 | -19.068 |
| CAIC | -64.8753 | | -58.1025 | | -38.6056 | -36.1057 | -18.9811 |
| BIC | -61.3996 | | -54.6267 | | -36.8814 | -34.3215 | -17.1968 |
| HQIC | -63.7277 | | -56.9549 | | -38.0455 | -35.4855 | -18.3609 |
| LL | 34.571 | | 31.1846 | | 20.3763 | 19.0963 | 10.534 |
| K-S | 0.1822 | | 0.1975 | | 0.316 | 0.3518 | 0.3033 |
| $H_0$ | Fail to reject | | Fail to reject | | reject | reject | reject |
| P-value | 0.0723 | | 0.0408 | | 0 | 0 | 0 |
| AD | 3.0872 | | 3.9213 | | 8.3144 | 9.6277 | 7.5927 |
| CVM | 0.5351 | | 0.6702 | | 1.6507 | 1.9901 | 1.4343 |
| determinant | 0.0588 | | 0.0381 | | - | - | - |

The analysis revealed that the Beta and Kumaraswamy distributions fitted the data well. MBUR , Topp Leaone and Unit Lindley did not fit the data. The GOMBUR-1& 2 fitted the data with values of AIC, CAIC, BIC, HQIC far exceeding the MBUR fitting. Both generalized forms outperform the Beta and the Kumaraswamy fitting. Both versions of generalization have higher value of the Log-likelihood than any other distribution. Moreover, the lowest value of the determinant of the variance covariance matrix after fitting the GOMBER-1 distribution renders it the most efficient distribution among all distributions. It also has the lower values of the test statistics AD, CVM & KS. Figure 49 shows the perfect alignment of GOMBUR-1 and Beta CDFs curves along with the Kumaraswamy CDF. This pattern is also obvious for the fitted PDFs in Figure 50. The PP plots and the QQ plots are shown in Figure 51 while in Figure 52 the enhanced alignment of the diagonal is spectacle after fitting the GOMBUR-1 than after fitting the MBUR. In Figure 53, the similarity of PP plots between the two versions of generalization is recognizable. Figure 54 discloses the theoretical CDFs and fitted PDFs of GOMBUR-1& 2 which are similar to each other.

Table (10) to be continued:

| | GOMBUR-1 | | GOMBUR-2 | |
|---|---|---|---|---|
| theta | $n = 6.7991$ | | $n = 14.5982$ | |
| | $\alpha = 1.469$ | | $\alpha = 1.469$ | |
| Variance | 2.4435 | 0.0097 | 9.7742 | 0.0194 |
| | 0.0097 | 0.0015 | 0.0194 | 0.0015 |
| SE(n) | 0.2256 | | 0.4513 | |
| SE(a) | 0.0057 | | 0.0057 | |
| AIC | -68.1668 | | -68.1668 | |
| CAIC | -67.9002 | | -67.9002 | |
| BIC | -64.4244 | | -64.4244 | |
| HQIC | -66.7526 | | -66.7526 | |
| LL | 36.0834 | | 36.0834 | |
| K-S Value | 0.1688 | | 0.1688 | |
| $H_0$ | Fail to reject | | Fail to reject | |
| P-value | 0.1152 | | 0.1152 | |
| AD | 2.7045 | | 2.7045 | |
| CVM | 0.4554 | | 0.4554 | |
| Determinant | 0.0037 | | 0.0147 | |
| Significant(n) | P<0.001 | | P<0.001 | |
| Significant(a) | P<0.001 | | P<0.001 | |

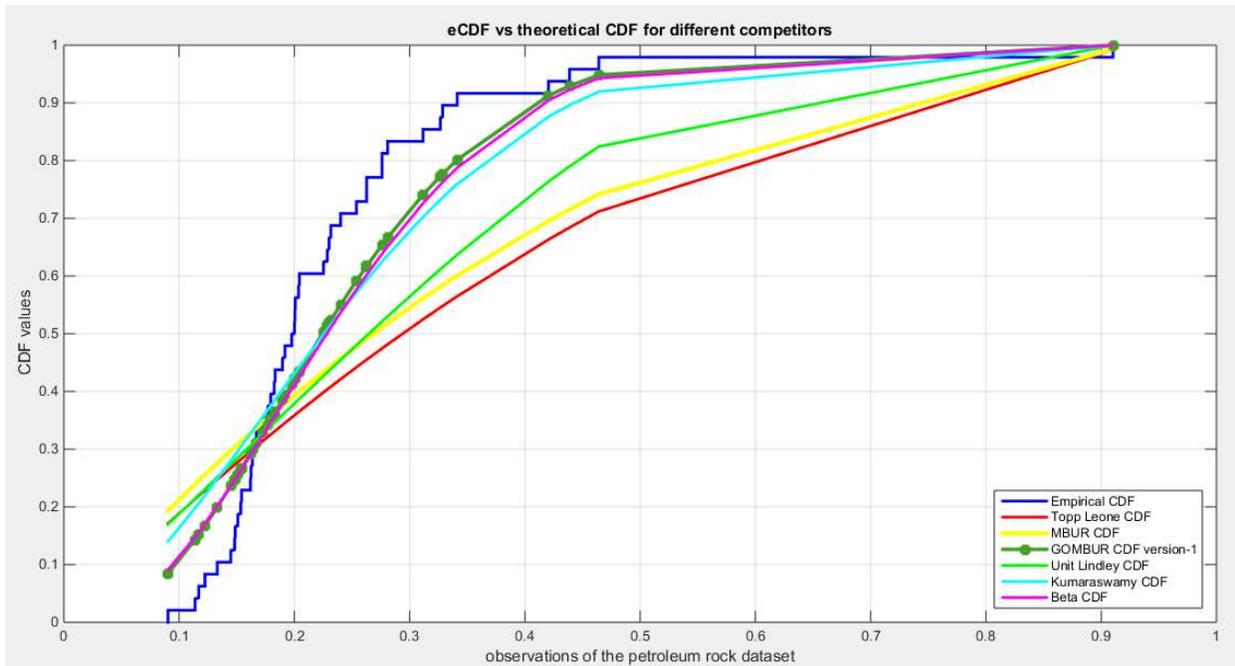

Fig. 49 shows the e-CDFs and the theoretical CDFs for the fitted distributions of petroleum rock data.

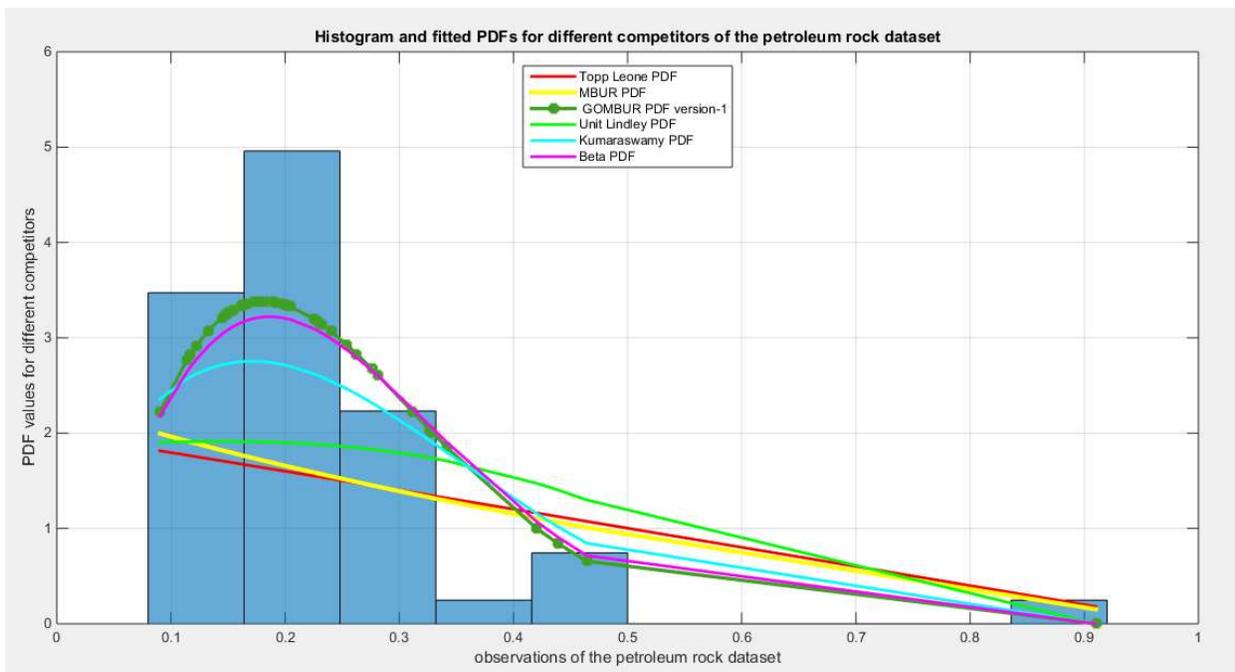

Fig. 50 shows the histogram of the petroleum rock data and the theoretical PDFs for the fitted distributions. The GOMBUR-1 shows near perfect alignments with Beta distribution. Kumaraswamy distribution fits the data. BMUR , Topp Leone, Unit Lindley distributions did not fit the data.

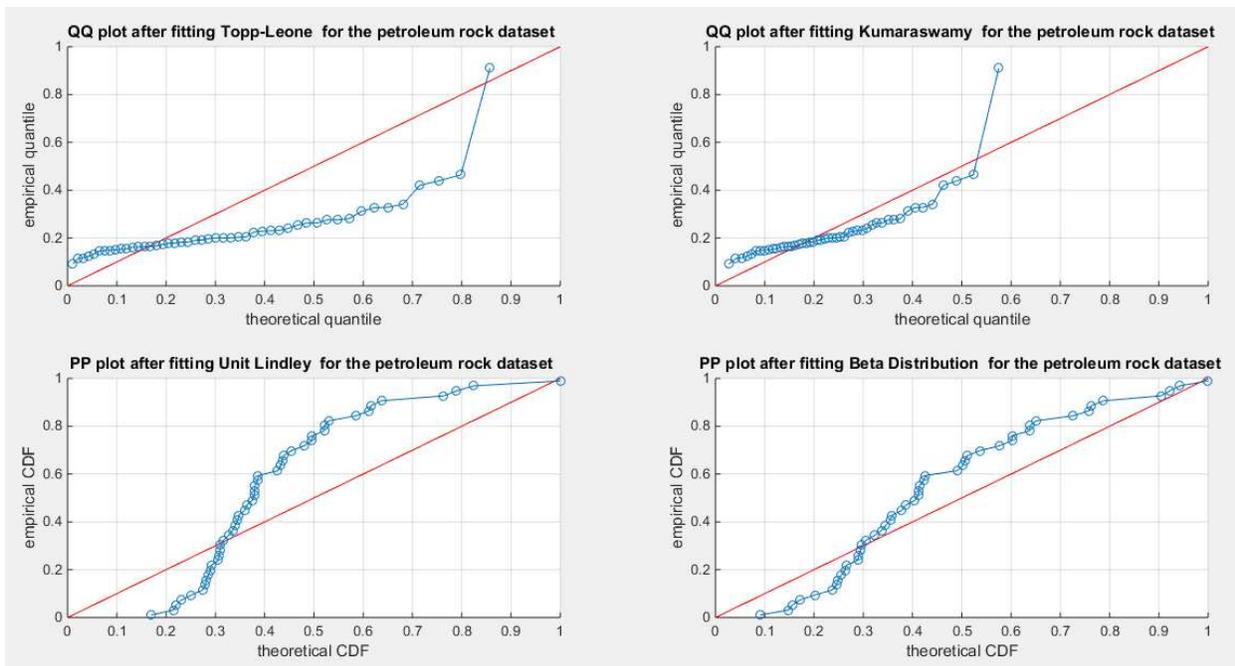

Fig. 51 shows the QQ plot for the fitted Topp Leone & Kumaraswamy distributions and the PP plot for the fitted Unit Lindley and Beta distribution for the petroleum rock dataset.

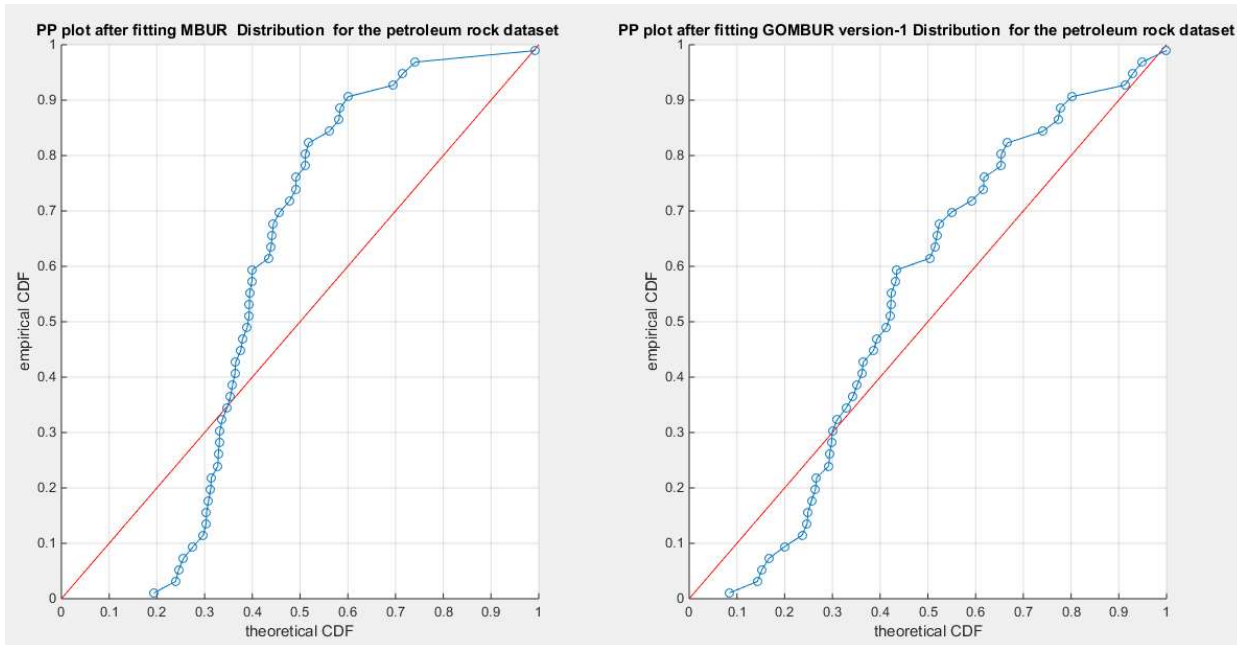

Fig. 52 shows the PP plot for the fitted MBUR & GOMBUR-1 for the petroleum rock dataset. The Generalization of MBUR enhances the diagonal alignment of GOMBUR-1 along both ends and along the center of the distribution.

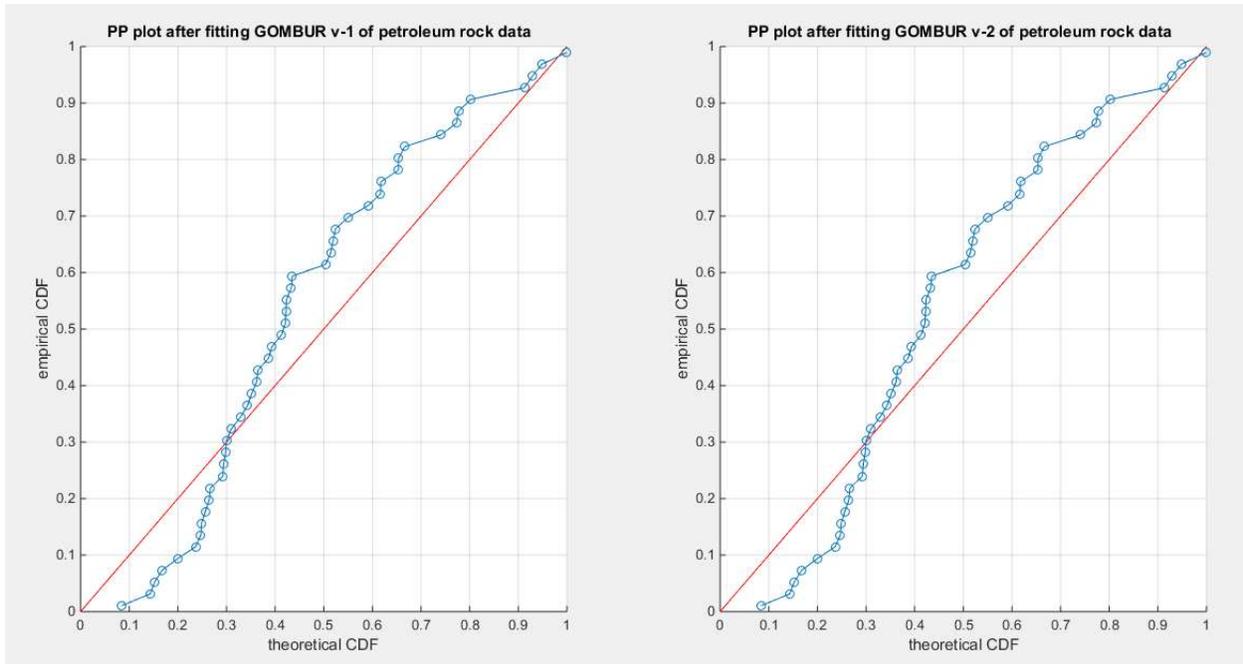

Fig. 53 shows the PP plot for the fitted GOMBUR-1 & GOMBUR-2 for petroleum rock dataset. They are identical

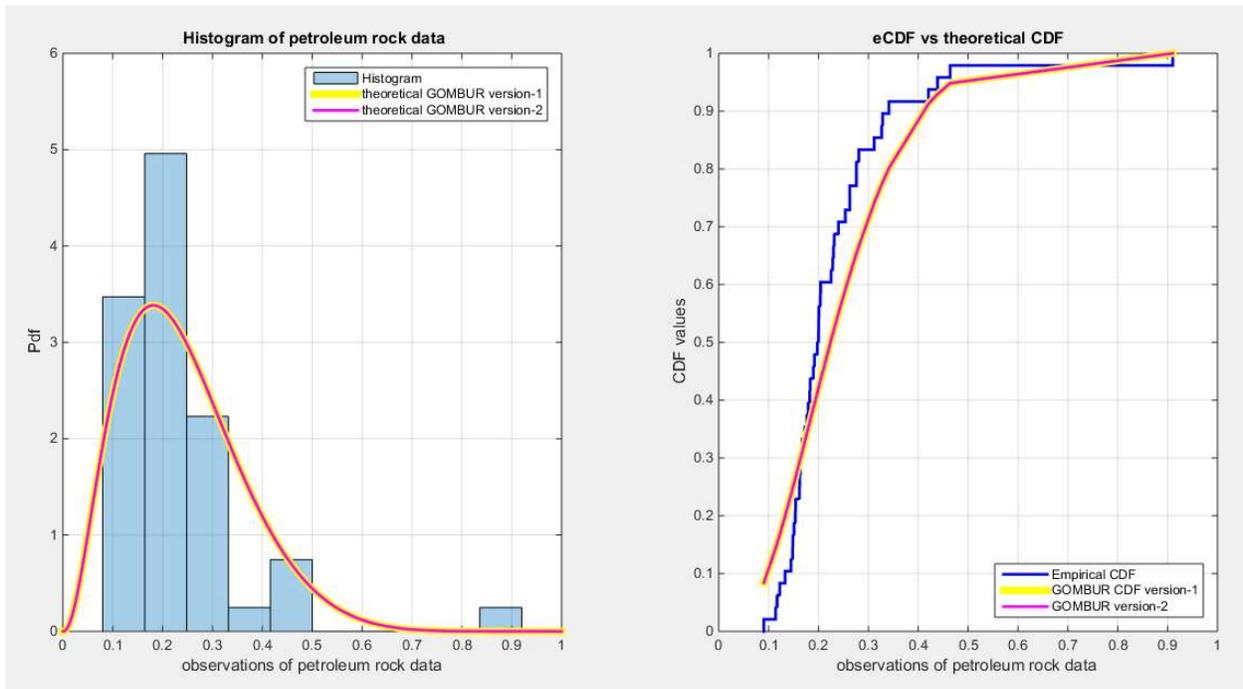

Fig. 54 shows on the left subplot the histogram of the petroleum rock and the fitted PDFs of both GOMBUR-1 & GOMBUR-2 and on the right subplot the e-CDFs and the theoretical CDFs for both distributions. Both the fitted CDFs and the fitted PDFs of both versions are identical. The PDF curve is right skewed.

Table (11) shows the result of analysis of the snow fall dataset

Table (11) shows the result of analysis of the snow fall dataset

|  | Beta | | Kumaraswamy | | MBUR | Topp-Leone | Unit-Lindley |
|---|---|---|---|---|---|---|---|
| theta | $\alpha = 0.8576$ | | $\alpha = 0.8615$ | | 1.8869 | 0.4352 | 8.43 |
|  | $\beta = 7.8057$ | | $\beta = 6.8358$ | | | | |
| Variance | 0.0684 | 0.4178 | 0.0189 | 0.2697 | 0.0153 | 0.0063 | 1.9726 |
|  | 0.4178 | 4.071 | 0.2697 | 5.3991 | | | |
| SE(a) | 0.0478 | | 0.0251 | | 0.0226 | 0.0145 | 0.2564 |
| SE(b) | 0.3684 | | 0.4242 | | - | - | - |
| AIC | -75.1218 | | -75.5952 | | -66.6947 | -60.8902 | -72.7696 |
| CAIC | -74.6773 | | -75.1508 | | -66.5518 | -60.7473 | -72.6268 |
| BIC | -72.3194 | | -72.7928 | | -65.2935 | -59.489 | -71.3684 |
| HQIC | -74.2253 | | -74.6987 | | -66.2464 | -60.4419 | -72.3214 |
| LL | 39.5609 | | 39.7976 | | 34.3474 | 31.4451 | 37.3848 |
| K-S | 0.1288 | | 0.1207 | | 0.2472 | 0.3015 | 0.1715 |
| $H_0$ | Fail to reject | | Fail to reject | | Fail to reject | reject | Fail to reject |
| P-value | 0.6554 | | 0.7294 | | 0.0422 | 0.0065 | 0.3045 |
| AD | 0.5715 | | 0.4728 | | 2.2335 | 3.4324 | 1.6564 |
| CVM | 0.0861 | | 0.0646 | | 0.396 | 0.6623 | 0.2918 |
| determinant | 0.1040 | | 0.0295 | | - | - | - |

The analysis discloses that the Beta, Kumaraswamy, MBUR and Unit Lindley distributions fit the data well but Topp Leone does not fit the data. The GOMBUR-1 & 2 fit the data better than any other distributions because both generalized forms have the highest negative values of AIC, CAIC, BIC & HQIC . They also have the highest Log-Likelihood value among them. In addition, they have the lowest values of the following tests: AD, CVM & KS. However, GOMBUR-1 has the lowest value of the determinant of variance covariance matrix. This solidifies GOMBUR-1 the most efficient distribution to fit the data well. Figure 55 illuminates the impeccable alignment of the theoretical CDF curves of the Beta and Kumaraswamy. The MBUR properly aligns with the two curves mainly at the lower end while the Unit Lindley unimpeachably aligns with the same curves mainly at the upper end. However, the GOMBUR-1 distribution faultlessly aligns along the ends and along the center of the two curves. Figure 56 exhibit the same manner. Figure 57 illustrates the PP plots and QQ plots for the competitor distributions.

Table (11) to be continued

|  | GOMBUR-1 | | GOMBUR-2 | |
|---|---|---|---|---|
| theta | $n = 4.6735$ | | $n = 10.3469$ | |
|  | $\alpha = 1.9954$ | | $\alpha = 1.9954$ | |
| Variance | 2.043 | 0.021 | 8.172 | 0.042 |
|  | 0.021 | 0.0063 | 0.042 | 0.0063 |
| SE(n) | 0.261 | | 0.5219 | |
| SE(a) | 0.0145 | | 0.0145 | |
| AIC | -77.3883 | | -77.3883 | |
| CAIC | -76.9439 | | -76.9439 | |
| BIC | -74.5889 | | -74.5889 | |
| HQIC | -76.4918 | | -76.4918 | |
| LL | 40.6942 | | 40.6942 | |
| K-S Value | 0.1011 | | 0.1011 | |
| H₀ | Fail to reject | | Fail to reject | |
| P-value | 0.8889 | | 0.8889 | |
| AD | 0.3453 | | 0.3453 | |
| CVM | 0.0443 | | 0.0443 | |
| Determinant | 0.0125 | | 0.0498 | |
| Significant(n) | P<0.001 | | P<0.001 | |
| Significant(a) | P<0.001 | | P<0.001 | |

Figure 58 high-lights the enhancing effect of generalization in sense of proper diagonal alignment at both ends and center after fitting the GOMBUR-1 in comparison with MBUR, although the latter distribution fits the data. Figure 59 depicts the resemblance of the PP plots of both versions of GOMBUR. Figure 60 emphasizes the indistinguishable curves of theoretical CDFs and fitted PDFs of both versions. The PDF curve is highly right skewed.

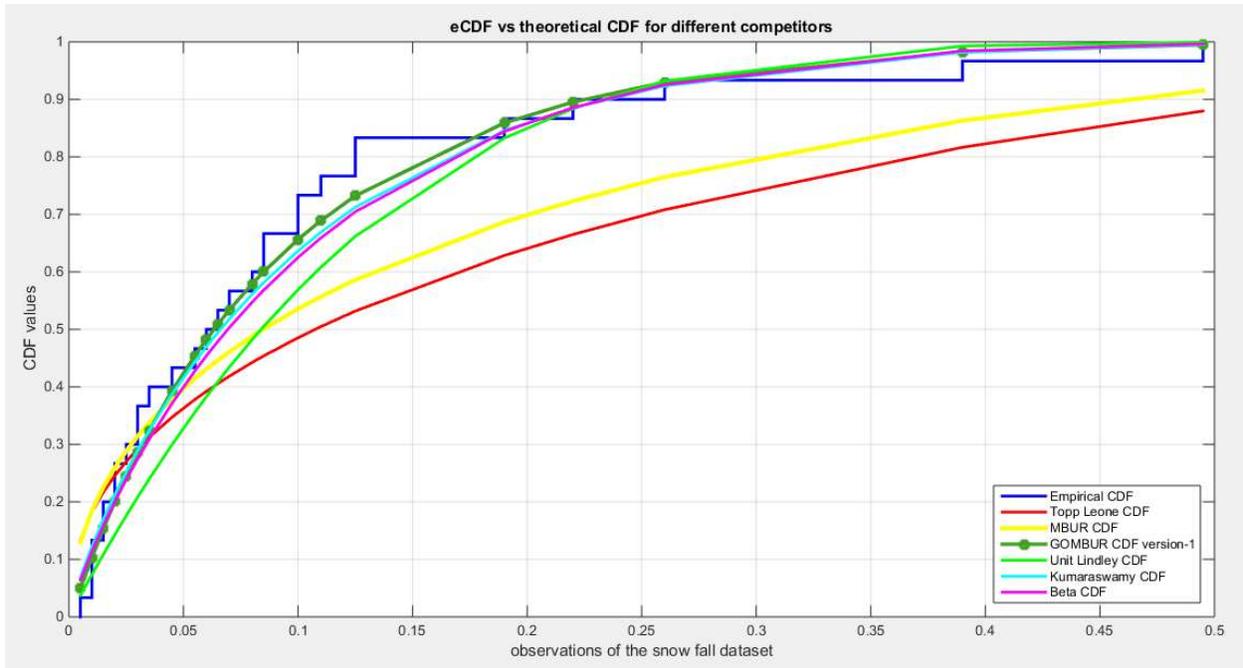

Fig. 55 shows the e-CDFs and the theoretical CDFs for the fitted distributions of snow fall data.

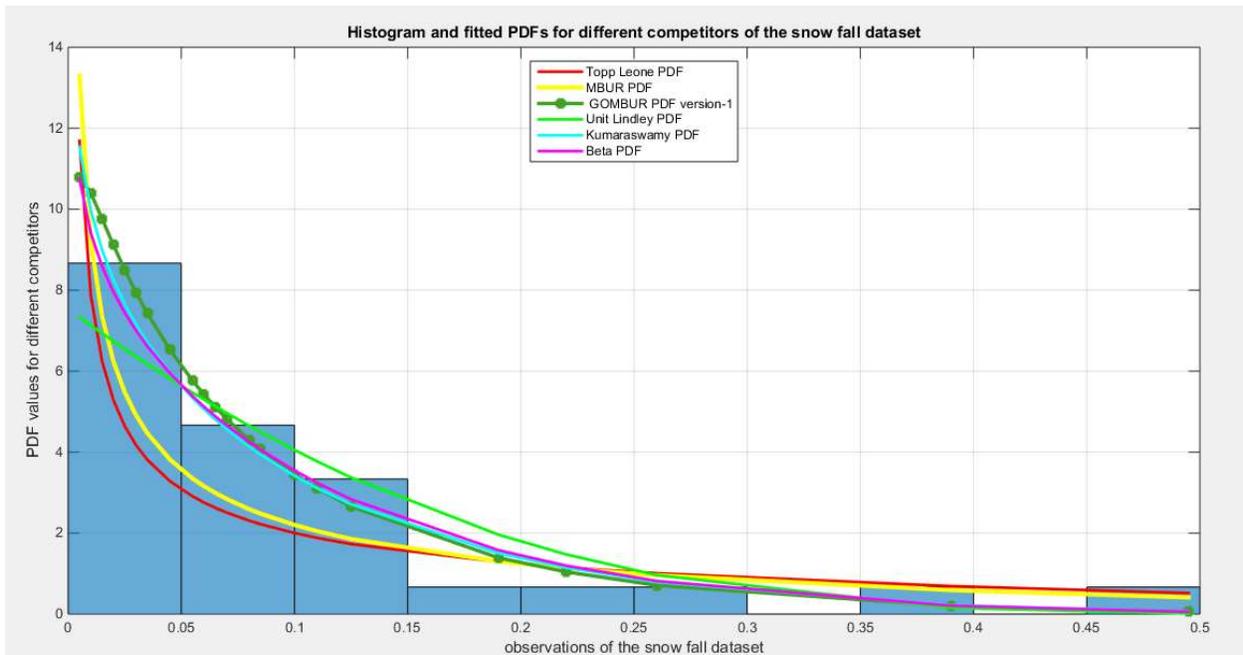

Fig. 56 shows the histogram of the snow fall data and the theoretical PDFs for the fitted distributions. The GOMBUR-1 shows near perfect alignments with Beta distribution. Kumaraswamy distribution fits the data. BMUR and Unit Lindley distributions also fit the data but Topp Leone does not fit the data.

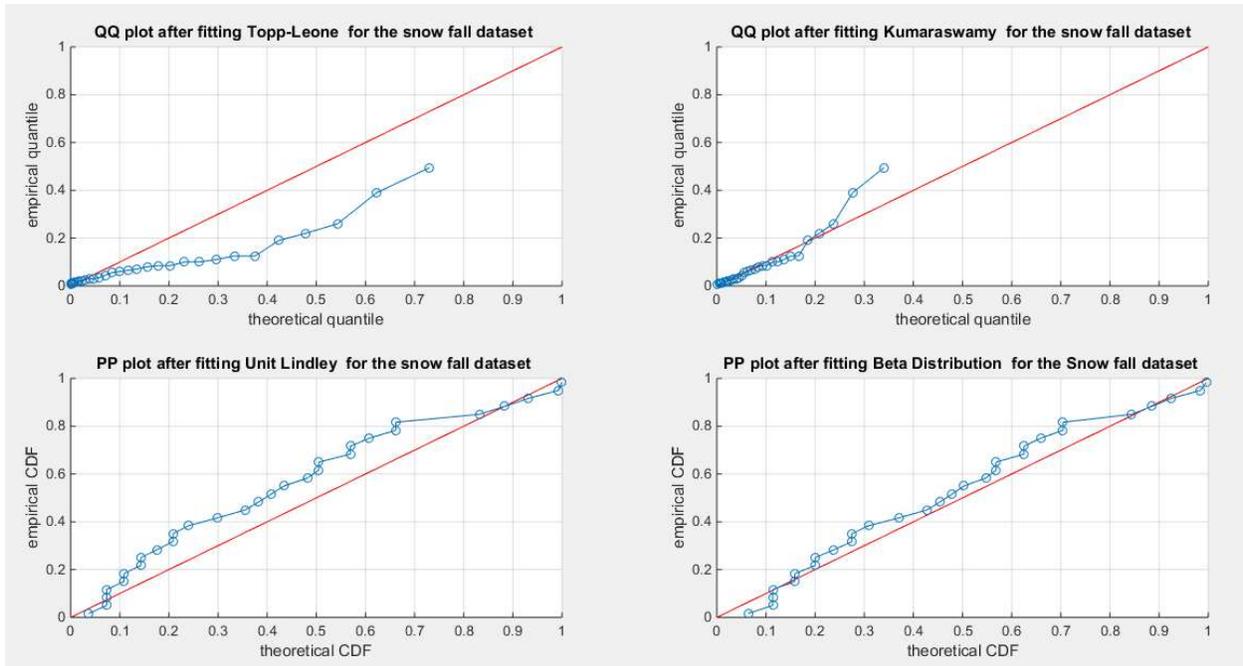

Fig. 57 shows the QQ plot for the fitted Topp Leone & Kumaraswamy distributions and the PP plot for the fitted Unit Lindley and Beta distribution for the snow fall dataset.

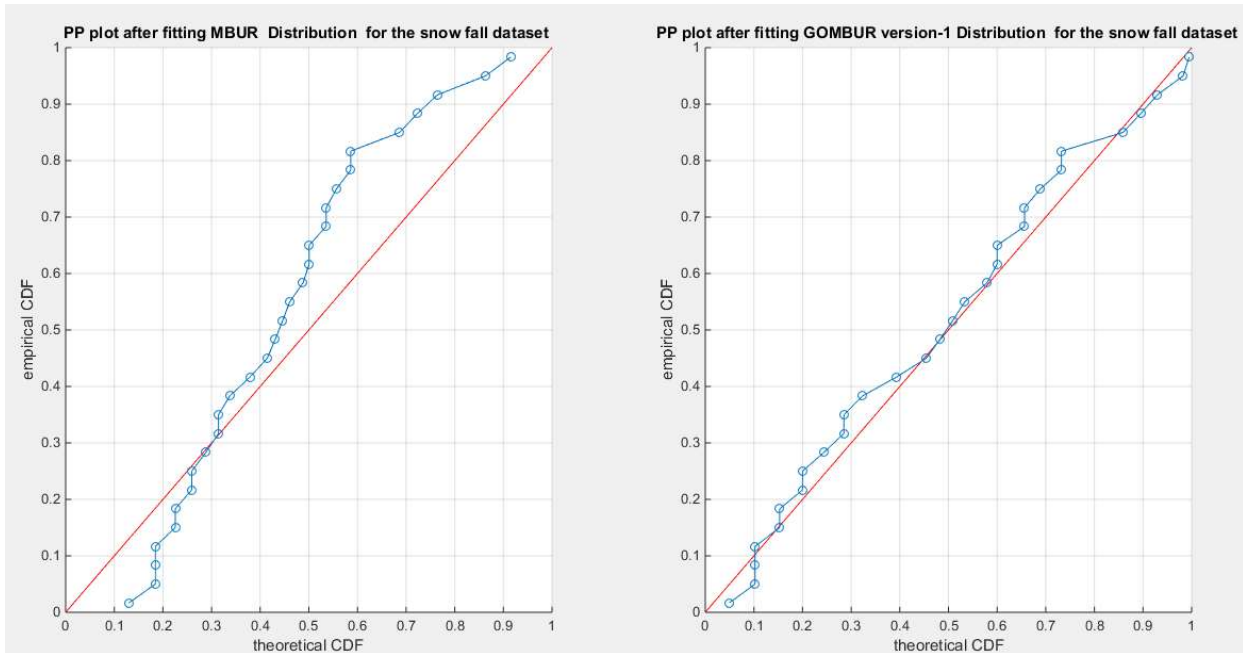

Fig. 58 shows the PP plot for the fitted MBUR & GOMBUR-1 for the snow fall dataset. Although the MBUR fits the data, the generalization of MBUR enhances the diagonal alignment of GOMBUR-1 along both ends and along the center of the distribution.

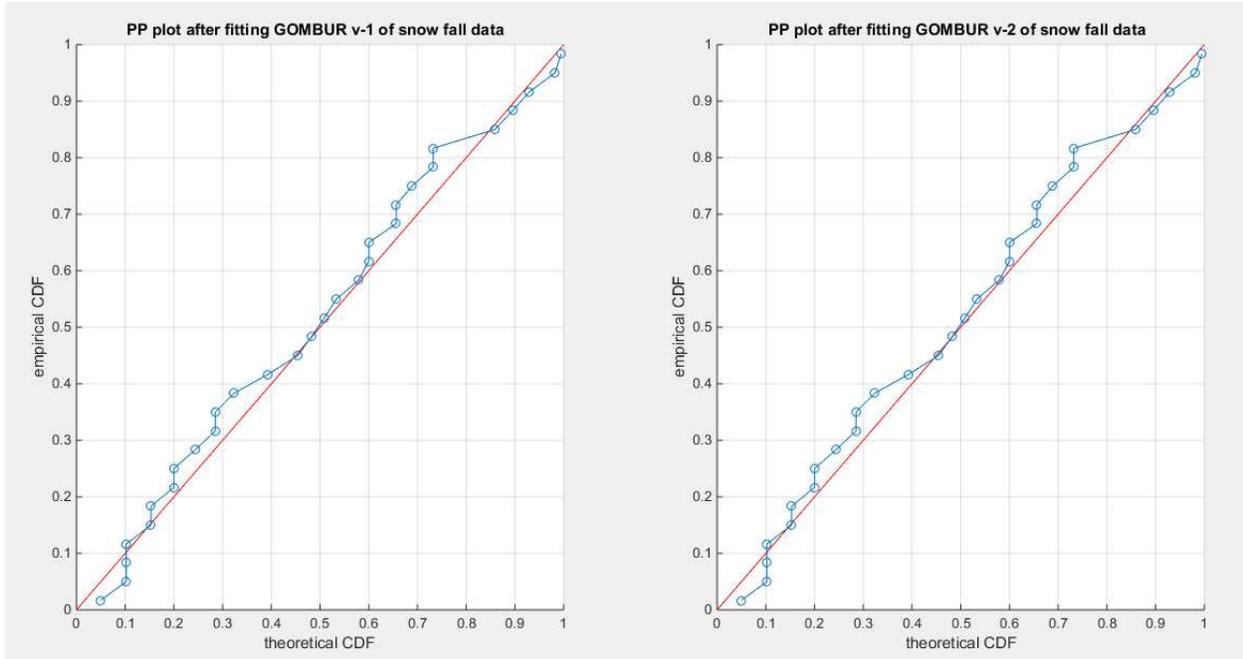

Fig. 59 shows the PP plot for the fitted GOMBUR-1 & GOMBUR-2 for snow fall dataset. They are identical

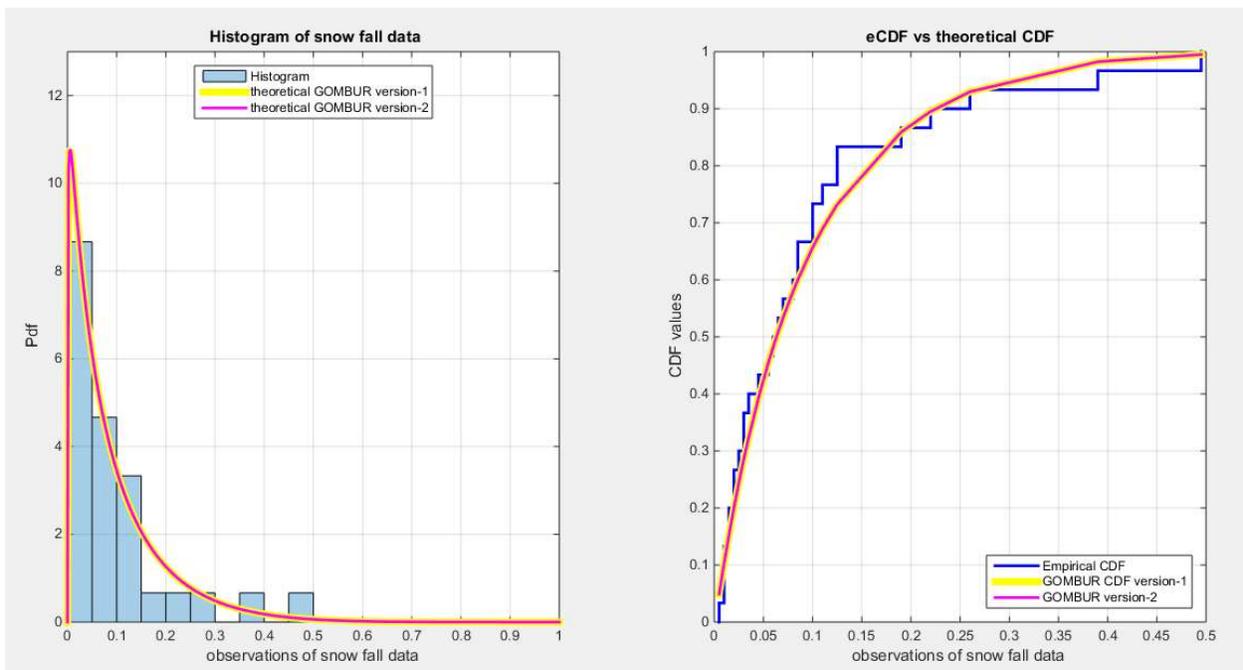

Fig. 60 shows on the left subplot the histogram of the snow fall and the fitted PDFs of both GOMBUR-1 & GOMBUR-2 and on the right subplot the e-CDFs and the theoretical CDFs for both distributions. Both the fitted CDFs and the fitted PDFs of both versions are identical. The PDF curve is highly right skewed.

Table (12) shows the result of analysis of the milk production dataset

Table (12) shows the result of analysis of the milk production dataset

|  | Beta | | Kumaraswamy | | MBUR | Topp-Leone | Unit-Lindley |
|---|---|---|---|---|---|---|---|
| theta | $\alpha = 2.4125$ | | $\alpha = 2.1949$ | | 1.039 | 2.0802 | 1.2001 |
|  | $\beta = 2.8297$ | | $\beta = 3.4363$ | | | | |
| Variance | 0.0631 | 0.0816 | 0.0495 | 0.1063 | 0.0013 | 0.0404 | 0.0079 |
|  | 0.0816 | 0.1643 | 0.1063 | 0.3388 | | | |
| SE(a) | 0.0243 | | 0.0215 | | 0.0035 | 0.0194 | 0.0086 |
| SE(b) | 0.0392 | | 0.0563 | | - | - | - |
| AIC | -43.5545 | | -46.7894 | | -39.3785 | -41.0524 | -48.7609 |
| CAIC | -43.4391 | | -46.6740 | | -39.3404 | -41.0143 | -48.7229 |
| BIC | -38.2088 | | -41.4437 | | -36.7057 | -38.3796 | -46.0881 |
| HQIC | -41.3874 | | -44.6223 | | -38.2949 | -39.9689 | -47.6774 |
| LL | 23.7772 | | 25.3947 | | 20.6892 | 21.5262 | 25.3805 |
| K-S | 0.0816 | | 0.0669 | | 0.1080 | 0.093 | 0.1002 |
| $H_0$ | Fail to reject | | Fail to reject | | Fail to reject | Fail to reject | Fail to reject |
| P-value | 0.3189 | | 0.5372 | | 0.1531 | 0.2473 | 0.1421 |
| AD | 1.3853 | | 1.003 | | 2.2535 | 1.8813 | 1.3116 |
| CVM | 0.2282 | | 0.1522 | | 0.3496 | 0.2848 | 0.2286 |
| determinant | 0.0037 | | 0.0055 | | - | - | - |

The analysis revealed that all the distributions fit the data. But the Unit Lindley is the best among them followed by Kumaraswamy, Beta , Topp Leone and lastly BMUR distribution. Unit Lindley has the highest negative values of AIC, CAIC, BIC & HQIC. The case is not true in sense of having the lowest values of AD, CVM & KS. But, Kumaraswamy distribution has the lowest values of AD, CVM & KS and the highest Log-likelihood value that is slightly larger than that of the Unit Lindley. The two parameters Kumaraswamy have, may account for the reduction in AIC, CAIC, BIC, & HQIC in comparison with the Unit Lindley which has one parameter. The GOMBUR 1 & 2 increases the log-likelihood value. The generalization increases the negativity of the AIC, CAIC, BIC, & HQIC, lower the AD, CVM & KS and increases log-likelihood value. Also, the generalization lowers the variance of the alpha parameter among all distribution. The determinant of the variance covariance matrix obtained after fitting GOMBUR-1 is the lowest among all distributions having two parameters. It is a matter of convenience to choose any distributions according to the need of the research. While the Unit Lindley is the best to fit the data, MBUR is the worst to fit the data. Generalization enhances its capability for fitting the data to approach the level of the Beta distribution fitting the data.

Table (12) to be continued

|  | GOMBUR-1 | | GOMBUR-2 | |
|---|---|---|---|---|
| theta | $n = 1.7903$ | | $n = 4.5807$ | |
|  | $\alpha = 1.0644$ | | $\alpha = 1.0664$ | |
| Variance | 0.1327 | 0.0031 | 0.531 | 0.0061 |
|  | 0.0031 | 0.0011 | 0.0061 | 0.0011 |
| SE(n) | 0.0352 | | 0.0704 | |
| SE(a) | 0.0031 | | 0.0031 | |
| AIC | -43.262 | | -43.262 | |
| CAIC | -43.1466 | | -43.1466 | |
| BIC | -37.9163 | | -37.9163 | |
| HQIC | -41.0949 | | -41.0949 | |
| LL | 23..631 | | 23..631 | |
| K-S Value | 0.0803 | | 0.0803 | |
| H$_0$ | Fail to reject | | Fail to reject | |
| P-value | 0.3366 | | 0.3366 | |
| AD | 1.367 | | 1.367 | |
| CVM | 0.2191 | | 0.2191 | |
| Determinant | 0.00013161 | | 0.00052642 | |
| Significant(n) | P<0.001 | | P<0.001 | |
| Significant(a) | P<0.001 | | P<0.001 | |

Figure 61 illustrates the theoretical CDFs of different competitors. Figure 62 discloses the fitted PDFs for the same distributions. Figure 63 depict PP plots and QQ plots for the competitors. Figure 64 exposes the PP plots of MBUR and the GOMBUR-1. Figure 65 exemplifies the PP plots for the both versions. Figure 66 elucidates the fitted PDFs and CDFs of both versions.

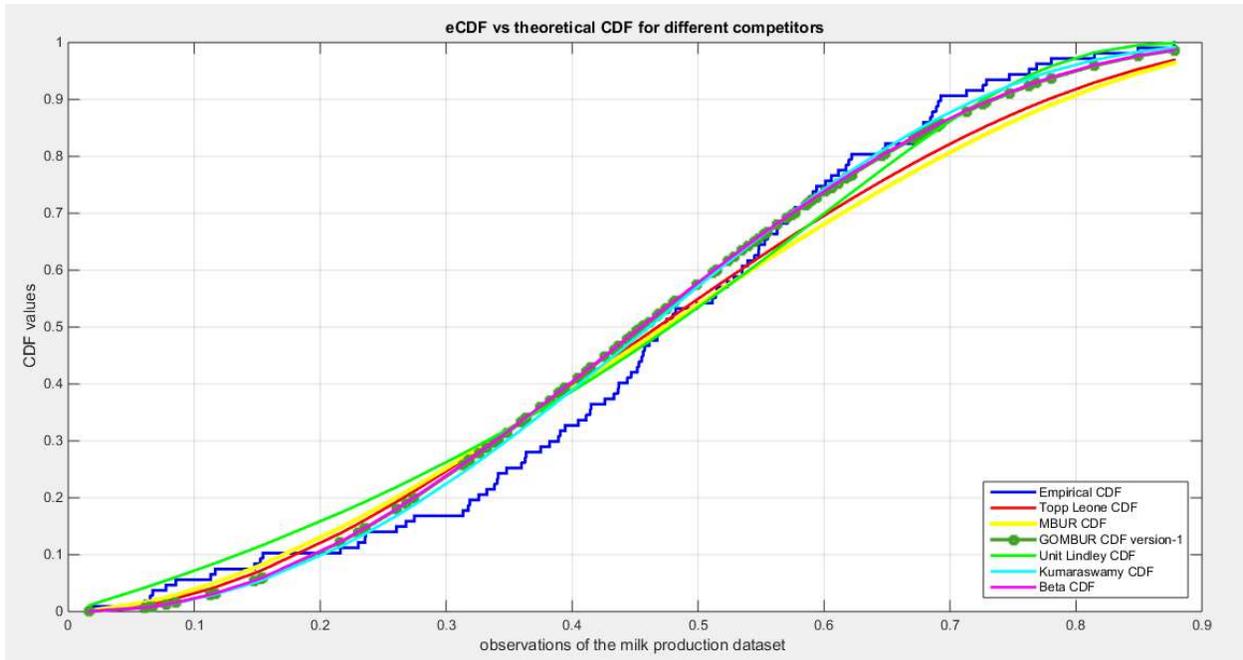

Fig. 61 shows the e-CDFs and the theoretical CDFs for the fitted distributions of the milk production data.

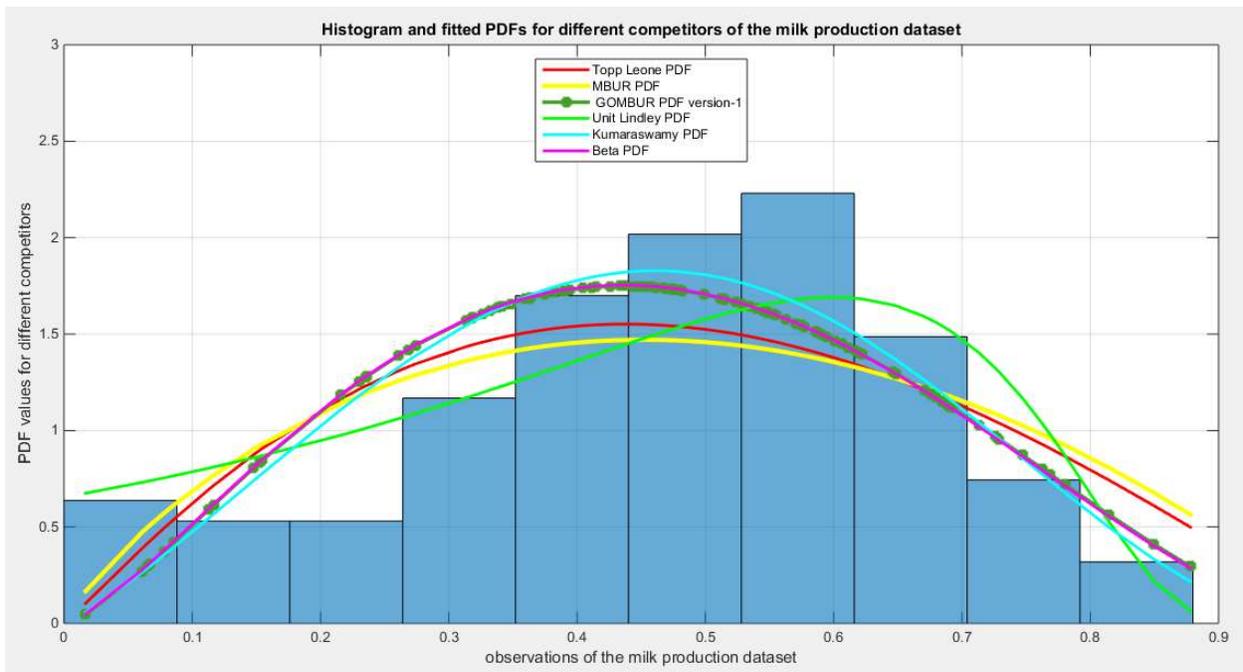

Fig. 62 shows the histogram of the milk production data and the theoretical PDFs for the fitted distributions. All the competitor distributions fit the data well but the Unit Lindley and the Kumaraswamy outperform other competitors. Beta distribution is the third distribution to fit the data, and after generalization of the MBUR , GOMBUR-1 properly aligns with Beta.

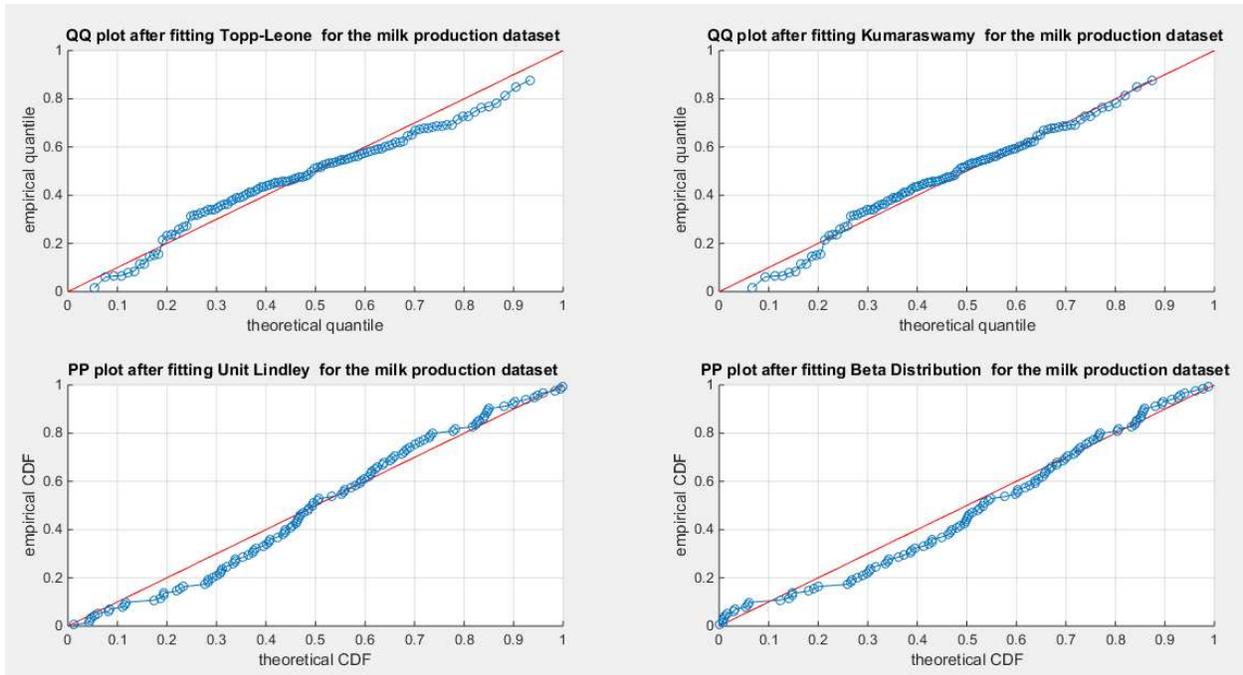

Fig. 63 shows the QQ plot for the fitted Topp Leone & Kumaraswamy distributions and the PP plot for the fitted Unit Lindley and Beta distribution for the milk production dataset.

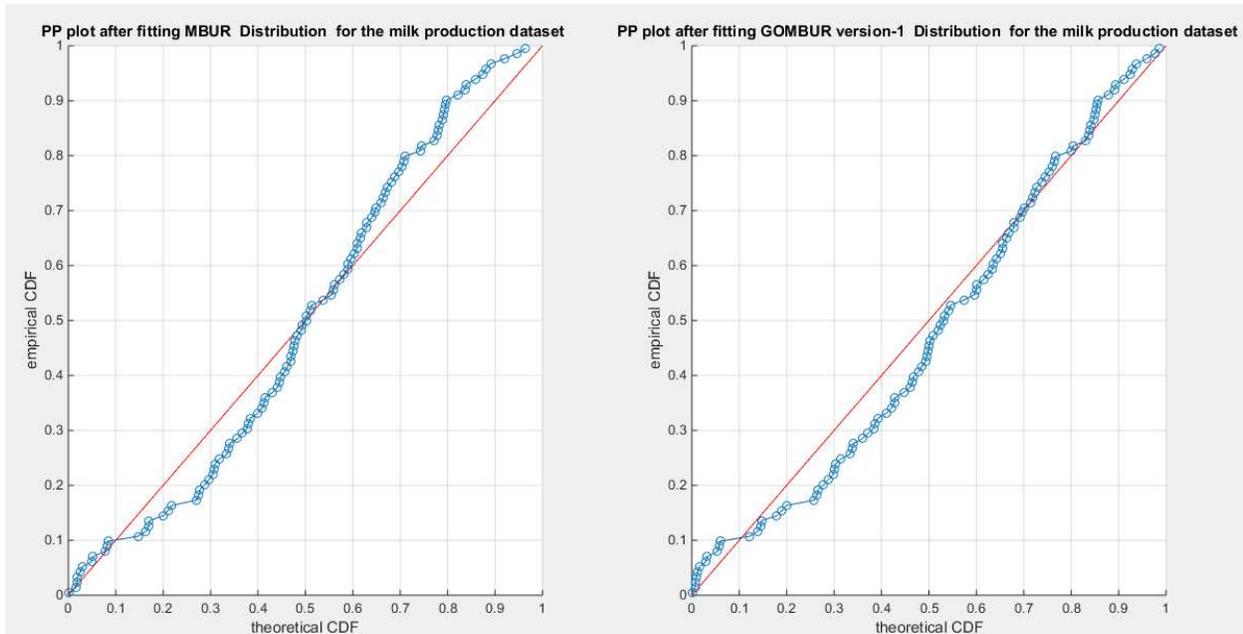

Fig. 64 shows the PP plot for the fitted MBUR & GOMBUR-1 for the milk production dataset. Since the MBUR is the least distribution to fit the data among all other competitors, generalization of MBUR enhances the diagonal alignment of GOMBUR-1 along both ends and along the center of the distribution.

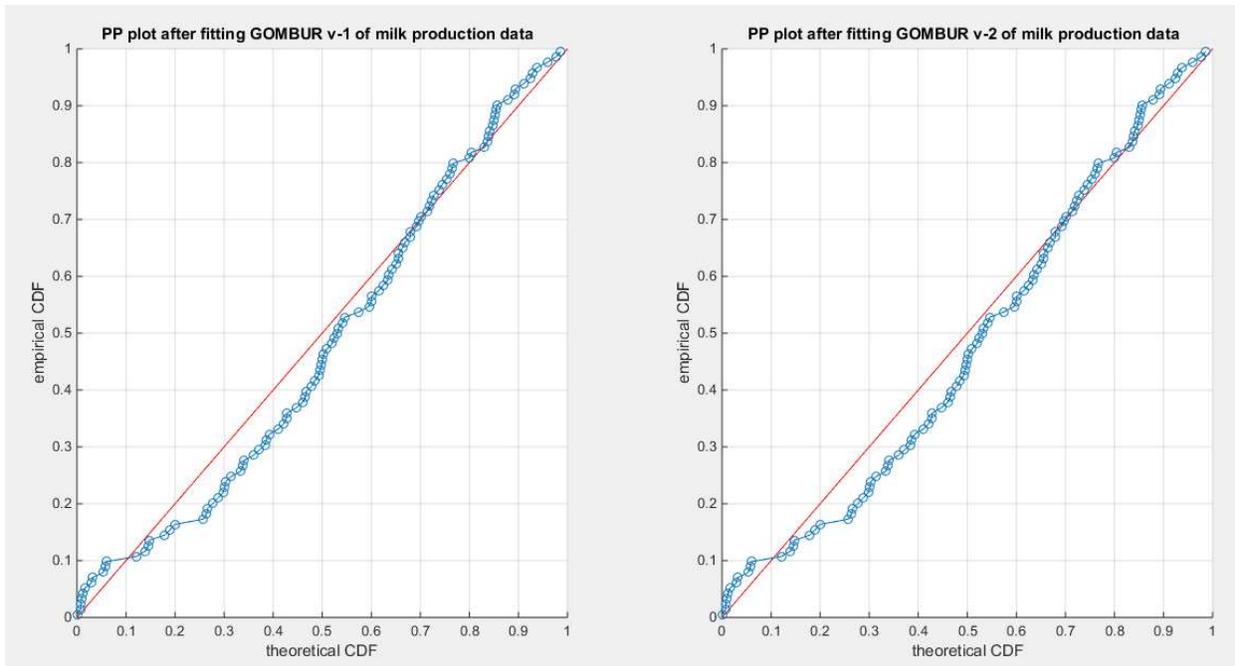

Fig. 65 shows the PP plot for the fitted GOMBUR-1 & GOMBUR-2 for milk production dataset. They are identical

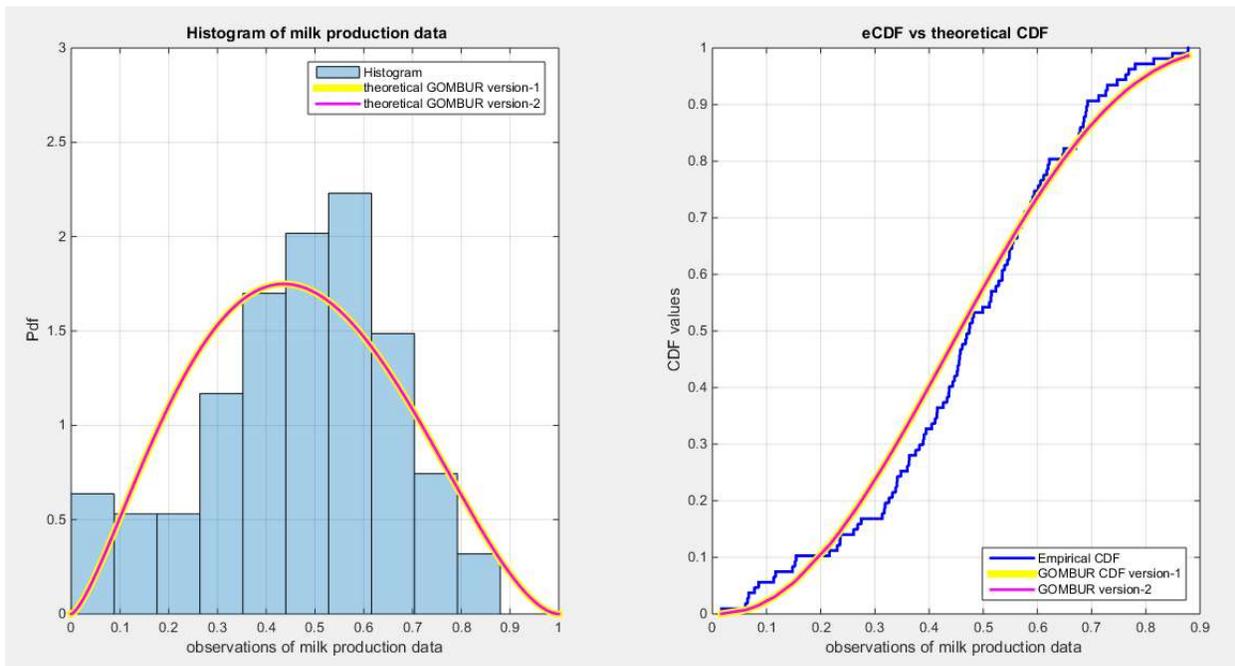

Fig. 66 shows on the left subplot the histogram of the milk production and the fitted PDFs of both GOMBUR-1 & GOMBUR-2 and on the right subplot the e-CDFs and the theoretical CDFs for both distributions. Both the fitted CDFs and the fitted PDFs of both versions are identical.

Table (13) shows analysis results of COVID-19 recovery rate in Spain dataset.

Table (13) shows analysis results of COVID-19 recovery rate in Spain dataset.

|  | Beta | | Kumaraswamy | | MBUR | Topp-Leone | Unit-Lindley |
|---|---|---|---|---|---|---|---|
| theta | $\alpha = 12.7943$ | | $\alpha = 8.0782$ | | 0.6366 | 10.5284 | 0.52 |
|  | $\beta = 4.8994$ | | $\beta = 7.7382$ | | | | |
| Variance | 8.088 | 3.2731 | 0.8967 | 1.6853 | 0.00079 | 1.6795 | 0.0022 |
|  | 3.2731 | 1.4278 | 1.6853 | 4.0747 | | | |
| SE(a) | 0.3501 | | 0.1166 | | 0.0035 | 0.1595 | 0.0057 |
| SE(b) | 0.1471 | | 0.2485 | | - | - | - |
| AIC | -111.1486 | | -113.6686 | | -90.4861 | -101.1088 | -90.2298 |
| CAIC | -110.9581 | | -113.4782 | | -90.4236 | -101.0463 | -90.1673 |
| BIC | -106.7692 | | -109.2893 | | -88.2964 | -98.9191 | -88.0402 |
| HQIC | -109.4181 | | -111.9382 | | -89.6208 | -100.2435 | -89.3646 |
| LL | 57.5743 | | 58.8343 | | 46.243 | 51.5544 | 46.1149 |
| K-S | 0.0996 | | 0.0846 | | 0.2226 | 0.1813 | 0.1752 |
| $H_0$ | Fail to reject | | Fail to reject | | reject | reject | reject |
| P-value | 0.3243 | | 0.4966 | | 0.0024 | 0.0227 | 0.0144 |
| AD | 1.052 | | 0.9317 | | 3.5144 | 2.0475 | 4.2480 |
| CVM | 0.1783 | | 0.1519 | | 0.5281 | 0.2775 | 0.6736 |
| determinant | 0.8345 | | 0.8136 | | - | - | - |

The analysis exposes that the Kumaraswamy followed by Beta distributions are the distributions that fir the data well. This is because they have the best indices. GOMBUR-1&2 have indices better than that of the Beta rendering the generalized versions the second distribution to fit the after Kumaraswamy. In advantage, GOMBUR-1 has the least variance of alpha among all the distributions and has the least value of the determinant of variance covariance matrix. This makes it the most efficient distribution to fit the data. Also the covariance between the two parameters in the generalized form is far less than that obvious in Kumaraswamy and Beta distributions. Figure 67-68 reveal the theoretical CDFs and fitted PDFs for the competitors. Figure 69 uncovers the PP plots and QQ plots for the competitors. Figure 70-72 display the PP plot of the MBUR, GOMBUR-1 & 2, the fitted PDFs and CDFs for the two generalized versions.

Table (13) to be continued

|  | GOMBUR-1 | | GOMBUR-2 | |
|---|---|---|---|---|
| theta | $n = 4.0506$ | | $n = 9.1012$ | |
|  | $\alpha = 0.6705$ | | $\alpha = 0.6705$ | |
| Variance | 0.7317 | 0.0032 | 2.9267 | 0.0064 |
|  | 0.0032 | 0.00036616 | 0.0064 | 0.00036616 |
| SE(n) | 0.1053 | | 0.2106 | |
| SE(a) | 0.0024 | | 0.0024 | |
| AIC | -111.3714 | | -111.3714 | |
| CAIC | -111.1809 | | -111.1809 | |
| BIC | -106.9921 | | -106.9921 | |
| HQIC | -109.641 | | -109.641 | |
| LL | 57.6857 | | 57.6857 | |
| K-S Value | 0.0977 | | 0.0977 | |
| $H_0$ | Fail to reject | | Fail to reject | |
| P-value | 0.3436 | | 0.3436 | |
| AD | 1.0324 | | 1.0324 | |
| CVM | 0.1743 | | 0.1743 | |
| Determinant | 0.00025776 | | 0.001 | |
| Significant(n) | P<0.001 | | P<0.001 | |
| Significant(a) | P<0.001 | | P<0.001 | |

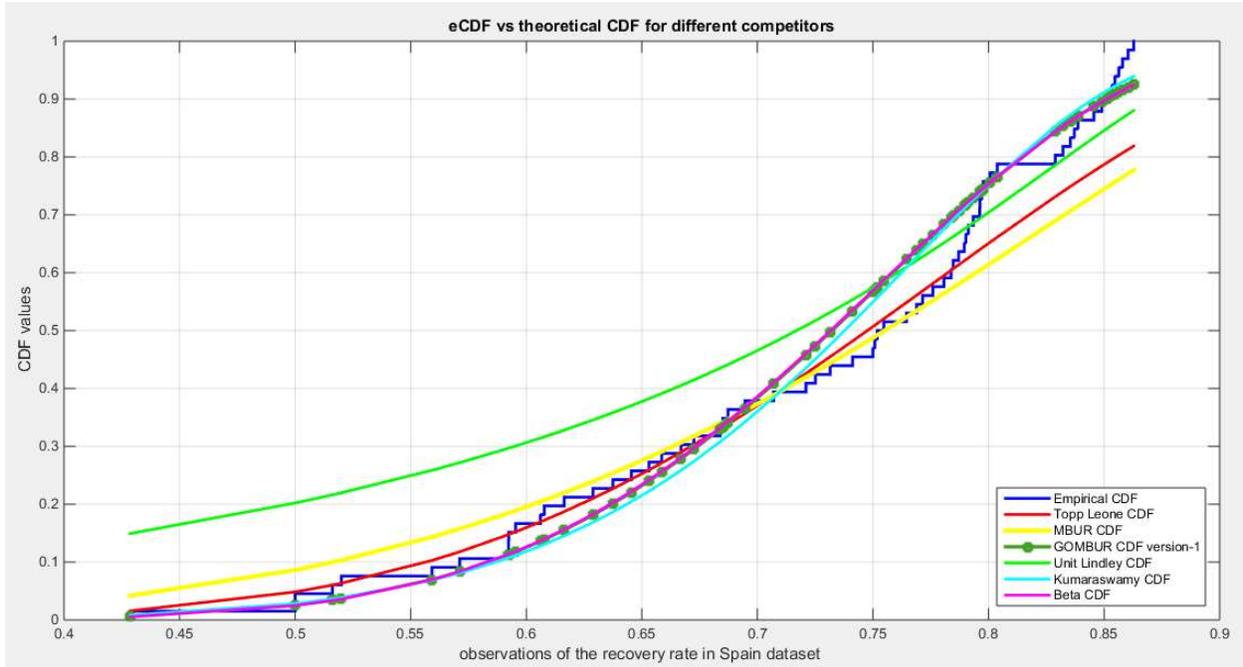

Fig. 67 shows the e-CDFs and the theoretical CDFs for the fitted distributions of the recovery rate in Spain data.

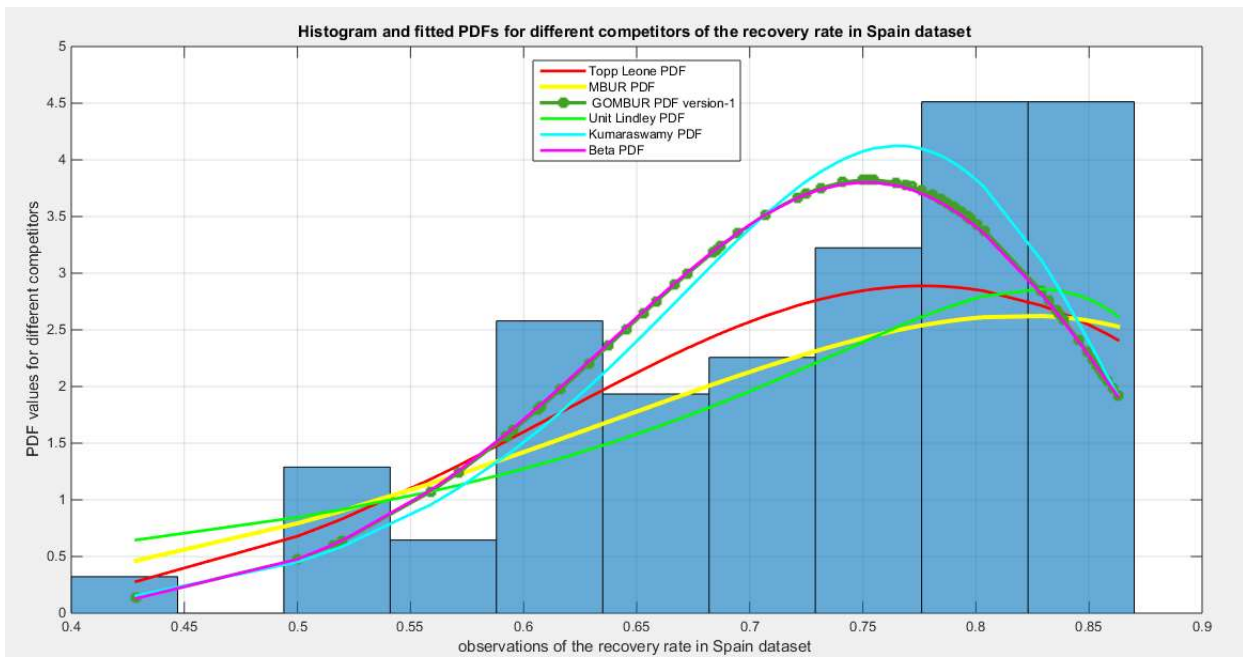

Fig. 68 shows the histogram of the recovery rate in Spain data and the theoretical PDFs for the fitted distributions. Kumaraswamy and Beta fit the data. After generalization the GOMBUR aligns appropriately with the Beta distribution and more or less parallels the Kumaraswamy distribution.

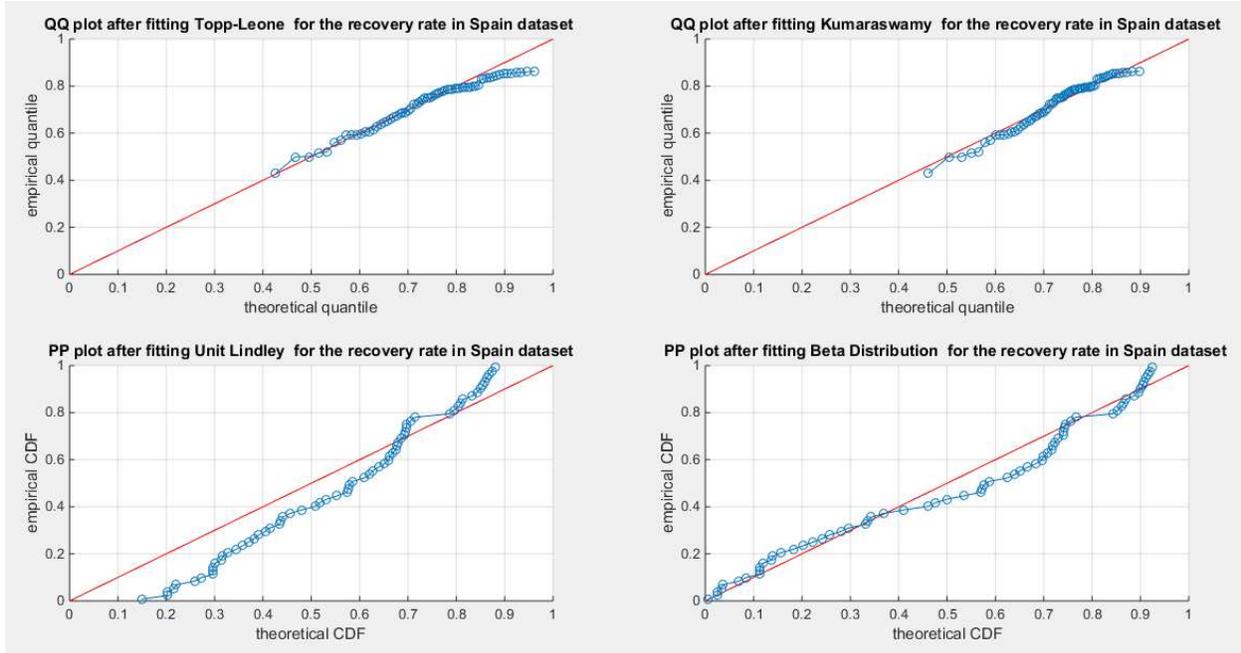

Fig. 69 shows the QQ plot for the fitted Topp Leone & Kumaraswamy distributions and the PP plot for the fitted Unit Lindley and Beta distribution for the recovery rate in Spain dataset.

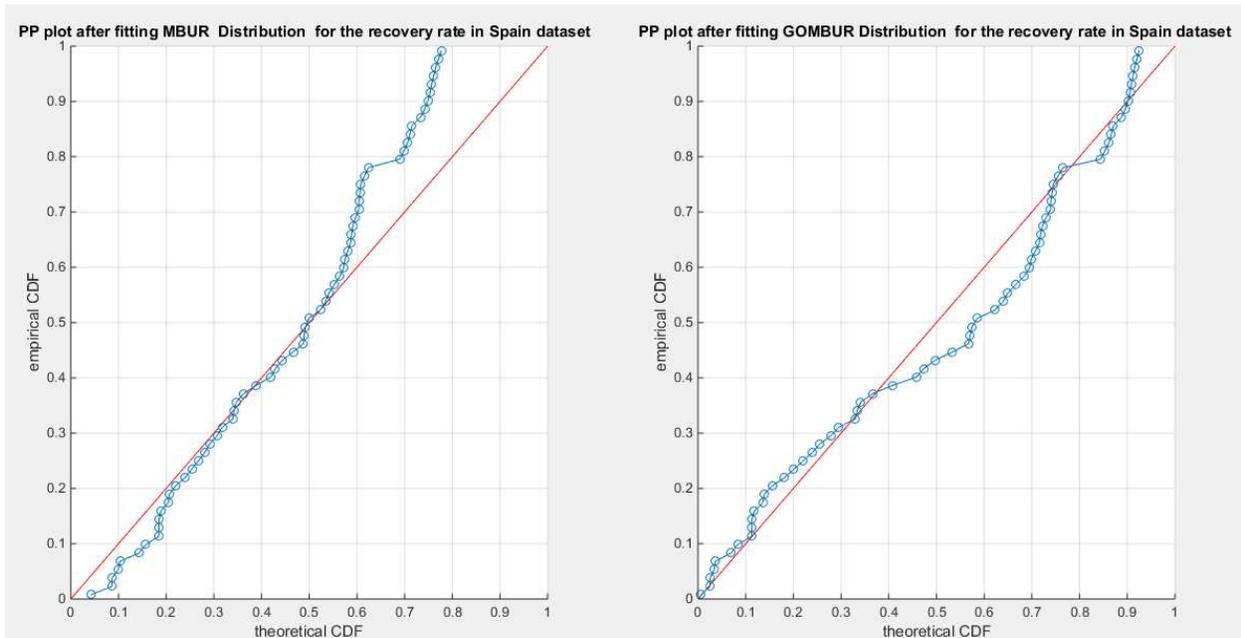

Fig. 70 shows the PP plot for the fitted MBUR & GOMBUR-1 for the recovery rate in Spain dataset. Generalization of the MBUR enhances the diagonal alignment along both ends.

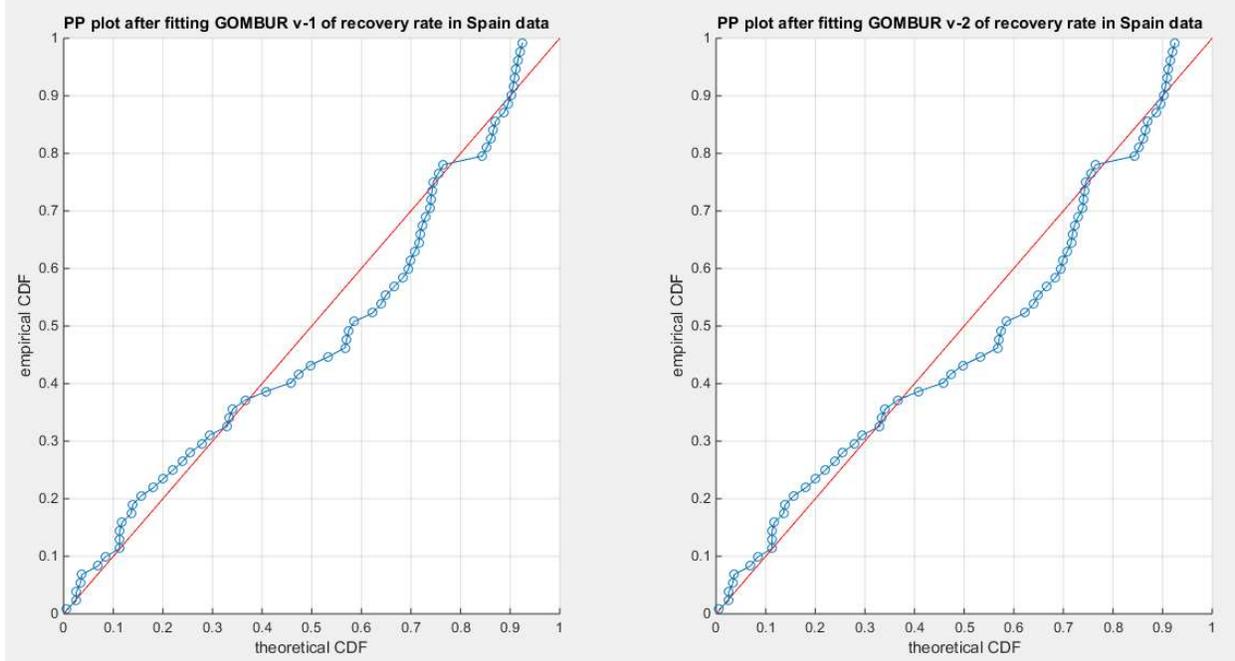

Fig. 71 shows the PP plot for the fitted GOMBUR-1 & GOMBUR-2 for recovery rate in Spain dataset. They are identical

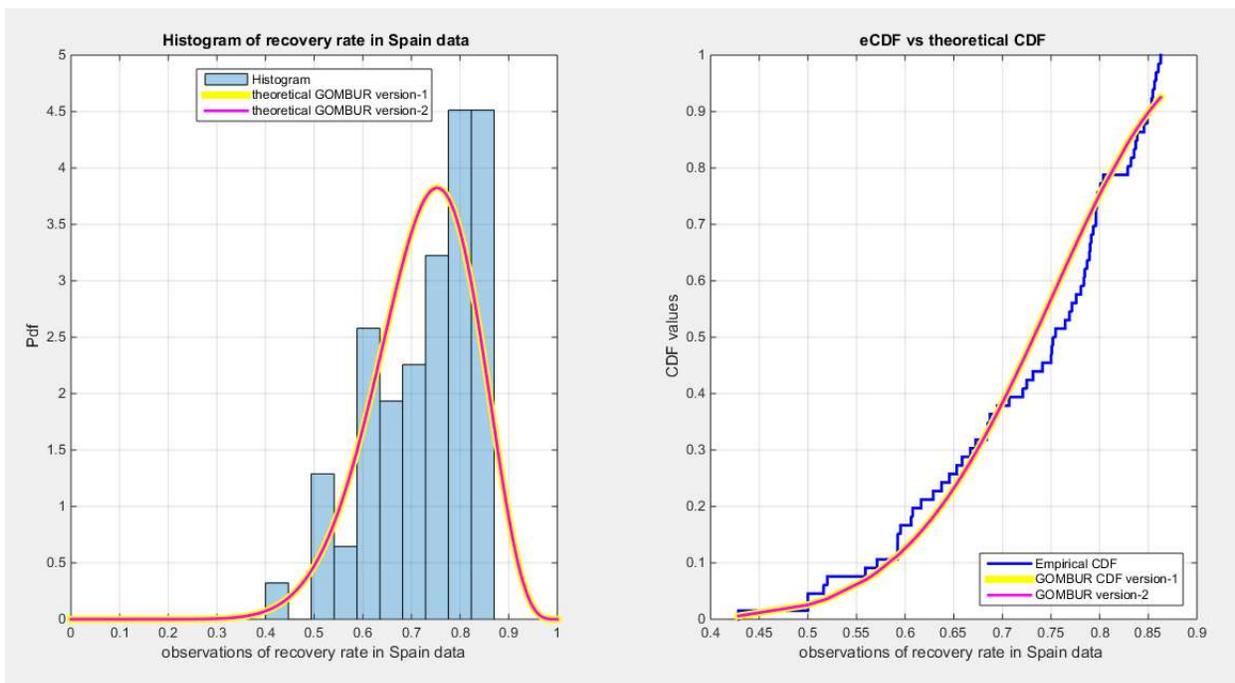

Fig. 72 shows on the left subplot the histogram of the recovery rate in Spain and the fitted PDFs of both GOMBUR-1 & GOMBUR-2 and on the right subplot the e-CDFs and the theoretical CDFs for both distributions. Both the fitted CDFs and the fitted PDFs of both versions are identical.

Table (14) shows analysis results of voter dataset.

Table (14): analysis results of voter dataset.

| | Beta | | Kumaraswamy | | MBUR | Topp-Leone | Unit-Lindley |
|---|---|---|---|---|---|---|---|
| theta | $\alpha = 8.6959$ | | $\alpha = 5.6224$ | | 0.6832 | 8.2339 | 0.5406 |
| | $\beta = 3.8673$ | | $\beta = 4.5073$ | | | | |
| Variance | 6.4427 | 2.3588 | 0.7462 | 0.9005 | 0.0016 | 1.7841 | 0.0041 |
| | 2.3588 | 0.9809 | 0.9005 | 1.6212 | | | |
| SE(a) | 0.4118 | | 0.1401 | | 0.0065 | 0.2167 | 0.0104 |
| SE(b) | 0.1607 | | 0.2066 | | | | |
| AIC | -47.9451 | | -46.8575 | | -42.1377 | -47.4106 | -36.517 |
| CAIC | -47.6022 | | -46.5146 | | -42.0266 | -47.2995 | -36.4059 |
| BIC | -44.6699 | | -43.5823 | | -40.5001 | -45.7731 | -34.8794 |
| HQIC | -46.7798 | | -45.6922 | | -41.555 | -46.828 | -35.9344 |
| LL | 25.9725 | | 25.4287 | | 22.0688 | 24.7053 | 19.2585 |
| K-S | 0.0938 | | 0.1048 | | 0.1364 | 0.119 | 0.1467 |
| $H_0$ | Fail to reject | | Fail to reject | | Fail to reject | Fail to reject | Fail to reject |
| P-value | 0.8605 | | 0.7596 | | 0.44 | 0.6129 | 0.1828 |
| AD | 0.2721 | | 0.3464 | | 1.3262 | 0.6861 | 1.4767 |
| CVM | 0.0416 | | 0.0559 | | 0.2193 | 0.1198 | 0.2266 |
| determinant | 0.7556 | | 0.3989 | | - | - | - |

The analysis reveals the Beta followed by Topp Leone, Kumaraswamy, MBUR and lastly Unit Lindley distributions fit the data well. This conclusion is guided by the indices as illustrated in Table (14). GOMBUR-1, in comparison to MBUR, has more negative values of AIC, CAIC, BIC & HQIC. It also has lower values of AD, CVM & KS. In addition, it has higher Log-likelihood value than that of the MBUR. All these enhancement of the indices renders the GOMBUR-1 to be the second distribution to fit the data after the Beta distribution. Taking into account that GOMBUR-1 has the least variance of the estimated alpha among all the distributions and has far less value of determinant of its estimated variance covariance matrix than that of the Beta distribution; this leverages it to be the most efficient distribution to fit the data. The covariance between the parameters of GOMBUR-1 is far less than that evident between the parameters of the Beta distribution. Figure 73-74 describe the theoretical CDFs and the fitted PDFs of the competitors respectively. Figure 75 clarifies the PP plots and QQ plots of the competitors. Figure 76-77 depict the comparison of the PP plots between MBUR, GOMBUR-1 & 2. Figure 78 explicates the fitted PDFs and CDFs of the two versions of generalization.

Table (14) to be continued

|  | GOMBUR-1 | | GOMBUR-2 | |
|---|---|---|---|---|
| theta | $n = 2.9469$ | | $n = 6.8938$ | |
|  | $\alpha = 0.713$ | | $\alpha = 0.713$ | |
| Variance | 0.7663 | 0.0059 | 3.0651 | 0.0118 |
|  | 0.0059 | 0.0009294 | 0.0118 | 0.0009294 |
| SE(n) | 0.142 | | 0.284 | |
| SE(a) | 0.0049 | | 0.0049 | |
| AIC | -47.8278 | | -47.8278 | |
| CAIC | -47.4849 | | -47.4849 | |
| BIC | -44.5526 | | -44.5526 | |
| HQIC | -46.6625 | | -46.6625 | |
| LL | 25.9139 | | 25.9139 | |
| K-S Value | 0.0949 | | 0.0949 | |
| H₀ | Fail to reject | | Fail to reject | |
| P-value | 0.8519 | | 0.8519 | |
| AD | 0.2797 | | 0.2797 | |
| CVM | 0.0428 | | 0.0428 | |
| Determinant | 0.00067749 | | 0.0027 | |
| Significant(n) | P<0.001 | | P<0.001 | |
| Significant(a) | P<0.001 | | P<0.001 | |

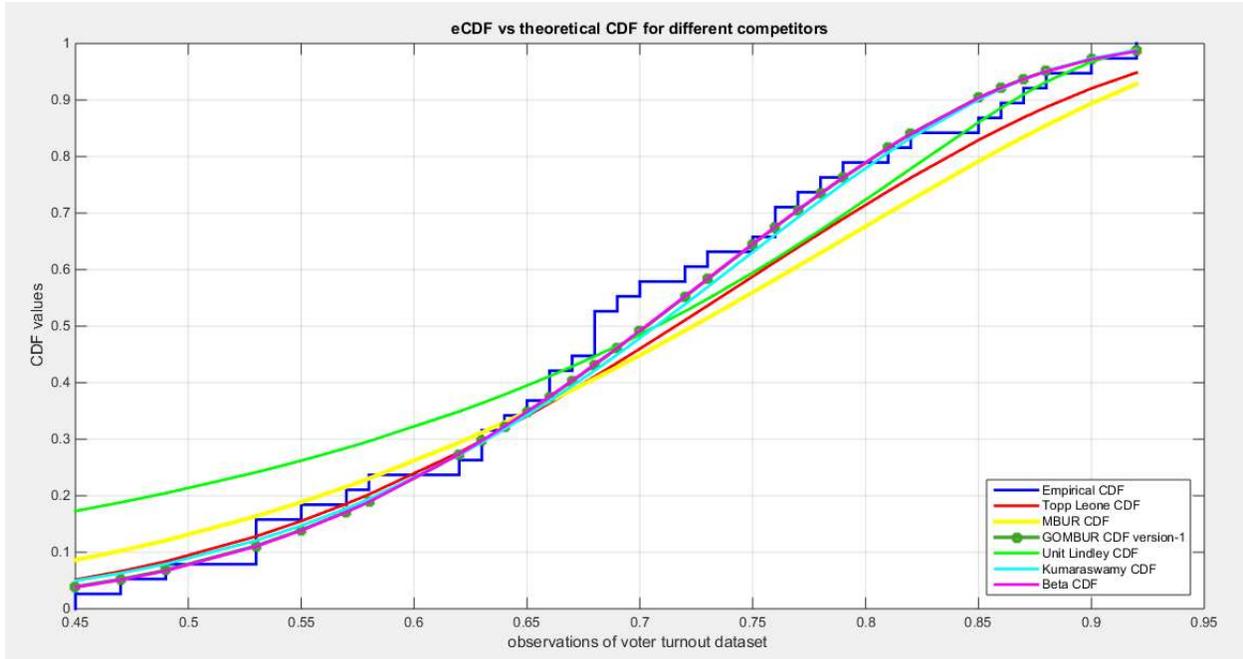

Fig. 73 shows the e-CDFs and the theoretical CDFs for the fitted distributions of the voter turnout data

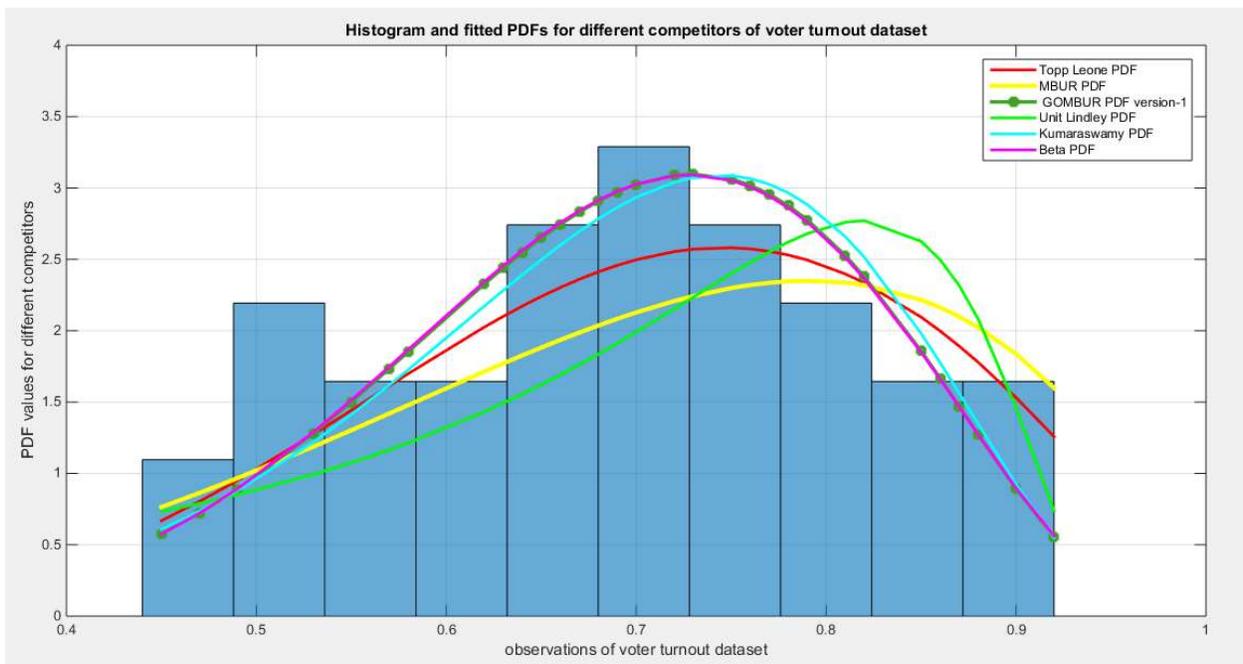

Fig. 74 shows the histogram of the voter turnout data and the theoretical PDFs for the fitted distributions. Beta, Topp Leone, Kumaraswamy, MBUR and lastly Unit Lindley fit the data. After generalization the GOMBUR aligns appropriately with the Beta distribution and more or less parallels the Kumaraswamy distribution.

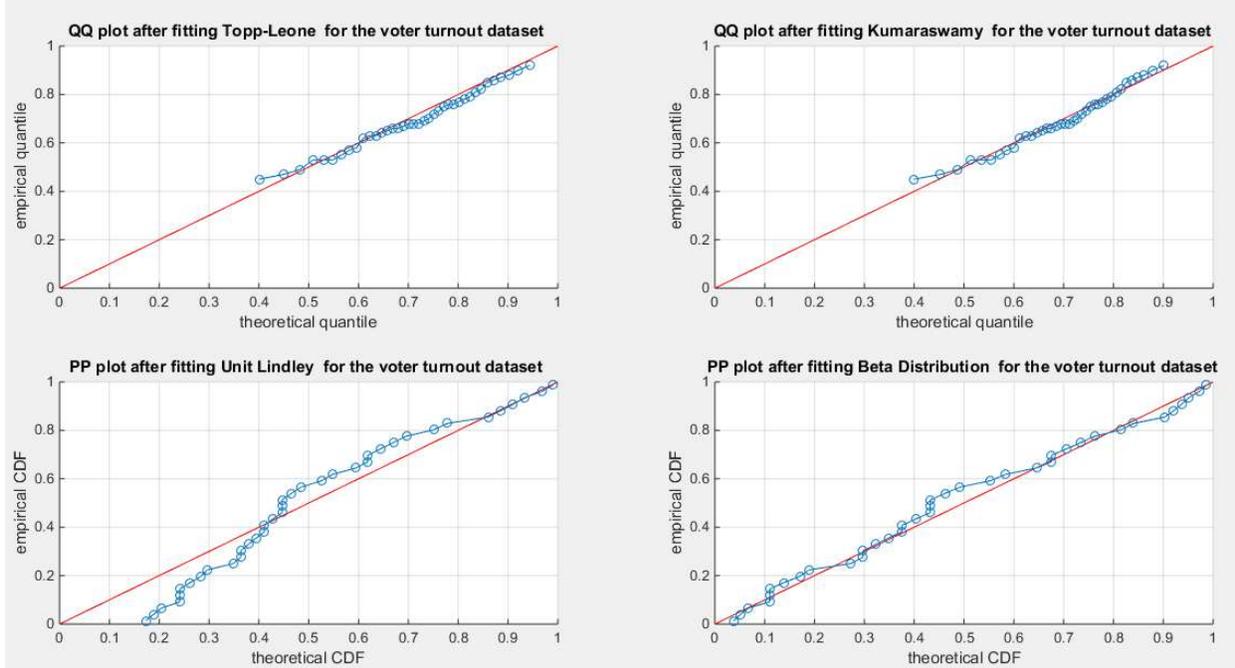

Fig. 75 shows the QQ plot for the fitted Topp Leone & Kumaraswamy distributions and the PP plot for the fitted Unit Lindley and Beta distribution for the voter turnout dataset.

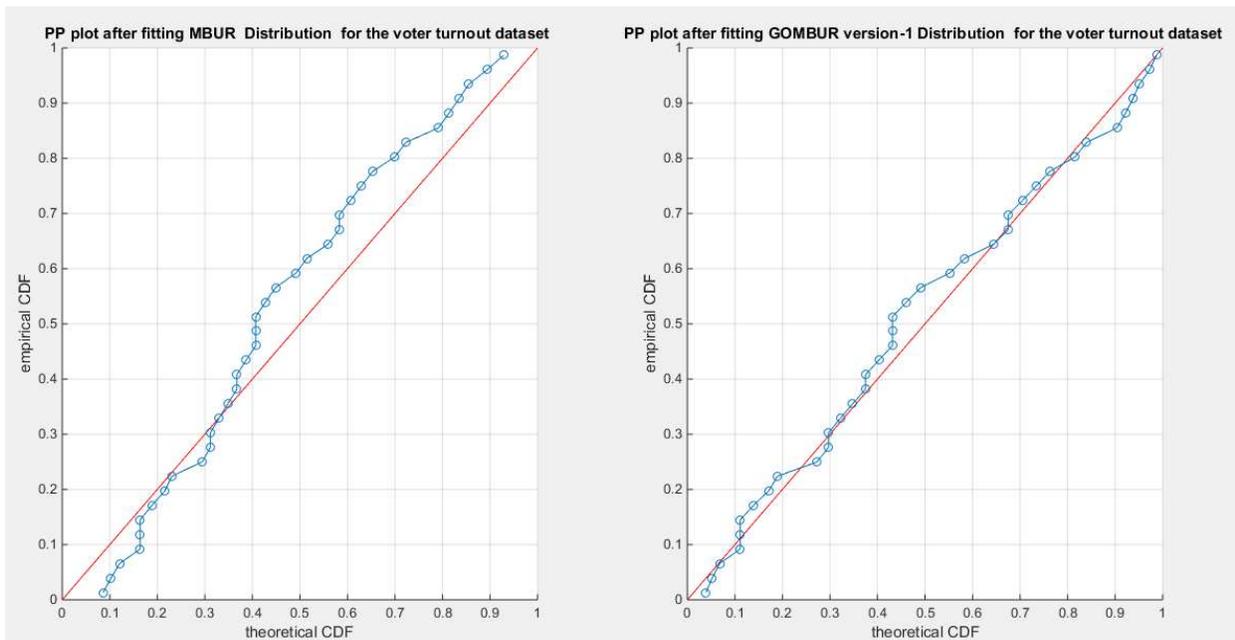

Fig. 76 shows the PP plot for the fitted MBUR & GOMBUR-1 for the voter turnout dataset. Although, MBUR fit the data and the PP plot shows an acceptable diagonal alignment the generalization of the MBUR enhances and refines this diagonal alignment along both ends and along the center of the distribution.

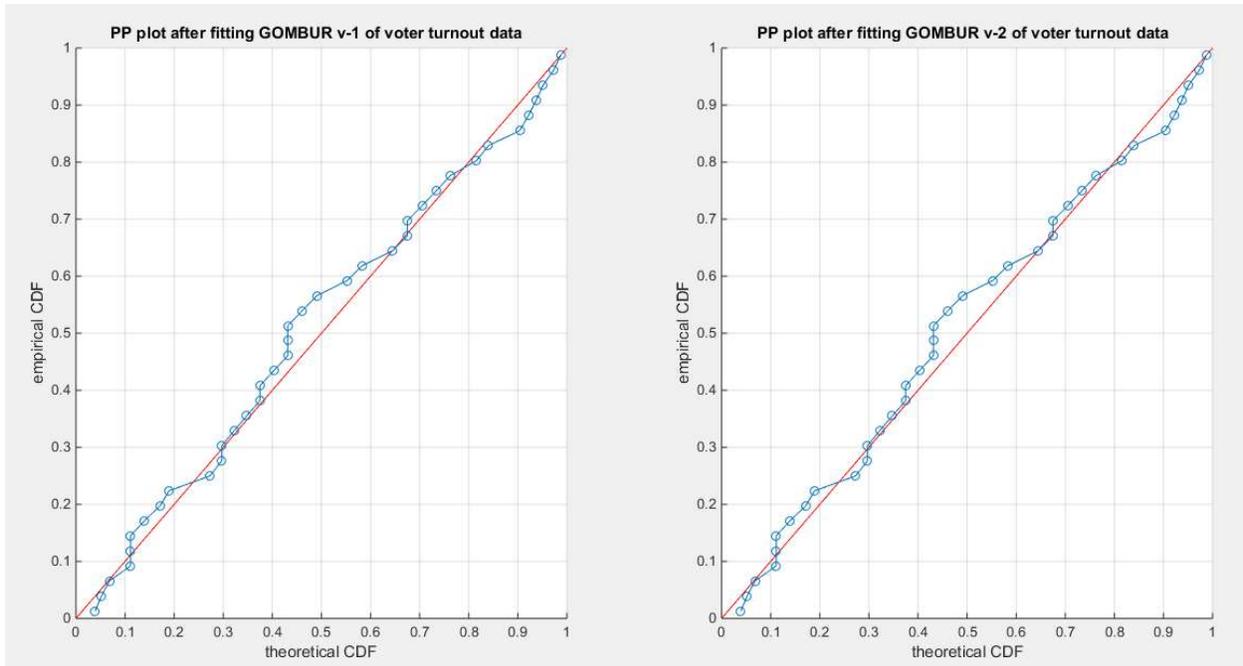

Fig. 77 shows the PP plot for the fitted GOMBUR-1 & GOMBUR-2 for voter turnout dataset. They are identical

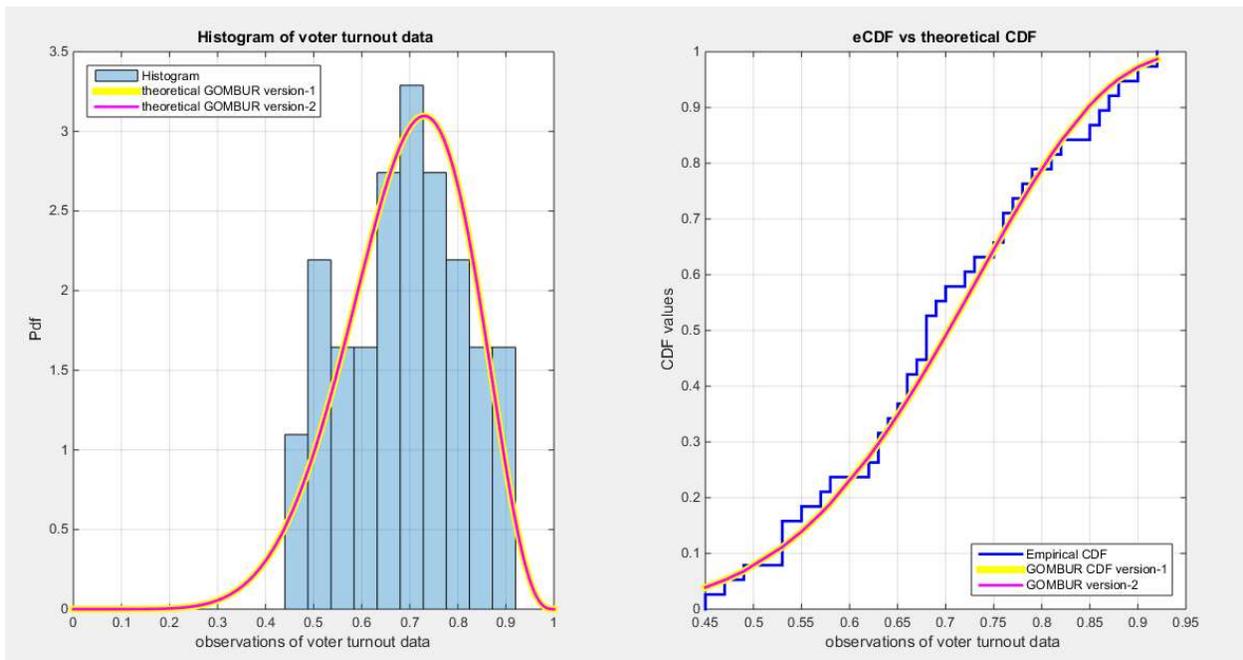

Fig. 78 shows on the left subplot the histogram of the voter turnout data and the fitted PDFs of both GOMBUR-1 & GOMBUR-2 and on the right subplot the e-CDFs and the theoretical CDFs for both distributions. Both the fitted CDFs and the fitted PDFs of both versions are identical.

Table (15) shows analysis results of unit capacity factors dataset.

Table (15): analysis results of unit capacity factors dataset.

| | Beta | | Kumaraswamy | | MBUR | Topp-Leone |
|---|---|---|---|---|---|---|
| theta | $\alpha = 0.4869$ | | $\alpha = 0.5044$ | | 1.6243 | 0.5943 |
| | $\beta = 1.1679$ | | $\beta = 1.1862$ | | | |
| Variance | 0.0482 | 0.0960 | 0.0166 | 0.0274 | 0.0149 | 0.0154 |
| | 0.0960 | 0.2919 | 0.0274 | 0.1066 | | |
| SE(a) | 0.0458 | | 0.0269 | | | |
| SE(b) | 0.1127 | | 0.0681 | | | |
| AIC | -15.2149 | | -15.3416 | | -13.2158 | -14.2302 |
| CAIC | -14.6149 | | -14.7416 | | -13.0253 | -14.0398 |
| BIC | -12.9439 | | -13.0706 | | -12.0803 | -13.0943 |
| HQIC | -14.6438 | | -14.7704 | | -12.9302 | -13.9447 |
| LL | 9.6075 | | 9.6708 | | 7.6079 | 8.1151 |
| K-S | 0.1836 | | 0.179 | | 0.1518 | 0.1602 |
| $H_0$ | Fail to reject | | Fail to reject | | Fail to reject | Fail to reject |
| P-value | 0.3742 | | 0.4051 | | 0.4074 | 0.4762 |
| AD | 0.6998 | | 0.6975 | | 1.9075 | 1.5271 |
| CVM | 0.1189 | | 0.1159 | | 0.2033 | 0.1735 |
| determinant | 0.0049 | | 0.001 | | - | - |

The analysis revealed that the Kumaraswamy followed by beta , Topp Leone and lastly BMUR distributions fit the data well as indicated by the indices. The GOMBUR-1 distribution also fit the data with better indices. It has slightly larger negative values of AIC, CAIC, BIC, & HQIC and minimally larger value of Log-Likelihood than those of the Kumaraswamy distribution. It has a marginally lower value of KS than that of the Kumaraswamy distribution. The variance of the estimated alpha after fitting the Kumaraswamy distribution is less than that after the GOMBUR-1. The determinant of the variance covariance matrix obtained after fitting the Kumaraswamy distribution is less than that after fitting the GOMBUR-1 which justifies it to be more efficient distribution to fit the data rather than GOMBUR distribution. Figure 79-80 show the fitted CDFs and PDF s of the competitors. Figure 81-83 illustrates the PP plot of the different competitors. Figure 84 displays the fitted CDFs and PDFs of the GOMBUR-1 & 2 distributions.

Table (15) to be continued

|  | GOMBUR-1 | | GOMBUR-2 | |
|---|---|---|---|---|
| theta | $n = 0.1959$ | | $n = 1.3917$ | |
|  | $\alpha = 1.5298$ | | $\alpha = 1.5298$ | |
| Variance | 0.1087 | 0.0214 | 0.435 | 0.0428 |
|  | 0.0214 | 0.0258 | 0.0428 | 0.0258 |
| SE(n) | 0.0688 | | 0.1375 | |
| SE(a) | 0.0335 | | 0.0335 | |
| AIC | -15.365 | | -15.365 | |
| CAIC | -14.765 | | -14.765 | |
| BIC | -13.094 | | -13.094 | |
| HQIC | -14.7939 | | -14.7939 | |
| LL | 9.6825 | | 9.6825 | |
| K-S Value | 0.1782 | | 0.1782 | |
| H₀ | Fail to reject | | Fail to reject | |
| P-value | 0.4103 | | 0.4103 | |
| AD | 0.7026 | | 0.7026 | |
| CVM | 0.1161 | | 0.1161 | |
| Determinant | 0.0024 | | 0.0094 | |
| Significant(n) | P<0.001 | | P<0.001 | |
| Significant(a) | P<0.001 | | P<0.001 | |

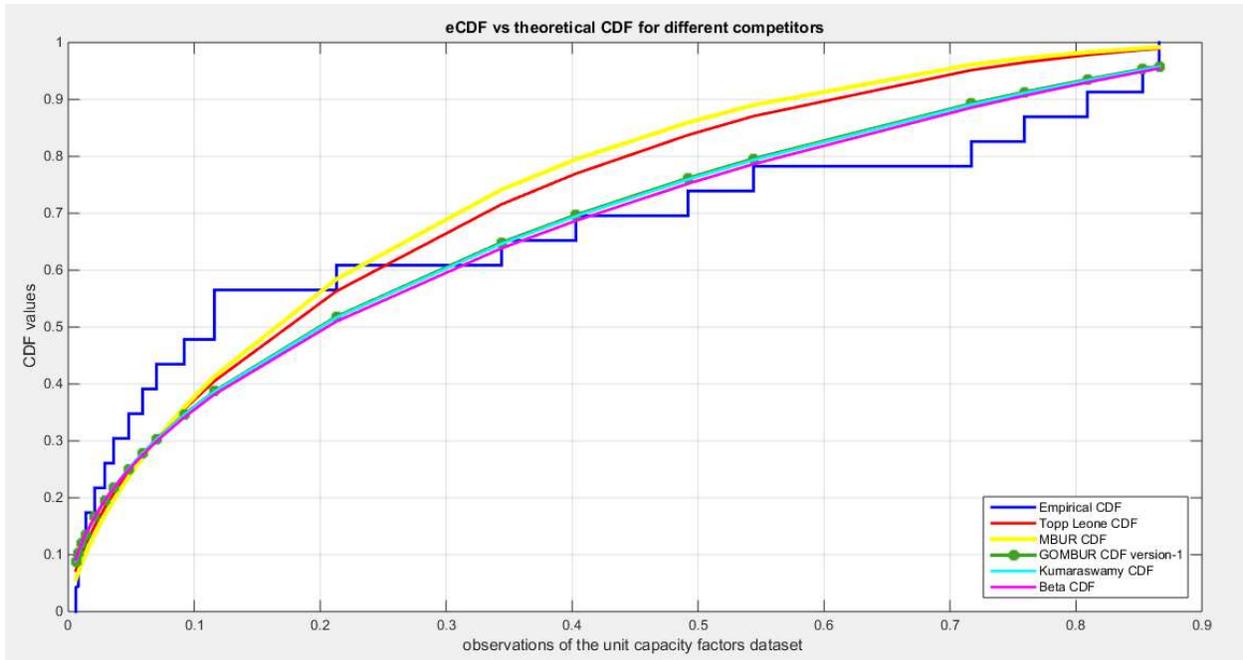

Fig. 79 shows the e-CDFs and the theoretical CDFs for the fitted distributions of the unit capacity factors data

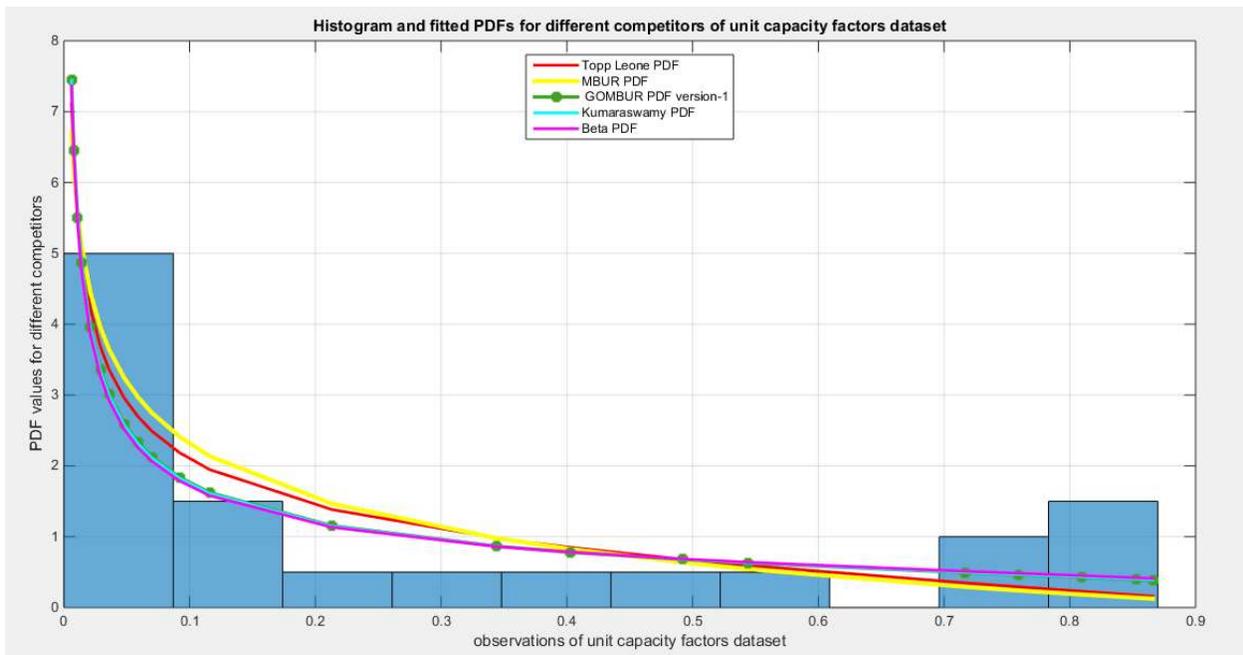

Fig. 80 shows the histogram of the unit capacity factors data and the theoretical PDFs for the fitted distributions. Kumaraswamy, Beta, Topp Leone and lastly MBUR distributions fit the data. After generalization the GOMBUR aligns appropriately with the Beta distribution and more or less parallels the Kumaraswamy distribution.

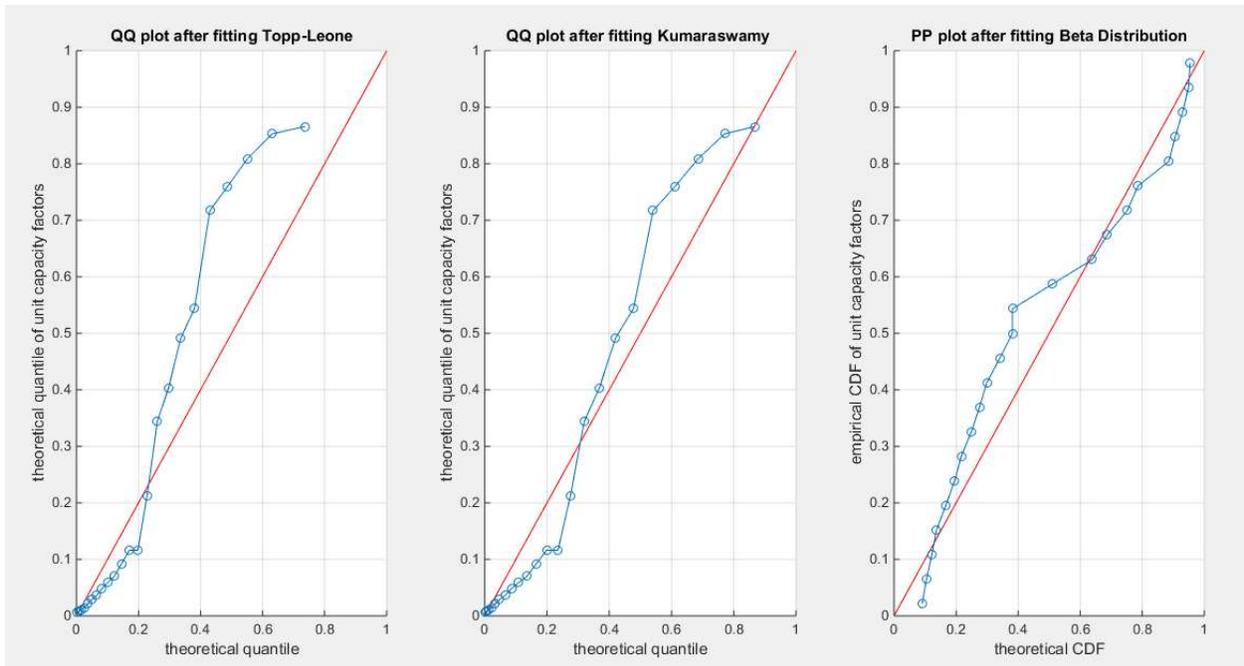

Fig. 81 shows the QQ plot for the fitted Topp Leone & Kumaraswamy distributions and the PP plot for the fitted Unit Lindley and Beta distribution for the unit capacity factors dataset.

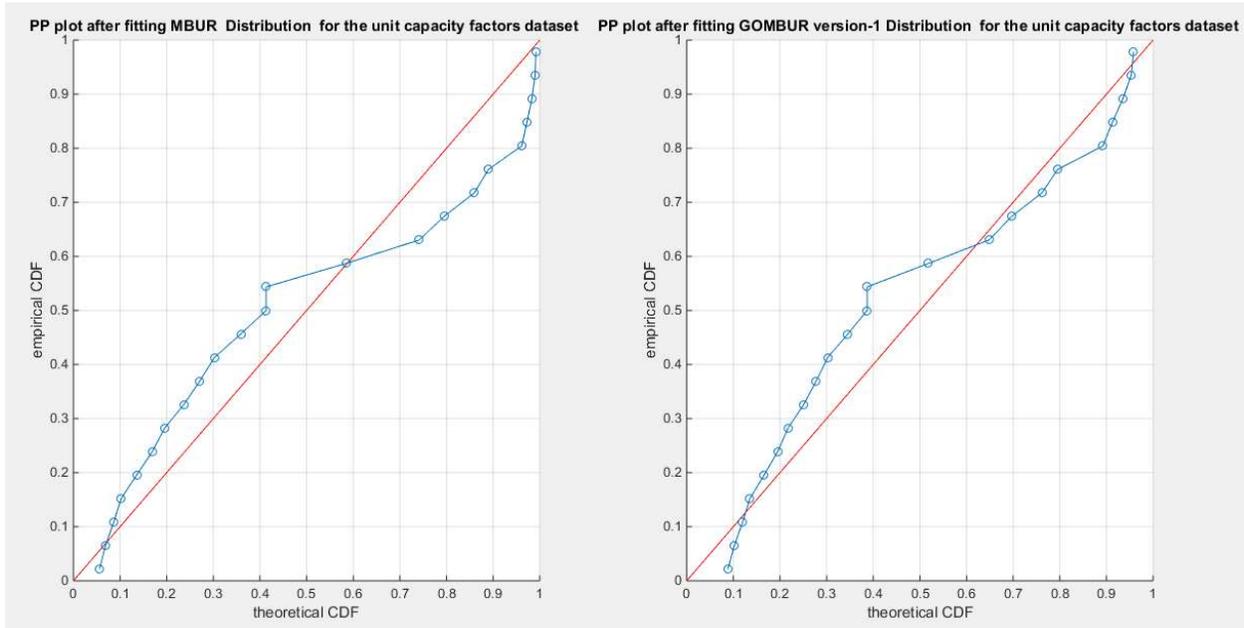

Fig. 82 shows the PP plot for the fitted MBUR & GOMBUR-1 for the voter turnout dataset. Although, MBUR fits the data and the PP plot shows an acceptable diagonal alignment; the generalization of the MBUR enhances and refines this diagonal alignment along the upper end and along the center of the distribution.

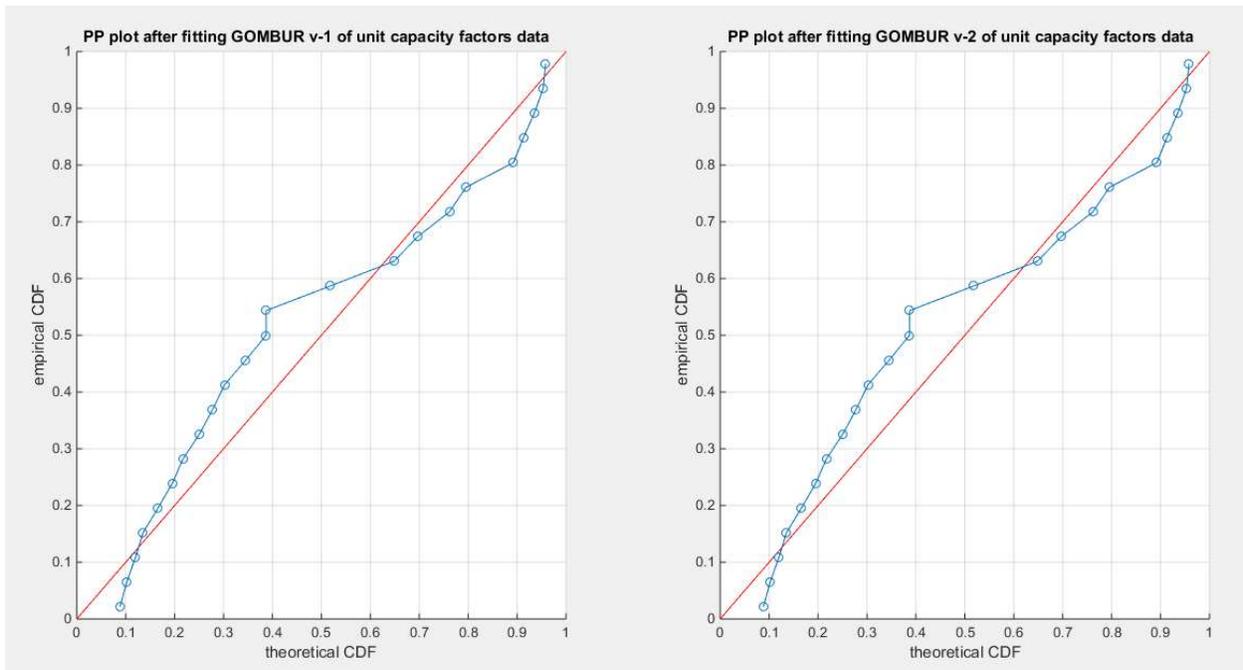

Fig. 83 shows the PP plot for the fitted GOMBUR-1 & GOMBUR-2 for unit capacity factors dataset. They are identical

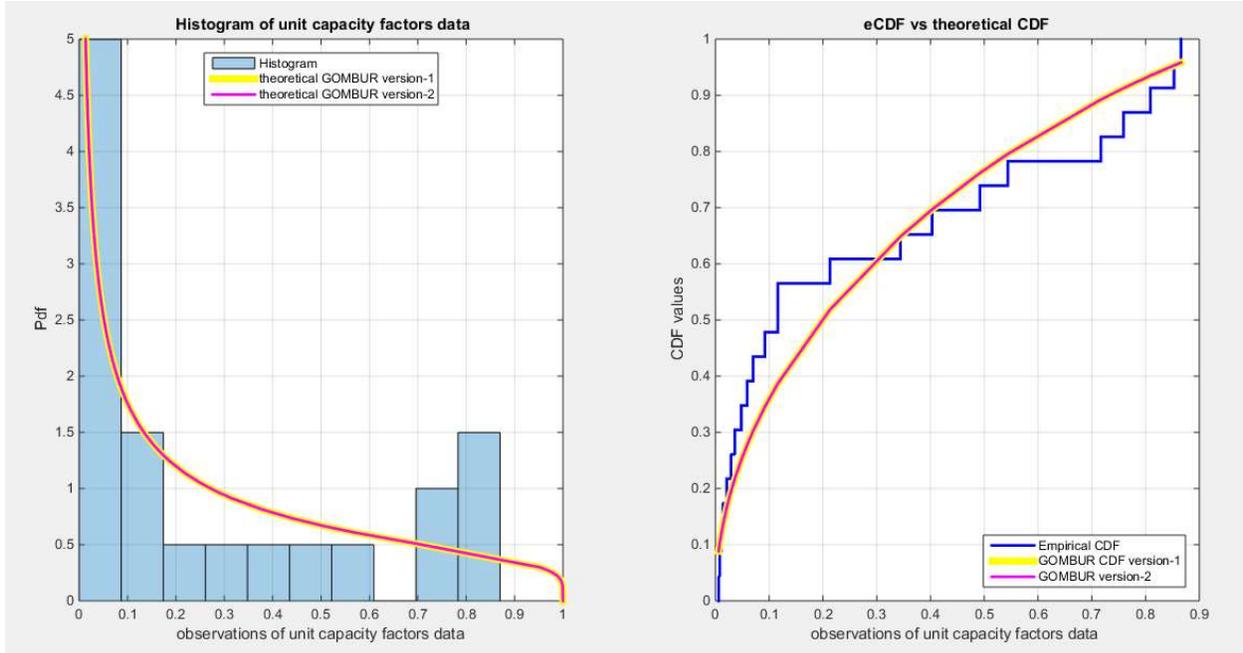

Fig. 84 shows on the left subplot the histogram of the unit capacity factors data and the fitted PDFs of both GOMBUR-1 & GOMBUR-2 and on the right subplot the e-CDFs and the theoretical CDFs for both distributions. Both the fitted CDFs and the fitted PDFs of both versions are identical.

# Section 4

## Conclusion

As illustrated from the analysis of the datasets adding new parameters to the MBUR increases its flexibility to accommodate different shapes of data with different characteristics like skewness , kurtosis and different tail behaviors.   This newly introduced parameter improves the estimation process by increasing validity indices like AIC, CAIC, BIS & HQIC. It also enhances the goodness of fit by reducing test statistics like AD, CVM & KS tests. It also increases the value of the Log-Likelihood. The two versions of the generalization yield different values for the n parameter but equal values for the alpha parameter. The variances of the estimated alpha obtained from the two versions are identical. The covariance between the two parameters is minimal. It is far less than the covariance obtained from fitting distributions like the Beta and the Kumaraswamy. The determinant of the estimated variance covariance matrix obtained from fitting GOMBUR-1 is minimal and almost with a least values in comparison with that obtained after  fitting Beta and Kumaraswamy distribution to the data. There is an exception for this that is obvious in the unit capacity factors. The determinant of the matrix obtained after fitting the Kunmaraswamy distribution is less than that obtained from the GOMBUR-1. Otherwise, there is a pattern of minimal value of the determinant of the matrix obtained by GOMBUR fitting to data. It is a matter of tradeoff to weigh the benefits of increased validity indices against increased efficiency of estimators which has a minimal variance. Generalization renders the MBUR perfectly aligns with the Beta distribution and more or less parallels the Kumaraswamy in most datasets.

# Section 5

In this paper Nelder Mead optimizer was used for MLE estimation of the parameters. Bayesian inference procedures may be used in future works.

 Declarations:  
**Ethics approval and consent to participate**  
Not applicable.  
**Consent for publication**  
Not applicable  
**Availability of data and material**  
Not applicable. Data sharing does not apply to this article as no datasets were generated or analyzed during the current study.  
**Competing interests**  
The author declares no competing interests of any type.  
**Funding**  
No funding resources. No funding roles in the design of the study and collection, analysis,

and interpretation of data and in writing the manuscript are declared.

**Authors' contribution**

AI (Attia Iman) carried the conceptualization by formulating the goals, and aims of the research article, formal analysis by applying the statistical, mathematical, and computational techniques to synthesize and analyze the hypothetical data, carried the methodology by creating the model, software programming and implementation, supervision, writing, drafting, editing, preparation, and creation of the presenting work.

**Acknowledgment**

Not applicable

Appendix A, table (A) of the datasets:

| 1- Dwelling without basic facilities ||||||
|---|---|---|---|---|---|
| 0.008 | 0.007 | 0.002 | 0.094 | 0.123 | 0.023 |
| 0.005 | 0.005 | 0.057 | 0.004 | 0.005 | 0.001 |
| 0.004 | 0.035 | 0.002 | 0.006 | 0.064 | 0.025 |
| 0.112 | 0.118 | 0.001 | 0.259 | 0.001 | 0.023 |
| 0.009 | 0.015 | 0.002 | 0.003 | 0.049 | 0.005 |
| 0.001 | | | | | |
| 2- Quality support network ||||||
| 0.98 | 0.96 | 0.95 | 0.94 | 0.93 | 0.8 |
| 0.82 | 0.85 | 0.88 | 0.89 | 0.78 | 0.92 |
| 0.92 | 0.9 | 0.96 | 0.96 | 0.94 | 0.77 |
| 0.95 | 0.91 | | | | |
| 3- Educational attainment ||||||
| 0.84 | 0.86 | 0.8 | 0.92 | 0.67 | 0.59 |
| 0.43 | 0.94 | 0.82 | 0.91 | 0.91 | 0.81 |
| 0.86 | 0.76 | 0.86 | 0.76 | 0.85 | 0.88 |
| 0.63 | 0.89 | 0.89 | 0.94 | 0.74 | 0.42 |
| 0.81 | 0.81 | 0.93 | 0.55 | 0.92 | 0.9 |
| 0.63 | 0.84 | 0.89 | 0.42 | 0.82 | 0.92 |
| 4- Flood data ||||||
| 0.26 | 0.27 | 0.3 | 0.32 | 0.32 | 0.34 |
| 0.38 | 0.38 | 0.39 | 0.4 | 0.41 | 0.42 |
| 0.42 | 0.42 | 045 | 0.48 | 0.49 | 0.61 |
| 0.65 | 0.74 | | | | |
| 5- Time between failures ||||||
| 0.216 | 0.015 | 0.4082 | 0.0746 | 0.0358 | 0.0199 |
| 0.0402 | 0.0101 | 0.0605 | 0.0954 | 0.1359 | 0.0273 |
| 0.0491 | 0.3465 | 0.007 | 0.656 | 0.106 | 0.0062 |
| 0.4992 | 0.0614 | 0.532 | 0.0347 | 0.1921 | 0.4992 |
| 6- COVID-19 death rate in Canada ||||||
| 0.1622 | 0.1159 | 0.1897 | 0.126 | 0.3025 | 0.219 |
| 0.2075 | 0.2241 | 0.2163 | 0.1262 | 0.1627 | 0.2591 |
| 0.1989 | 0.3053 | 0.2170 | 0.2241 | 0.2174 | 0.2541 |
| 0.1997 | 0.3333 | 0.2594 | 0.223, | 0.229 | 0.1536 |
| 0.2024 | 0.2931 | 0.2739 | 0.2607 | 0.2736 | 0.2323 |
| 0.1563 | 0.2677 | 0.2181 | 0.3019 | 0.2136 | 0.2281 |
| 0.2346 | 0.1888 | 0.2729 | 0.2162 | 0.2746 | 0.2936 |
| 0.3259 | 0.2242 | 0.181 | 0.2679 | 0.2296 | 0.2992 |
| 0.2464 | 0.2576 | 0.2338 | 0.1499 | 0.2075 | 0.1834 |
| 0.3347 | 0.2362 | | | | |

Table (A) to be continued

| 7- COVID-19 death rate in Spain | | | | | |
|---|---|---|---|---|---|
| 0.333 | 0.5 | 0.5 | 0.5714 | 0.25 | 0.3469 |
| 0.4839 | 0.2105 | 0.2311 | 0.3127 | 0.48 | 0.2749 |
| 0.3625 | 0.3922 | 0.3414 | 0.3711 | 0.4288 | 0.4077 |
| 0.3939 | 0.4076 | 0.4079 | 0.4408 | 0.4046 | 0.3836 |
| 0.3545 | 0.3275 | 0.3162 | 0.315 | 0.3053 | 0.293 |
| 0.279 | 0.2685 | 0.2588 | 0.2492 | 0.2481 | 0.2453 |
| 0.2355 | 0.2285 | 0.2241 | 0.2193 | 0.2162 | 0.2153 |
| 0.2129 | 0.2098 | 0.2037 | 0.2066, | 0.2087 | 0.2038 |
| 0.2029 | 0.2023 | 0.1993 | 0.1962 | 0.1711 | 0.1678 |
| 0.1646 | 0.1629 | 0.1613 | 0.1544 | 0.151 | 0.1484 |
| 0.1465 | 0.1453 | 0.1436 | 0.142 | 0.1396 | 0.1372 |
| 8- COVID-19 death rate in United Kingdom | | | | | |
| 0.1292 | 0.3805 | 0.4049 | 0.2564 | 0.3091 | 0.2413 |
| 0.139 | 0.1127 | 0.3547 | 0.3126 | 0.2991 | 0.2428 |
| 0.2942 | 0.0807 | 0.1285 | 0.2775 | 0.3311 | 0.2825 |
| 0.2559 | 0.2756 | 0.1652 | 0.1072 | 0.3383 | 0.3575 |
| 0.2708 | 0.2649 | 0.0961 | 0.1565 | 0.158 | 0.1981 |
| 0.4154 | 0.399 | 0.2483 | 0.1762 | 0.176 | 0.1543 |
| 0.3238 | 0.3771 | 0.4132 | 0.4602 | 0.352 | 0.1882 |
| 0.1742 | 0.4033 | 0.4999 | 0.393 | 0.3963 | 0.396 |
| 0.2029 | 0.1791 | 0.4768 | 0.5331 | 0.3739 | 0.4015 |
| 0.3828 | 0.1718 | 0.1657 | 0.4542 | 0.4772 | 0.3402 |
| 9- shape perimeter by squared area from measurement on petroleum rock samples | | | | | |
| 0.0903296 | 0.203654 | 0.204314 | 0.280887 | 0.197653 | 0.328641 |
| 0.148622 | 0.162394 | 0.262727 | 0.179455 | 0.326635 | 0.230081 |
| 0.183312 | 0.150944 | 0.200071 | 0.191802 | 0.154192 | 0.464125 |
| 0.117063 | 0.148141 | 0.14481 | 0.133083 | 0.276016 | 0.420477 |
| 0.122417 | 0.228595 | 0.113852 | 0.225214 | 0.176969 | 0.200744 |
| 0.167045 | 0.231623 | 0.91029 | 0.341273 | 0.438712 | 0.262651 |
| 0.189651 | 0.172567 | 0.240077 | 0.311646 | 0.163586 | 0.182453 |
| 0.164127 | 0.153481 | 0.161865 | 0.276016 | 0.253832 | 0.200447 |
| 10- snow fall dataset | | | | | |
| 0.03 | 0.02 | 0.015 | 0.045 | 0.1 | 0.1 |
| 0.125 | 0.19 | 0.39 | 0.11 | 0.07 | 0.01 |
| 0.055 | 0.22 | 0.08 | 0.005 | 0.125 | 0.035 |
| 0.085 | 0.06 | 0.01 | 0.065 | 0.02 | 0.26 |
| 0.03 | 0.015 | 0.025 | 0.01 | 0.495 | 0.085 |

Table (A) to be continued

| 11- milk production ||||||
|---|---|---|---|---|---|
| 0.4365 | 0.426 | 0.514 | 0.6907 | 0.7471 | 0.2605 |
| 0.6196 | 0.8781 | 0.499 | 0.6058 | 0.6891 | 0.577 |
| 0.5394 | 0.1479 | 0.2356 | 0.6012 | 0.1525 | 0.5483 |
| 0.6927 | 0.7261 | 0.3323 | 0.0671 | 0.2361 | 0.48 |
| 0.5707 | 0.7131 | 0.5853 | 0.6768 | 0.535 | 0.4151 |
| 0.6789 | 0.4576 | 0.3259 | 0.2303 | 0.7687 | 0.4371 |
| 0.3383 | 0.6114 | 0.348 | 0.4564 | 0.7804 | 0.3406 |
| 0.4823 | 0.5912 | 0.5744 | 0.5481 | 0.1131 | 0.729 |
| 0.0168 | 0.5529 | 0.453 | 0.3891 | 0.4752 | 0.3134 |
| 0.3175 | 0.1167 | 0.675 | 0.5113 | 0.5447 | 0.4143 |
| 0.5627 | 0.515 | 0.0776 | 0.3945 | 0.4553 | 0.447 |
| 0.5285 | 0.5232 | 0.6465 | 0.065 | 0.8492 | 0.8147 |
| 0.3627 | 0.3906 | 0.4438 | 0.4612 | 0.3188 | 0.216 |
| 0.6707 | 0.622 | 0.5629 | 0.4675 | 0.6844 | 0.3413 |
| 0.4332 | 0.0854 | 0.3821 | 0.4694 | 0.3635 | 0.4111 |
| 0.5349 | 0.3751 | 0.1546 | 0.4517 | 0.2681 | 0.4049 |
| 0.5553 | 0.5878 | 0.4741 | 0.3598 | 0.7629 | 0.5941 |
| 0.6174 | 0.686 | 0.0609 | 0.6488 | 0.2747 | |
| 12- COVID-19 recovery rate in Spain ||||||
| 0.667 | 0.5 | 0.5 | 0.4286 | 0.75 | 0.6531 |
| 0.5161 | 0.7895 | 0.7689 | 0.6873 | 0.52 | 0.7251 |
| 0.6375 | 0.6078 | 0.6289 | 0.5712 | 0.5923 | 0.6061 |
| 0.5924 | 0.5921 | 0.5592 | 0.5954 | 0.6164 | 0.6455 |
| 0.6725 | 0.6838 | 0.685 | 0.6947 | 0.721 | 0.7315 |
| 0.7412 | 0.7508 | 0.7519 | 0.7547 | 0.7645 | 0.7715 |
| 0.7759 | 0.7807 | 0.7838 | 0.7847 | 0.7871 | 0.7902 |
| 0.7934 | 0.7913 | 0.7962 | 0.7971 | 0.7977 | 0.8007 |
| 0.8038 | 0.8289 | 0.8322 | 0.8354 | 0.8371 | 0.8387 |
| 0.8456 | 0.849 | 0.8535 | 0.8547 | 0.8564 | 0.858 |
| 0.8604 | 0.8628 | 0.6586 | 0.707 | 0.7963 | 0.8516 |
| 13- voter turnout data set ||||||
| 0.92 | 0.76 | 0.88 | 0.68 | 0.47 | 0.53 |
| 0.66 | 0.62 | 0.85 | 0.64 | 0.69 | 0.75 |
| 0.76 | 0.58 | 0.70 | 0.81 | 0.63 | 0.67 |
| 0.73 | 0.53 | 0.77 | 0.55 | 0.57 | 0.90 |
| 0.63 | 0.79 | 0.82 | 0.78 | 0.68 | 0.49 |
| 0.66 | 0.53 | 0.72 | 0.87 | 0.45 | 0.86 |
| 0.68 | 0.65 | | | | |

Table (A) to be continued

| 14- unit capacity factors ||||||
|---|---|---|---|---|---|
| 0.853 | 0.759 | 0.866 | 0.809 | 0.717 | 0.544 |
| 0.492 | 0.403 | 0.344 | 0.213 | 0.116 | 0.116 |
| 0.092 | 0.070 | 0.059 | 0.048 | 0.036 | 0.029 |
| 0.021 | 0.014 | 0.011 | 0.008 | 0.006 | |